\documentclass[preprint,12pt,authoryear]{elsarticle}

\usepackage{amssymb}
\usepackage{amsmath}
\usepackage{tabularx}
\usepackage{booktabs}
\usepackage{subcaption}
\usepackage{lineno}
\usepackage{setspace}
\usepackage{xcolor}
\newcommand{\journalName}{Ocean engineering} % For all three files

\journal{\journalName}
\myfooter[C]{Accepted manuscript, not the publisher's version of record. \copyright\ 2026, CC BY-NC-ND 4.0. Ocean Engineering 350 (2026) 124202.}
\myfooterfont{\scriptsize}
\makeatletter
\def\ps@elsaccepted{%
  \let\@oddhead\@empty
  \let\@evenhead\@empty
  \def\@oddfoot{\hbox to \textwidth{\hfil\@myfooterfont\@elsarticlemyfooter\hfil}}%
  \let\@evenfoot\@oddfoot
}
\g@addto@macro\ps@pprintTitle{\def\@oddfoot{\hbox to \textwidth{\hfil\@myfooterfont\@elsarticlemyfooter\hfil}}\let\@evenfoot\@oddfoot}
\makeatother
\AtBeginDocument{\pagestyle{elsaccepted}}
\newif\ifrevision
\revisionfalse 
\newcommand{\rev}[1]{
  \ifrevision
    {\color{red}#1}
  \else
    #1
  \fi
}
\begin{document}

\begin{frontmatter}

  \title{\rev{AUWave: A Data-Driven Model for Reconstructing Significant Wave Heights Using Sparse Observations}}

  \author[0,1,2]{Hongyuan Shi} %% Author name
  \author[1]{Yilin Zhai \corref{cor1}}
  \author[1]{Ping Dong}
  \author[1,3]{Zaijin You}
  \author[1,2]{Chao Zhan}
  \author[1,2]{Qing Wang}
  \cortext[cor1]{Corresponding author}

  %% Author affiliation
  \affiliation[0]{
        organization={Shandong Key Laboratory Of Estuary And Coast \& Nuclear Environment},
        addressline={Ludong University},
        city={Shandong},
        postcode={264025},
        state={Yantai},
        country={China}
    }
  \affiliation[1]{
        organization={School of Hydraulic and Civil Engineering},
        addressline={Ludong University},
        city={Shandong},
        postcode={264025},
        state={Yantai},
        country={China}
    }
  
  \affiliation[2]{
        organization={Institute of Coastal Research},
        addressline={Ludong University},
        city={Yantai},
        postcode={264025},
        state={Shandong},
        country={China}
    }
    \affiliation[3]{
        organization={Navigation College},
        addressline={Dalian Maritime University},
        city={Dalian},
        postcode={116026},
        state={Liaoning},
        country={China}
    }
  %% Abstract
  \begin{abstract}
    Reconstructing high-resolution regional significant wave height (SWH) fields from sparse buoy observations is a critical challenge for ocean monitoring. We introduce AUWave, a hybrid deep learning framework that fuses a station-wise encoder with a multi-scale U-Net enhanced by self-attention to recover regional SWH fields. Trained and validated using NDBC buoy observations and ERA5 reanalysis over the Hawaii region, AUWave achieves high accuracy. It consistently outperforms a representative baseline, especially in configurations with more than a single buoy, demonstrating the benefit of its multi-scale architecture. Spatial error analysis shows performance is highest near observation sites, as expected. Further, buoy ablation studies identify critical anchor stations whose removal disproportionately degrades performance, offering actionable guidance for observational network design. AUWave provides a scalable pathway for gap-filling, creating high-resolution priors for data assimilation, and contingency reconstruction. Cross-basin evaluations in the Atlantic and Pacific confirm the model's robustness and portability, highlighting its potential for operational use across diverse ocean regimes.
  \end{abstract}

  %% Keywords

\end{frontmatter}

\section{Introduction}
\label{sec:introduction}
Significant wave height (SWH) is a fundamental descriptor of sea state and plays a critical role in maritime safety, coastal and offshore engineering, renewable energy siting, and climate-related studies \citep{CALOIERO2022110322,GAO2023120261,shi2023machine}. Accurate knowledge of SWH is essential for navigation risk assessment, offshore structural design, and wave energy exploitation, as well as for understanding ocean-atmosphere interactions and their contribution to climate variability \citep{hajinezhadian2023probabilistic,patra2024quantifying,seo2023ocean,azman2021structural,zhang2025novel,ibarra2023cmip6}.

Despite its importance, operational observations of ocean waves remain sparse and heterogeneous. In-situ buoy networks provide highly accurate measurements but are limited to discrete, irregularly distributed locations, often concentrated along coastlines or major shipping routes. Satellite altimeters offer broader spatial coverage; however, their sampling is constrained by orbital trajectories, resulting in discontinuous spatiotemporal coverage and relatively long revisit intervals at fixed locations \citep{le2025satellite,gao2020numerical,gao2017numerical}. Reanalysis products and global hindcasts, such as ERA5, provide spatially continuous wave fields but are generated by numerical models that may suffer from systematic errors and regional biases, particularly in areas with limited assimilated observations \citep{Wang04032022,10337180,riley2024ndbc}. Moreover, their inherent latency—ranging from several days to months—even for near-real-time products such as ERA5T, restricts their utility in rapidly evolving or operational marine environments \citep{hersbach2020era5}. Bridging the gap between sparse real-time observations and the need for spatially continuous, high-resolution wave fields therefore remains a practically important and scientifically challenging problem.
 
\rev{Beyond large-scale statistics, many coastal and offshore engineering applications are governed by highly localized and transient wave processes. Wave-structure and wave-topography interactions can induce pronounced spatial heterogeneity in wave fields, such as harbor resonance driven by incident wave groups, bathymetry-modulated long-wave transmission associated with Bragg scattering, and gap resonance occurring between closely spaced floating or fixed structures \citep{gao2023mechanism,gong2024hydrodynamics}.Under narrow-banded or focused wave group excitations, these mechanisms may lead to strong local amplification of wave height and energy, accompanied by nonlinear harmonic interactions and enhanced dissipation due to vortex shedding and wave-structure coupling\citep{gao2021investigation,mi2025gap,gao2026,gao2026influences}. Such extreme responses typically occur at specific locations and times rather than being spatially uniform, and their identification commonly relies on spatially dense wave gauge arrays or high-resolution numerical simulations to resolve the underlying resonance modes and transient evolution. In contrast, real-world monitoring systems usually consist of only a few sparsely distributed buoys, making it difficult to capture the spatial structure and extrema of the wave field and potentially leading to underestimation of critical conditions relevant to structural safety and operational risk.}

Classical mapping techniques, including spatial interpolation and data assimilation, attempt to mitigate observational sparsity but generally rely on prescribed covariance structures and linear assumptions. Their performance tends to deteriorate when observations are scarce, unevenly distributed, or located far from dynamically active regions \citep{asch2016data,fletcher2017data}. Physics-based spectral wave models provide dynamically consistent representations of wave evolution and are widely used for forecasting and hindcasting. However, their accuracy depends strongly on the quality of external forcing (e.g., wind fields), calibration of parameterizations, and substantial computational resources, which limits their applicability for rapid, high-resolution regional updates under sparse observational constraints \citep{gaffet2025new,yao2025numeric}.

In recent years, deep learning methods have emerged as promising alternatives for geophysical field reconstruction. By learning complex, nonlinear mappings directly from data, neural networks can bypass explicit assumptions about covariance or governing dynamics and instead exploit statistical regularities present in large datasets. In the geosciences, deep learning has demonstrated strong performance in image-to-image translation, super-resolution, and geophysical field reconstruction, indicating its ability to capture both local features and large-scale spatial dependencies beyond the capabilities of traditional methods \citep{huang2024underwater,ijjeh2023deep,song2021wavefield}.

Several studies have recently explored deep learning approaches for reconstructing spatially continuous wave fields from sparse buoy observations. \citet{han2023local} proposed DGWBNet, which combines conditional variational autoencoders with adversarial learning to jointly estimate SWH, wave period, and direction while providing uncertainty quantification. \citet{duan2024reconstruction} developed the RWR model using fully connected and convolutional neural networks to reconstruct regional SWH fields and examined the influence of boundary conditions and buoy density on reconstruction accuracy. \citet{lv2025new} further integrated U-Net and generative adversarial networks within an Actor–Critic framework to enable simultaneous observation-point optimization and multidimensional wave field reconstruction.

While these approaches have advanced the state of wave field reconstruction, several limitations remain. Models based on variational inference or reinforcement learning frameworks often involve high training complexity and computational cost, which may be unnecessary for deterministic reconstruction tasks. Conversely, simpler feedforward architectures, although computationally efficient, tend to struggle in capturing the non-local spatial dependencies inherent to wave propagation due to the localized nature of standard convolutional operations. In addition, observation layout optimization frameworks are difficult to deploy in many practical scenarios where sensor locations are fixed.

To address these challenges, this study proposes the AUWave model, which combines multilayer perceptron and U-Net architectures to more effectively capture both global and local spatial features under fixed buoy layouts and single-parameter reconstruction settings. Furthermore, this work introduces automated hyperparameter and architectural optimization using Optuna, enabling systematic exploration of model configurations to enhance reconstruction accuracy and stability while reducing reliance on manual, experience-based tuning.

The main contributions of this paper are as follows:
\begin{itemize}
	\item A novel network architecture is proposed, which effectively integrates Multilayer Perceptron (MLP) with U-Net. This design preserves the convolutional structure's ability to capture local features while compensating for the shortcomings in modeling non-local dependencies through MLP.
	\item Automated hyperparameter and architectural optimization is introduced. For the first time in wave field reconstruction tasks, the Optuna framework is employed to systematically search for optimal network hyperparameters and structural configurations, overcoming the limitations of manual tuning.
	\item Improved reconstruction accuracy and stability are achieved. Under identical data and observation conditions, AUWave significantly outperforms existing methods, demonstrating its effectiveness and generalizability in sparse observation scenarios.
	\item Cross-regional transfer validation is conducted. Beyond the Hawaii source domain, we evaluate AUWave under domain shift over two additional basins (Atlantic and Pacific) to assess out-of-domain generalization without task-specific re-engineering.
\end{itemize}

The remainder of the paper is organized as follows. Section~\ref{sec:methods} describes the study area, data, preprocessing, model architecture, and hyperparameter optimization, and outlines the cross-regional transfer protocol. Section~\ref{sec:results} presents ablation studies, spatial-temporal error analyses, buoy-configuration sensitivity, and transfer validation over Atlantic and Pacific, with comparisons to the RWR baseline where applicable. Section~\ref{sec:discussion} discusses implications, limitations, and future directions. Section~\ref{sec:conclusion} concludes.

\section{Methods}
\label{sec:methods}

\subsection{Study Area and Data}
\label{subsec:study_area_and_data}
\begin{figure}[htbp]
  \begin{subfigure}[b]{0.32\textwidth}
    \centering
    \includegraphics[width=\textwidth]{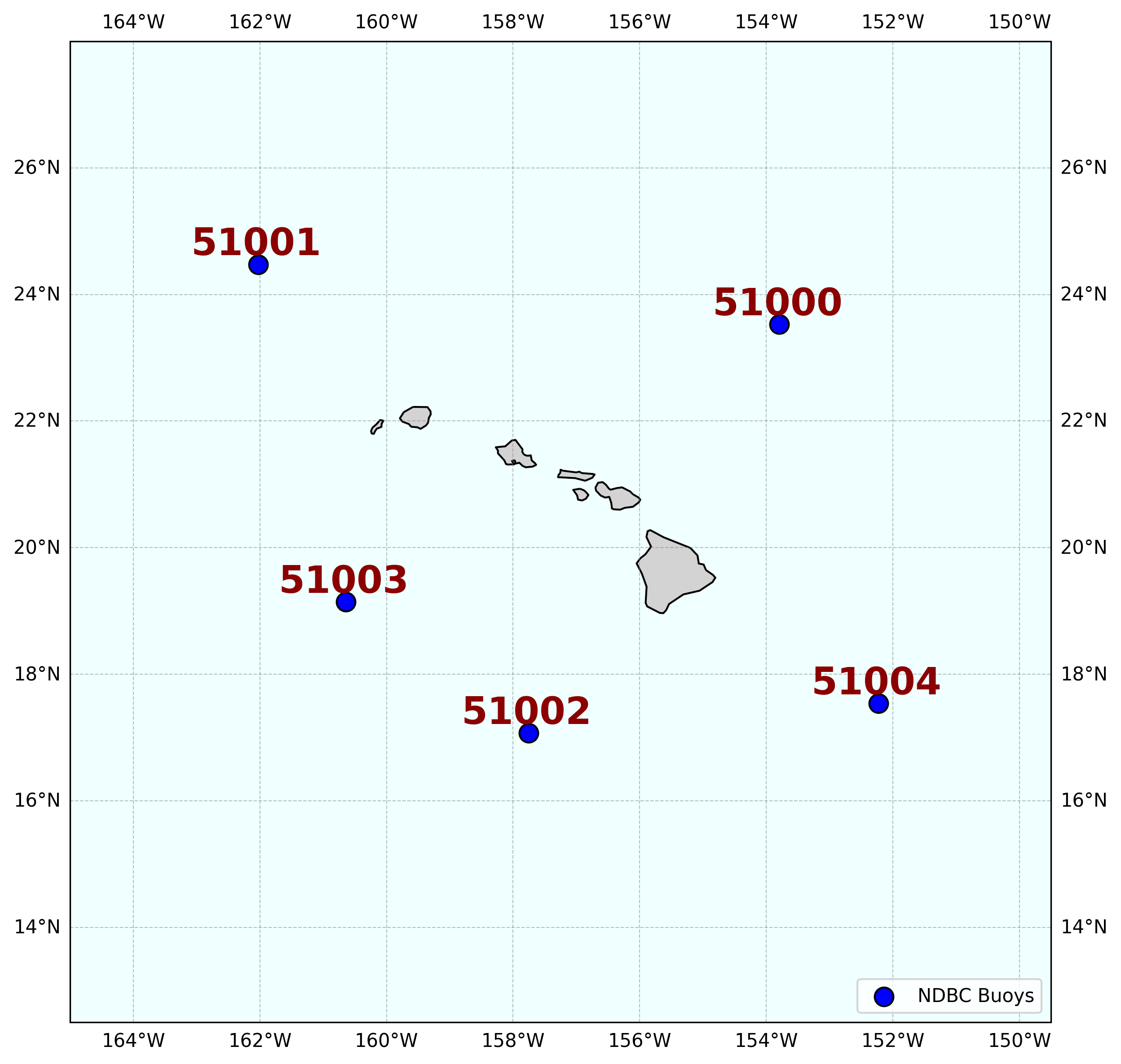}
    \caption{Hawaii}
    \label{fig:hawaii}
  \end{subfigure}
  \begin{subfigure}[b]{0.32\textwidth}
    \centering
    \includegraphics[width=\textwidth]{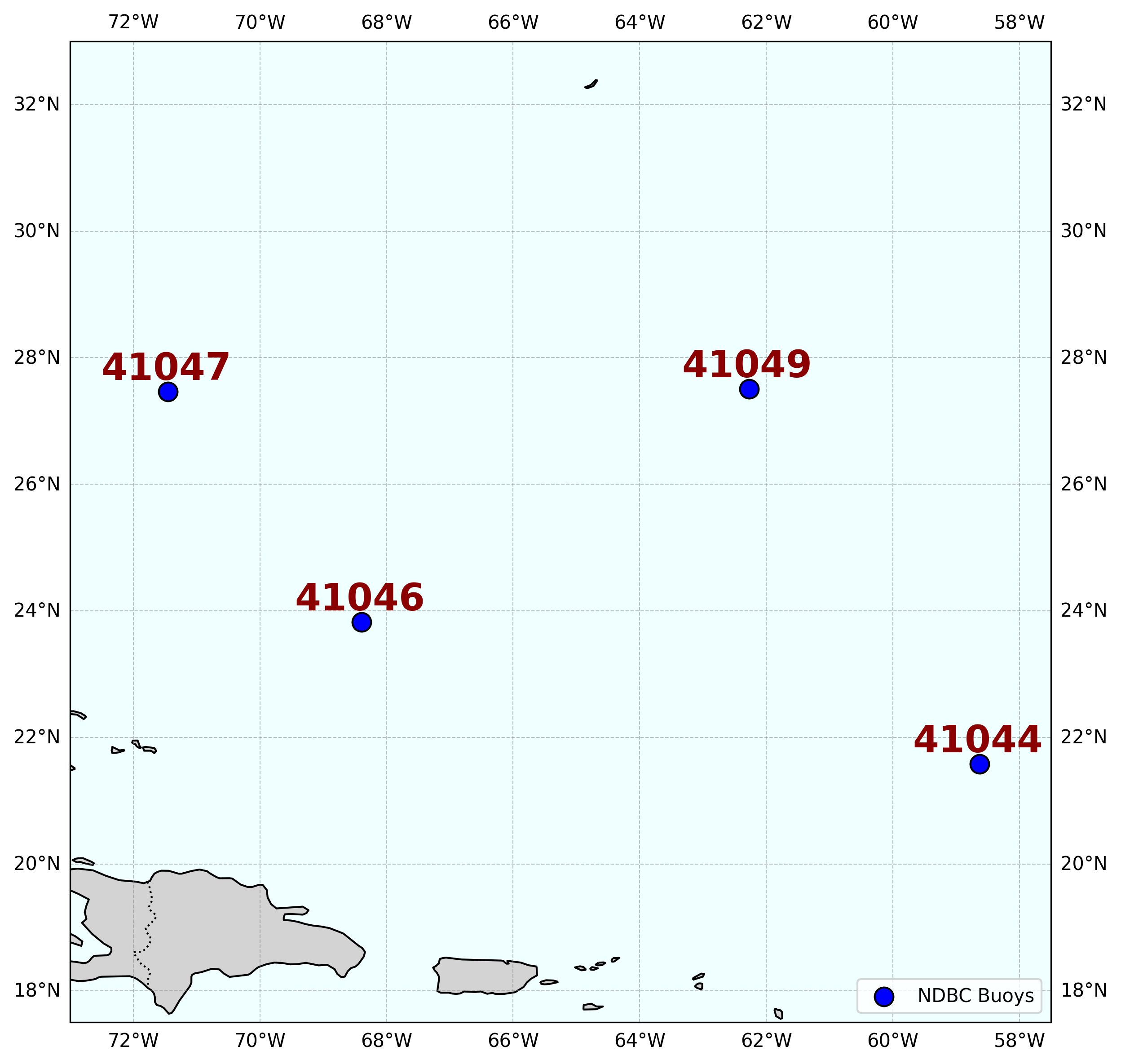}
    \caption{Atlantic}
    \label{fig:atlantic}
  \end{subfigure}
  \begin{subfigure}[b]{0.32\textwidth}
    \centering
    \includegraphics[width=\textwidth]{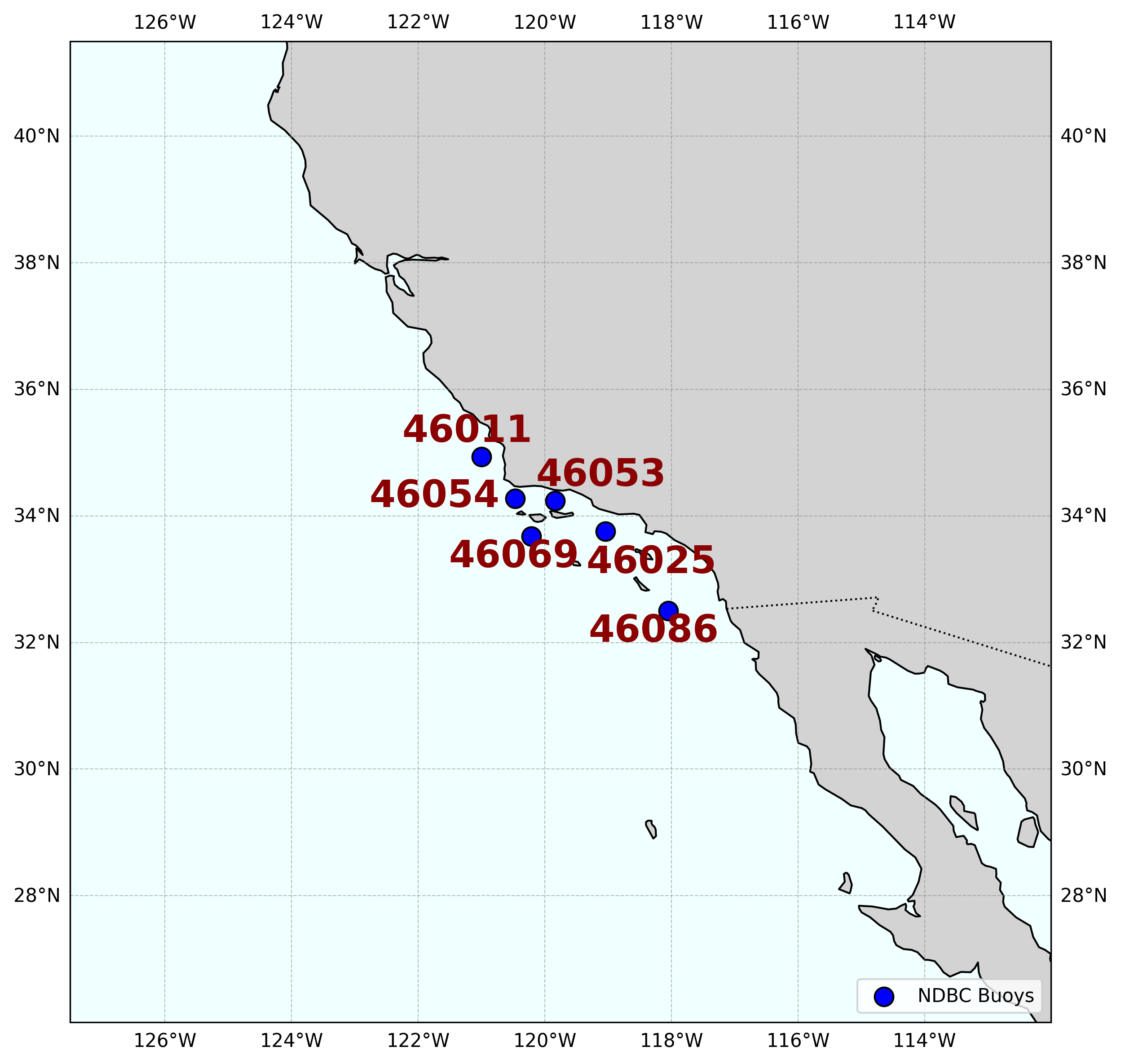}
    \caption{Pacific}
    \label{fig:pacific}
  \end{subfigure}
  \caption{Study Areas and Buoy Locations}
  \label{fig:study-areas}
\end{figure}

The dataset used for model training and evaluation comprises SWH measurements obtained from two primary sources: (1) sparse in-situ observations provided by the National Data Buoy Center (NDBC), and (2) gridded field data (32$\times$32) derived from the ERA5 reanalysis product. The ERA5 dataset offers a spatial resolution of 0.5° and a temporal resolution of 1 hour, ensuring consistency across study regions and enabling robust model evaluation (see Figure~\ref{fig:study-areas}).

We consider three regional domains: Hawaii (source region for model development), the mid-latitude North Atlantic, and the North Pacific. Unless otherwise stated, model selection and tuning are conducted on Hawaii only; Atlantic and Pacific are used to assess cross-regional transfer under domain shift without re-tuning. All regions use the same spatiotemporal sampling (ERA5 0.5° hourly products and co-located NDBC buoys) and are mapped to a unified 32$\times$32 grid per domain to ensure architectural compatibility.

\subsection{Data Preprocessing}
\label{sec:data-preprocessing}

The raw NDBC buoy records span the period from 2022 to 2023. To ensure temporal consistency and spatial comparability between in-situ observations and ERA5 reanalysis, both datasets were systematically processed to enable direct comparison and joint analysis. First, quality-controlled buoy measurements were extracted from the National Data Buoy Center (NDBC), and records with missing SWH values were removed. Since reporting times across stations are not perfectly synchronized (e.g., observations at different minutes within an hour), all timestamps were rounded to the nearest full hour. Duplicate entries were subsequently aggregated by \texttt{[timestamp, station\_id]}, retaining the first record.

ERA5 reanalysis data were simultaneously pre-processed to extract SWH values on a latitude-longitude grid. To align the two datasets, the intersection of timestamps common to both buoy records and ERA5 outputs was identified and used as the shared temporal index. Based on this unified time axis, ERA5 data were organized into a four-dimensional array with dimensions corresponding to \texttt{[time, variable, latitude, longitude]}, where each record was mapped to its grid cell according to the spatial indices of latitude and longitude. The buoy data were aligned to the same time index, forming a two-dimensional array with rows representing time steps and columns representing stations.  

For strict temporal consistency, only time steps with complete buoy observations across all stations were retained. ERA5 data were chronologically sorted, and missing values in numerical fields were filled with zeros to mitigate sparsity. 

Finally, the aligned dataset was split in chronological order: the first 70\% of samples were used as the training set, the middle 15\% as the testing set for hyperparameter tuning and model architecture search, and the final 15\% as the validation set for independent performance evaluation.

\subsection{AUWave Model Architecture}
\label{sec:model-architecture}

\begin{figure}[htbp]
  \centering
  \includegraphics[width=0.99\textwidth]{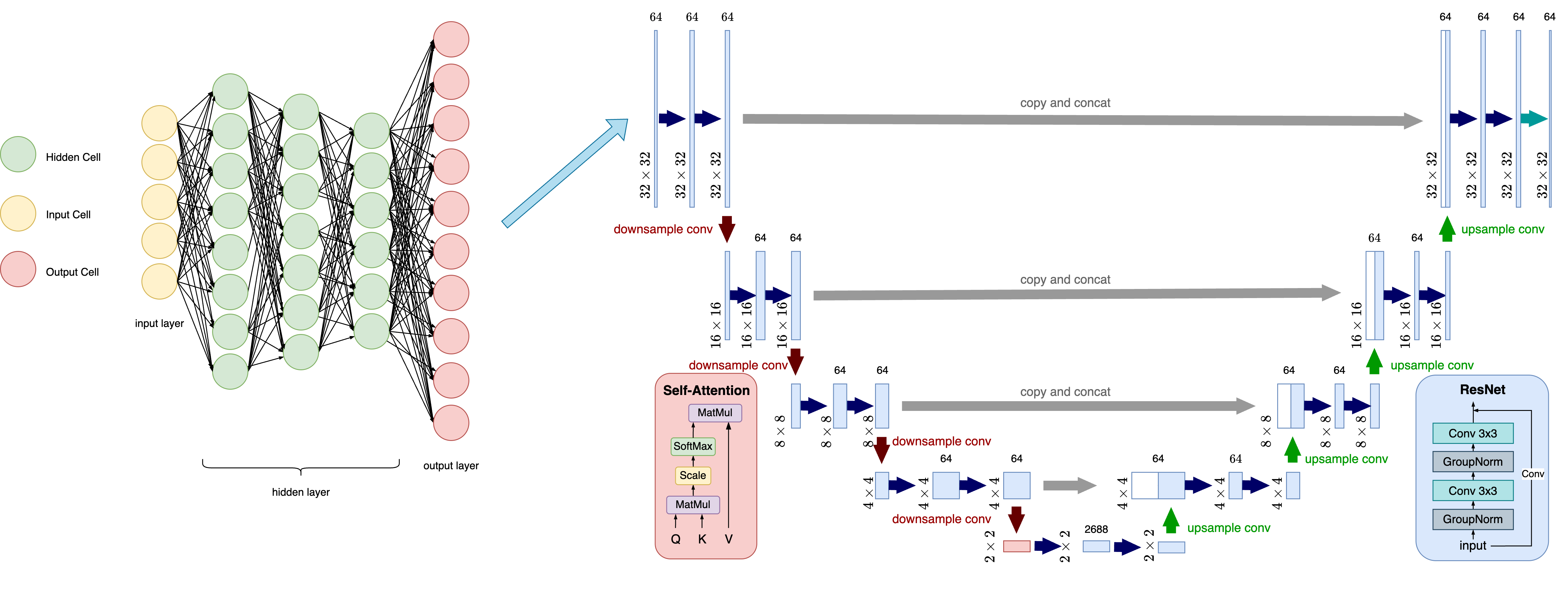}
  \caption{AUWave architecture}
  \label{fig:unet-architecture}
\end{figure}

To reconstruct 2D wave fields from sparse time series, we develop a deep learning model named AUWave (see Figure~\ref{fig:unet-architecture}), which integrates a MLP, a U-Net encoder-decoder, and self-attention mechanisms. The model is implemented using PyTorch and PyTorch Lightning frameworks. It takes as input a sparse wave sequence represented by a one-dimensional vector of size $1 \times n$, where $n$ denotes the total number of buoy stations in the study region and each entry corresponds to the SWH at a given station and time. The model outputs a $32 \times 32$ single-channel wave field grid.

The MLP module consists of three fully connected layers with hidden dimensions [896, 832, 800], projecting the input into a 1792-dimensional latent space. Batch normalization and LeakyReLU activation (slope = 0.2) are applied at each layer. A linear projection then reshapes the encoded vector into an initial 2D wave field of size 1x32x32 \citep{xu2020reluplex}.

The core network adopts a canonical U-Net-like architecture featuring a symmetric encoder-decoder topology with five hierarchical downsampling and upsampling stages.  This configuration enables progressive abstraction of spatial information during encoding and faithful reconstruction of high-resolution outputs during decoding.  The encoder channel widths are defined as [64, 64, 64, 64, 2688], gradually increasing representational capacity while reducing spatial resolution, whereas the decoder mirrors this configuration with [2688, 64, 64, 64, 64], facilitating the recovery of fine-scale features through successive upsampling \citep{he2016deep}.

Each stage is composed of three residual blocks based on the ResNet paradigm, which are known to alleviate vanishing gradient issues and allow for efficient training of deeper networks.  Within each residual block, 3x3 convolutional filters are employed to capture local spatial correlations, followed by group normalization with 32 groups to stabilize feature distributions and improve generalization across varying batch sizes.  The SiLU (Sigmoid Linear Unit) activation function is subsequently applied, offering smooth, non-monotonic nonlinearity that improves optimization dynamics compared to ReLU-like activations \citep{he2020resnet,wu2022learning,ramachandran2017searching}.

To enhance global context modeling, a self-attention mechanism is incorporated at the U-Net bottleneck \citep{vaswani2023attentionneed}.

The model is trained using the mean squared error (MSE) loss. To ensure numerical stability and improve model training efficiency, all input and output data were transformed using a log(1+x) function. After prediction, the results were denormalized to recover the original wave height values. Key hyperparameters, such as hidden layer dimensions and learning rate, are optimized using the Optuna framework (see Section~\ref{sec:hyperopt} for details).

\begin{equation}
\mathrm{RMSE} = \sqrt{\frac{1}{n} \sum_{i=1}^{n} \left( y_i - \hat{y}_i \right)^{2}}
\end{equation}

\begin{equation}
x' = \log(1+x)
\end{equation}

\begin{equation}
x = \exp(x') - 1
\end{equation}

\subsection{Hyperparameter Optimization}
\label{sec:hyperopt}
To determine the optimal configuration for the proposed wave field reconstruction model AUWave, we employed a systematic and automated hyperparameter optimization approach. The model's performance—particularly its generalization ability—is highly sensitive to key hyperparameters such as the learning rate and network architecture. Therefore, we used the Optuna framework to perform an efficient search over a well-defined space. The optimization objective was to minimize the MSE on the validation set.

The optimization process was implemented using the integration between PyTorch Lightning and Optuna, which efficiently manages training, evaluation, and resource allocation for each trial. A comprehensive and flexible search space was defined for key hyperparameters, as summarized in Table~\ref{tab:hyperparameters}. In particular, for the U-Net architecture, we adopted a dynamic strategy: both the depth (number of blocks) and the number of channels per layer were variable, and the possibility of incorporating attention mechanisms was also explored.

To balance exploration and exploitation, we adopted the Tree-structured Parzen Estimator (TPE) as the core sampling algorithm. TPE is a Bayesian optimization method that constructs probabilistic models from past trials to suggest promising hyperparameter sets. To further improve efficiency, we employed a Median Pruner, which compares the current trial's loss with the median of completed trials at the same step. Trials performing significantly worse than the median were terminated early to focus computational resources on more promising configurations \citep{watanabe2023tree}.

A total of 500 optimization trials were conducted. Each trial was trained for up to 100000 epochs, with early stopping enabled: if the validation loss did not improve within 100 consecutive epochs, the training was halted. After exhaustive tuning, we obtained the optimal hyperparameter combination, achieving a minimum validation loss of 0.043285. This configuration was adopted for all subsequent experiments and final model training.

\begin{table}[htbp]
  \centering
  \footnotesize
  \begin{tabularx}{\linewidth}{@{} l l l X @{}}
    \toprule

    \textbf{Category}  & \textbf{Hyperparameter} & \textbf{Search Space} & \textbf{Description}                                             \\
    \midrule
    \textbf{Optimizer} & Learning Rate           & $[10^{-5}, 10^{-3}]$  & Controls the update step size of model parameters.               \\
                       & Scheduler Gamma         & $[0.9, 0.99]$         & Decay factor for the StepLR scheduler.                           \\
    \midrule
    \textbf{MLP}       & Layers                  & $[1, 4]$              & Number of layers in MLP                                          \\
                       & Hidden Dimensions       & $[128, 4096]$         & Dimension of the fully connected layers in the sequence encoder. \\
                       & Latent Dimension        & $[256, 2048]$         & Dimension of the encoded features.                               \\
    \midrule
    \textbf{U-Net}     & Number of Blocks        & $[3, 5]$              & Number of down-sampling and up-sampling blocks.                  \\
                       & Layers per Block        & $[1, 3]$              & Number of convolutional layers within each U-Net block.          \\
                       & Output Channels         & $[32, 8192]$          & Output channels for each block in the U-Net.                     \\
                       & Attention Mechanism     & [True, False]         & Enables the use of attention modules within the U-Net.           \\
    \bottomrule
  \end{tabularx}
  \caption{Hyperparameter Search Space}
  \label{tab:hyperparameters}
\end{table}

\captionsetup[subfigure]{labelformat=simple,labelsep=space}
\renewcommand\thesubfigure{\alph{subfigure}}

\newcounter{innersubfigure}[subfigure]
\renewcommand\theinnersubfigure{\alph{subfigure}.\roman{innersubfigure}}

\newcommand{\innersubcaption}[2]{%
  \refstepcounter{innersubfigure}%
  \par\begingroup
    \scriptsize
    \makebox[\linewidth]{\textbf{(\theinnersubfigure)}~#1}%
    \label{#2}%
  \par\endgroup
}

\section{Results}
\label{sec:results}

\subsection{Temporal Variability of Reconstruction Errors}
\label{sec:temporal-variability}
To visually assess the model's performance, we compare the true and predicted wave fields for five randomly selected examples from the Hawaii dataset. Each example includes three subplots: (a) the true wave field, (b) the predicted wave field, and (c) the error map with the Root Mean Square Error (RMSE) value indicated in the title. The comparisons are presented in Figures~\ref{fig:hawaii_ex1} to \ref{fig:hawaii_ex5}. The locations of the sparse input buoys are marked with red crosses ('x'), and land areas are masked in white. And \rev{the time steps shown in Figures~\ref{fig:hawaii_all}, \ref{fig:atlantic_all}, and \ref{fig:pacific_all} are representative examples selected from the evaluation datasets to qualitatively illustrate model behavior across different reconstruction error levels, rather than being chosen to highlight specific events or optimal performance.}

These figures demonstrate  the model is generally successful in capturing the main spatial characteristics of the ocean wave field. For instance, in cases with low overall error, such as the examples with RMSE values of 0.1463 and 0.1647 (Figure~\ref{fig:hawaii_ex5} and Figure~\ref{fig:hawaii_ex3}), the predicted field exhibits a high degree of similarity to the ground truth. The model accurately reproduces the location, shape, and magnitude of the primary wave amplitude crests and troughs.

Conversely, the figure also includes cases where the reconstruction is less accurate. The example with the highest error (RMSE = 0.3574, Figure~\ref{fig:hawaii_e1_err}) demonstrates notable discrepancies. While the general large-scale pattern is still discernible, the error map reveals significant localized areas of both overestimation (red patches) and underestimation (dark blue patches). These larger errors predominantly occur in regions distant from the sparse buoy locations, underscoring the inherent challenge of interpolating complex oceanographic features over data-sparse areas. Intermediate cases, such as those with RMSE values of 0.2097 and 0.2183(Figure~\ref{fig:hawaii_ex2} and Figure~\ref{fig:hawaii_ex4}), represent a typical level of performance, where the main wave structures are correctly identified but with some inaccuracies in amplitude and fine-scale details.

\begin{figure}[htbp]
  \centering

  \begin{subfigure}[t]{\textwidth}
    \phantomsubcaption\label{fig:hawaii_ex1}
    \centering
    \begin{minipage}[t]{0.32\textwidth}
      \centering
      \includegraphics[width=\linewidth]{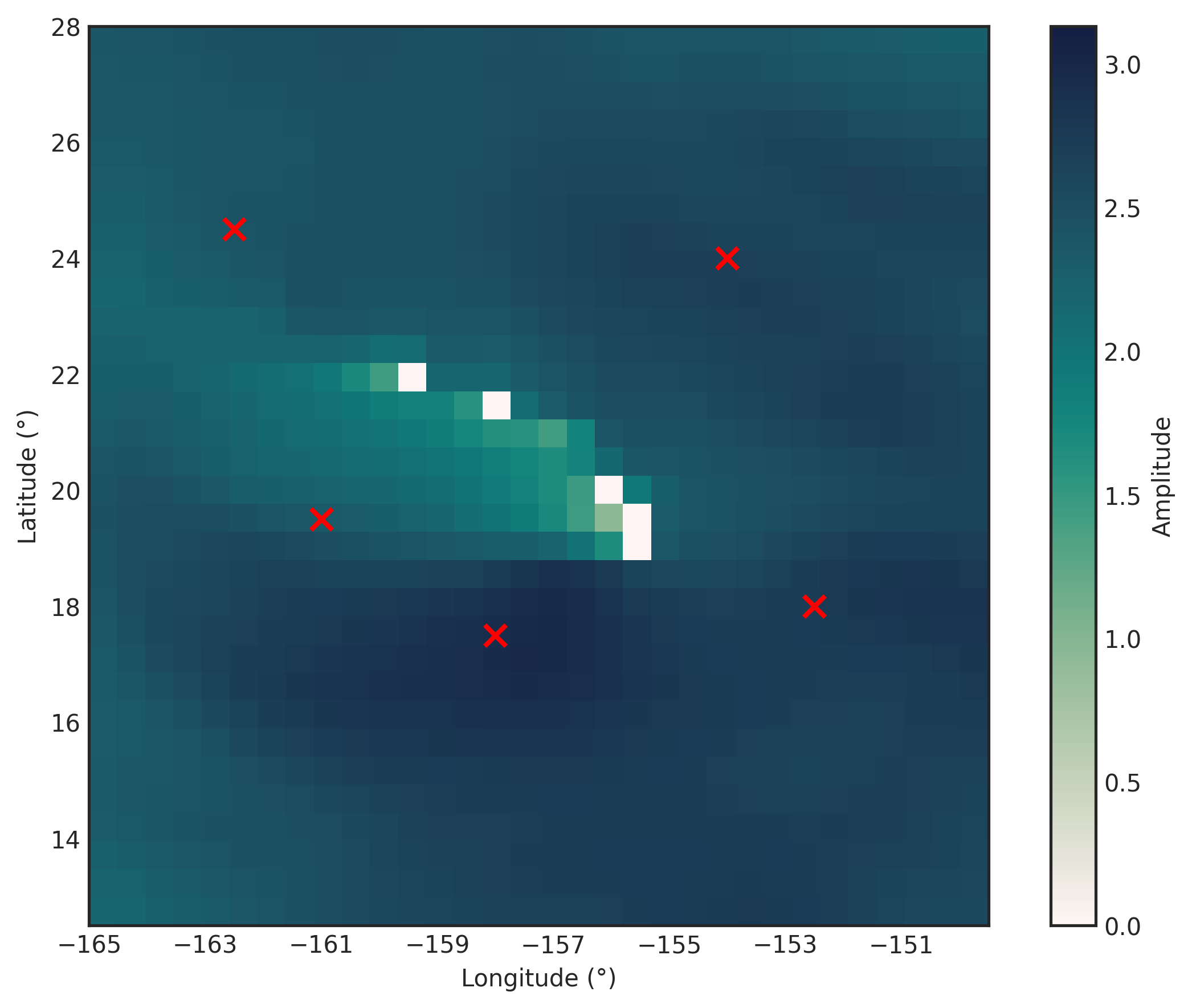}
      \innersubcaption{True}{fig:hawaii_e1_true}
    \end{minipage}\hfill
    \begin{minipage}[t]{0.32\textwidth}
      \centering
      \includegraphics[width=\linewidth]{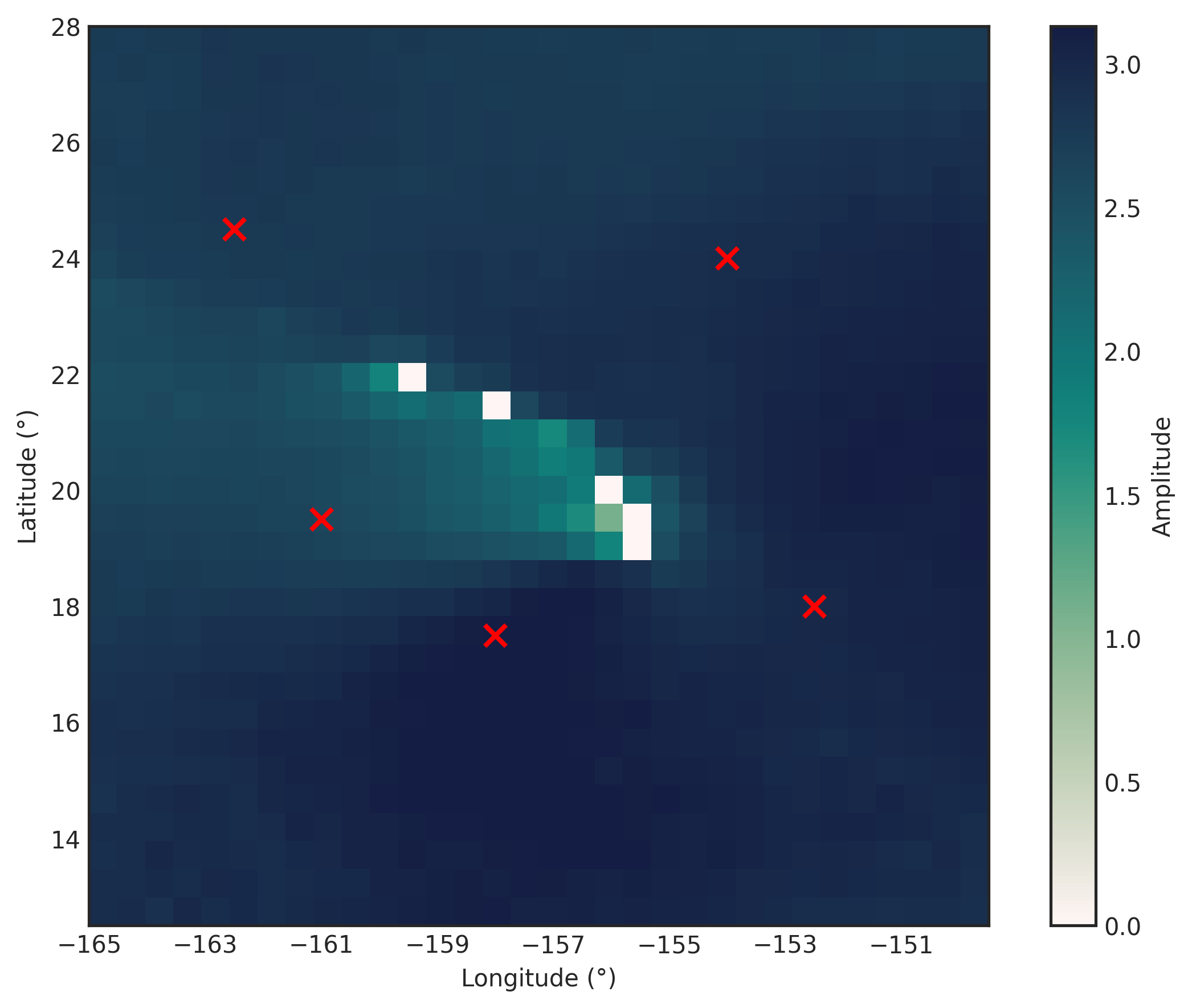}
      \innersubcaption{Predicted}{fig:hawaii_e1_pred}
    \end{minipage}\hfill
    \begin{minipage}[t]{0.32\textwidth}
      \centering
      \includegraphics[width=\linewidth]{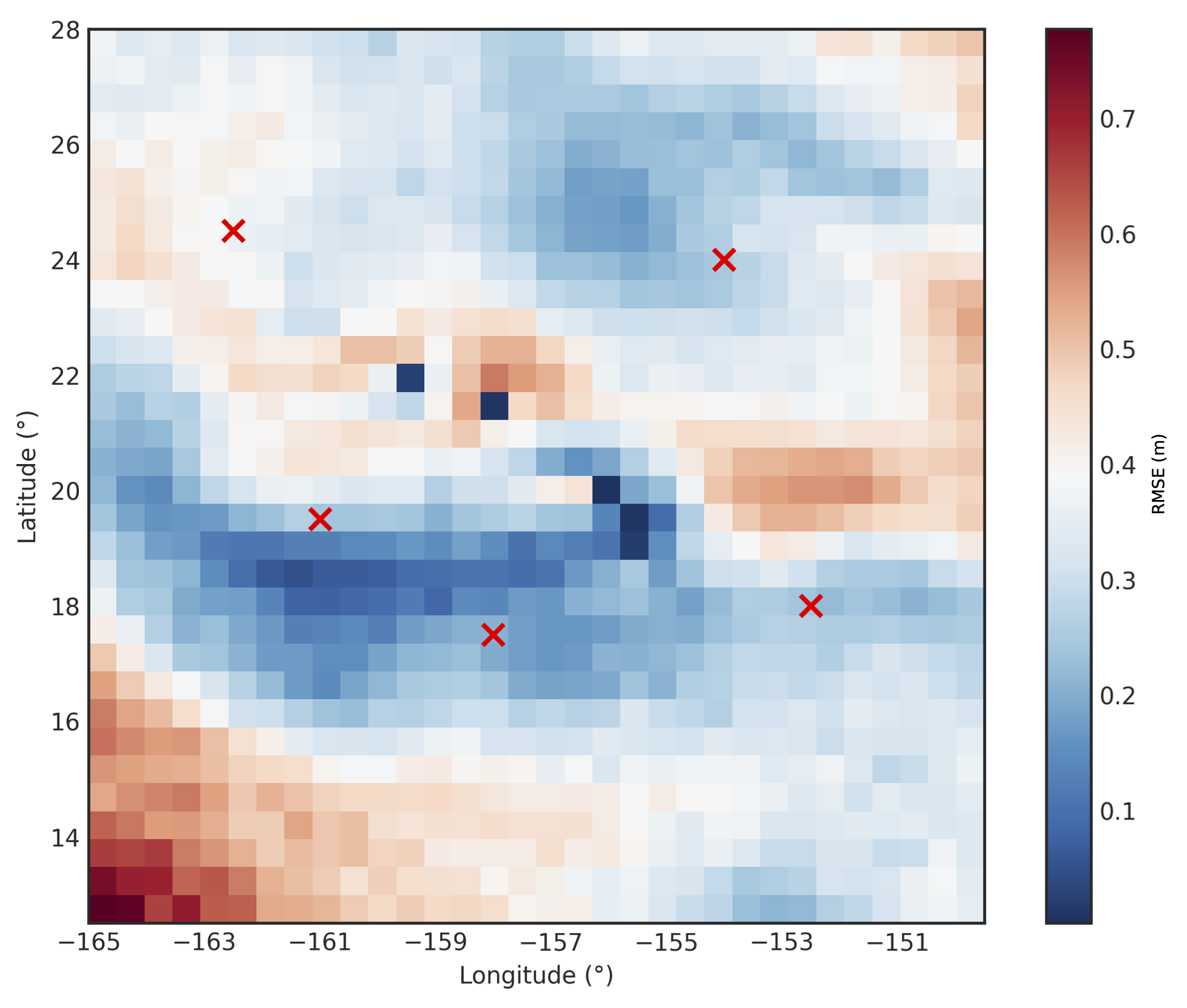}
      \innersubcaption{Error (RMSE=0.3574)}{fig:hawaii_e1_err}
    \end{minipage}

    \scriptsize \vspace{2pt}\par\itshape 	Snapshot at time step 1
  \end{subfigure}

  \begin{subfigure}[t]{\textwidth}
    \refstepcounter{subfigure}\label{fig:hawaii_ex2}
    \setcounter{innersubfigure}{0}
    \centering
    \begin{minipage}[t]{0.32\textwidth}
      \includegraphics[width=\linewidth]{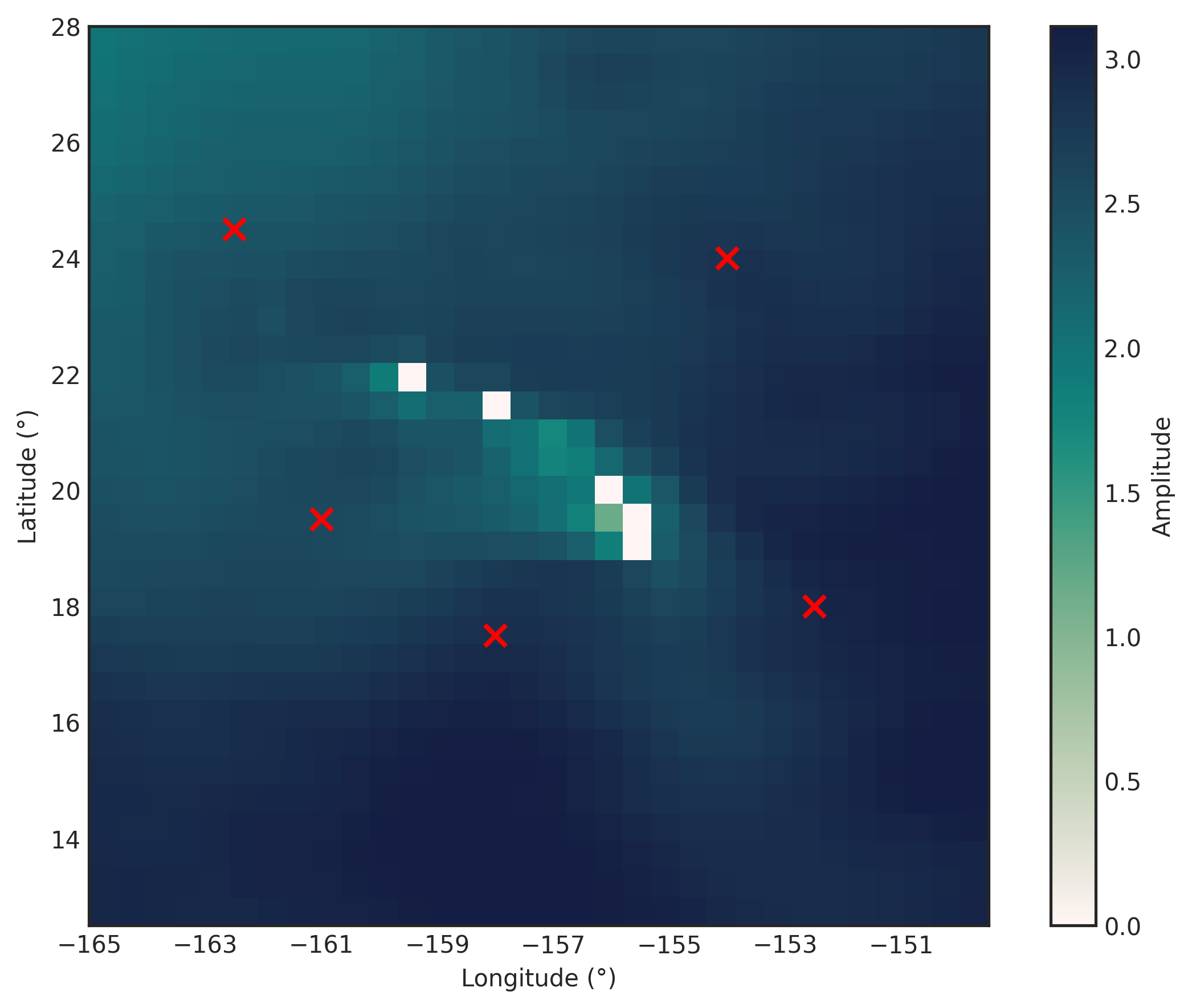}
      \innersubcaption{True}{fig:hawaii_e2_true}
    \end{minipage}\hfill
    \begin{minipage}[t]{0.32\textwidth}
      \includegraphics[width=\linewidth]{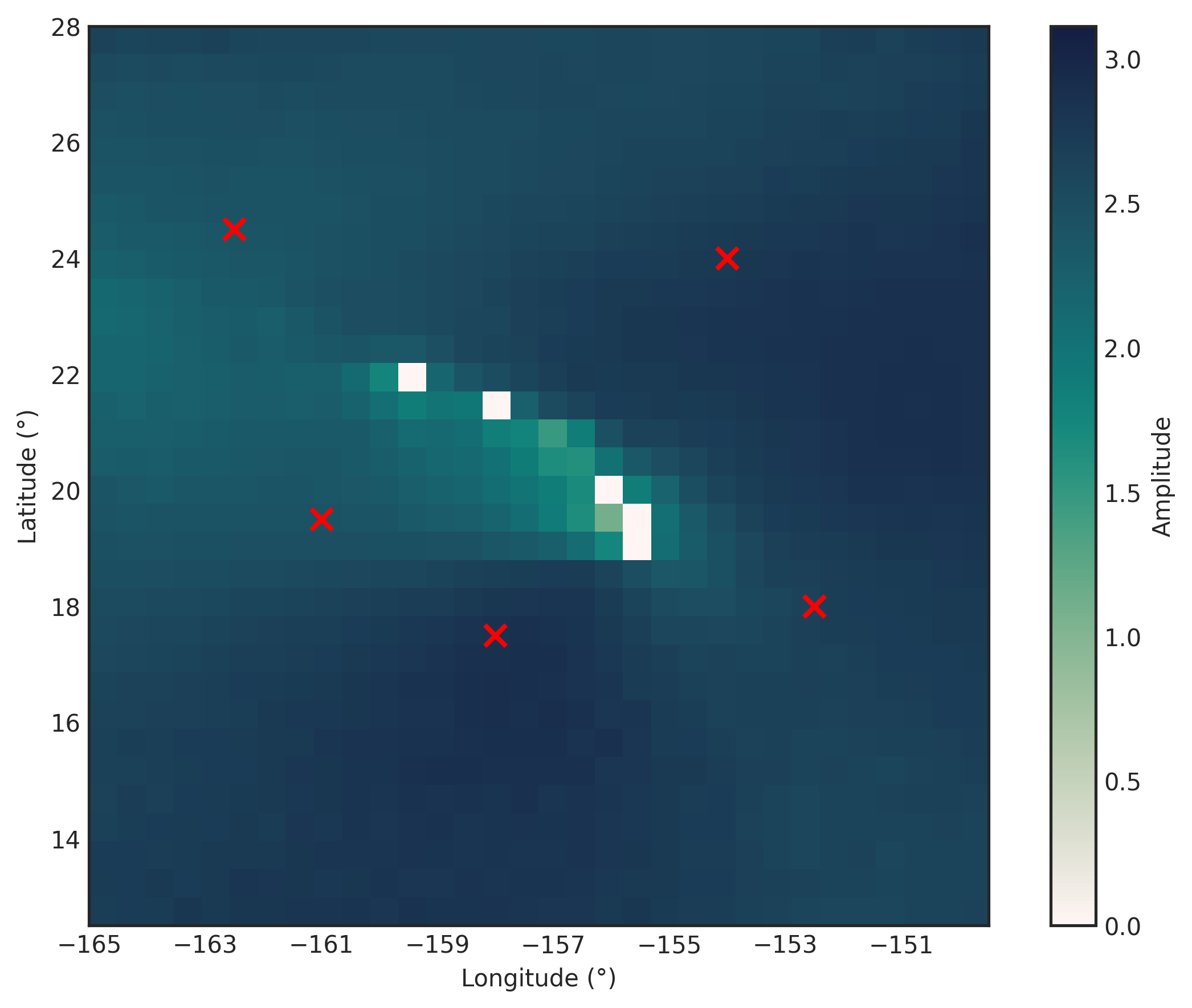}
      \innersubcaption{Predicted}{fig:hawaii_e2_pred}
    \end{minipage}\hfill
    \begin{minipage}[t]{0.32\textwidth}
      \includegraphics[width=\linewidth]{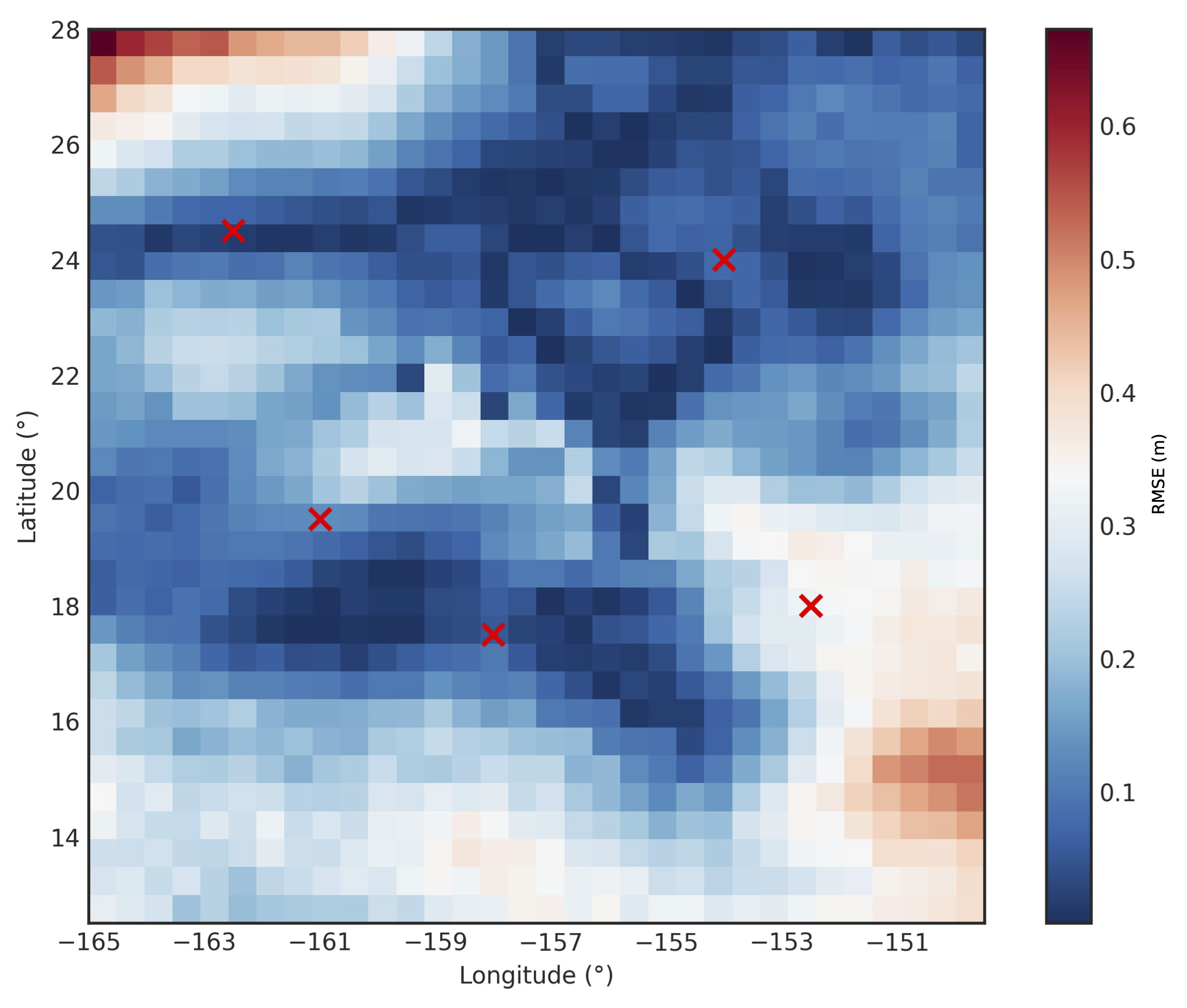}
      \innersubcaption{Error (RMSE=0.2097)}{fig:hawaii_e2_err}
    \end{minipage}
    \scriptsize \vspace{2pt}\par\itshape 	Snapshot at time step 2
  \end{subfigure}

  \begin{subfigure}[t]{\textwidth}
    \refstepcounter{subfigure}\label{fig:hawaii_ex3}
    \setcounter{innersubfigure}{0}
    \centering
    \begin{minipage}[t]{0.32\textwidth}
      \includegraphics[width=\linewidth]{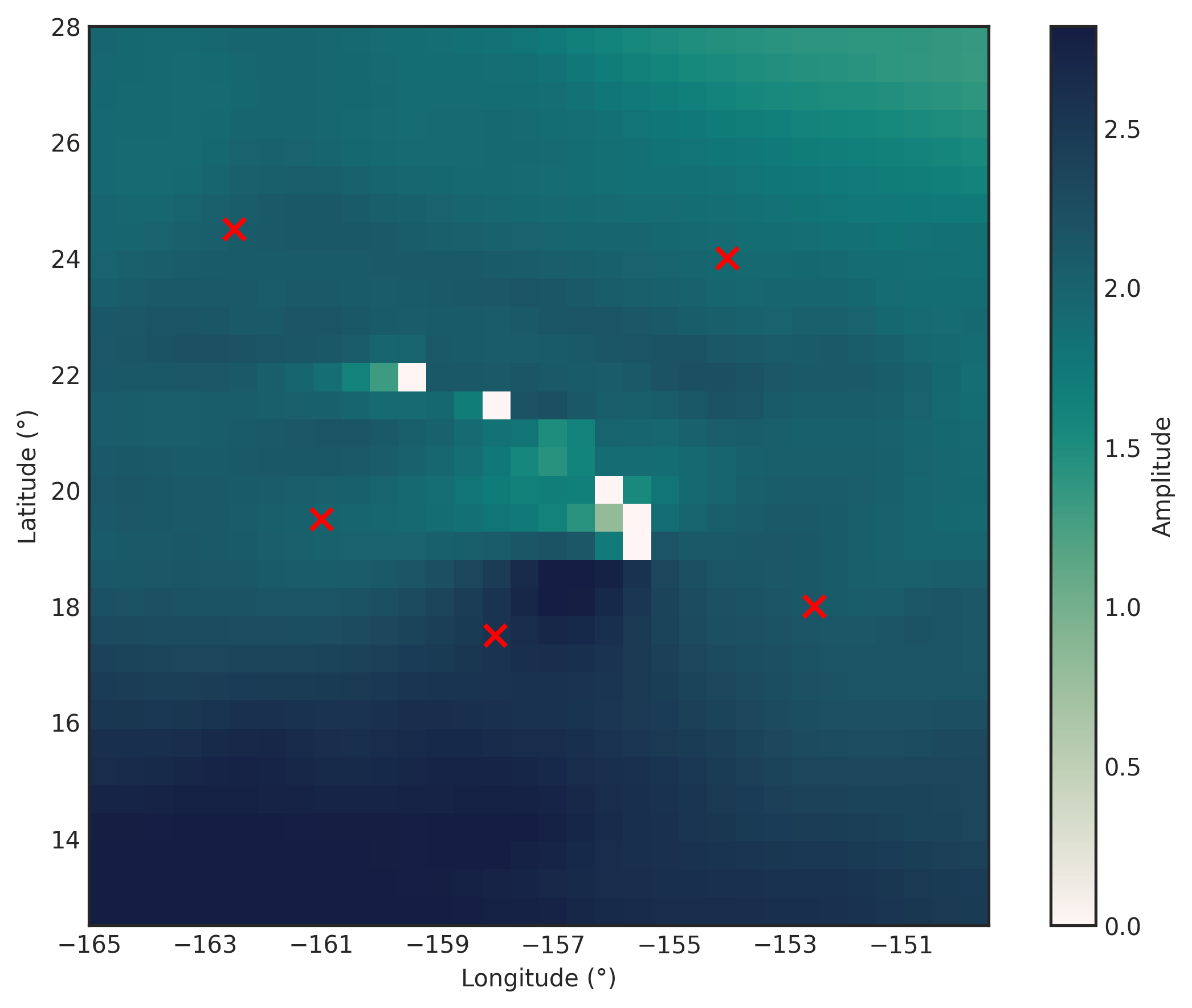}
      \innersubcaption{True}{fig:hawaii_e3_true}
    \end{minipage}\hfill
    \begin{minipage}[t]{0.32\textwidth}
      \includegraphics[width=\linewidth]{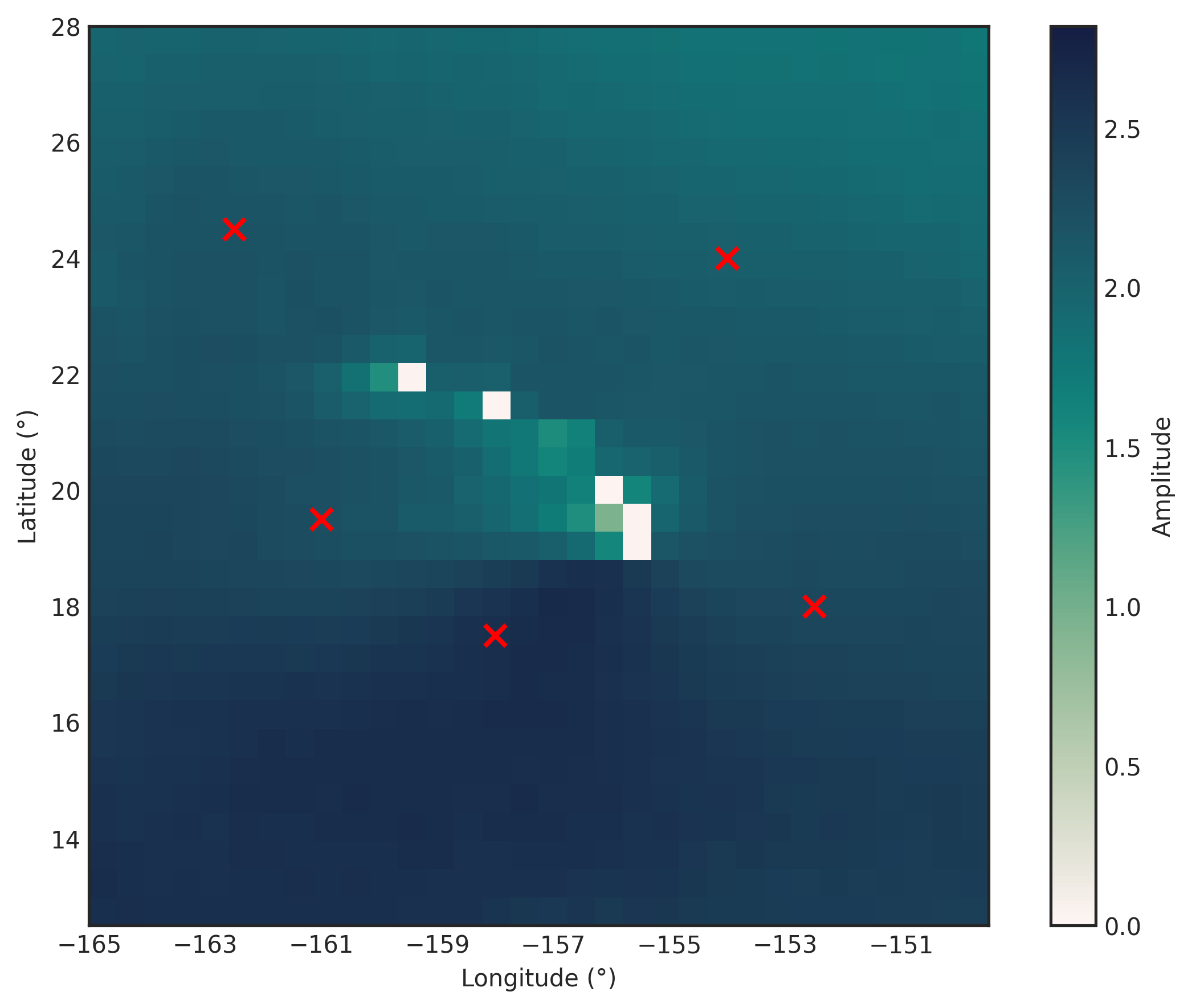}
      \innersubcaption{Predicted}{fig:hawaii_e3_pred}
    \end{minipage}\hfill
    \begin{minipage}[t]{0.32\textwidth}
      \includegraphics[width=\linewidth]{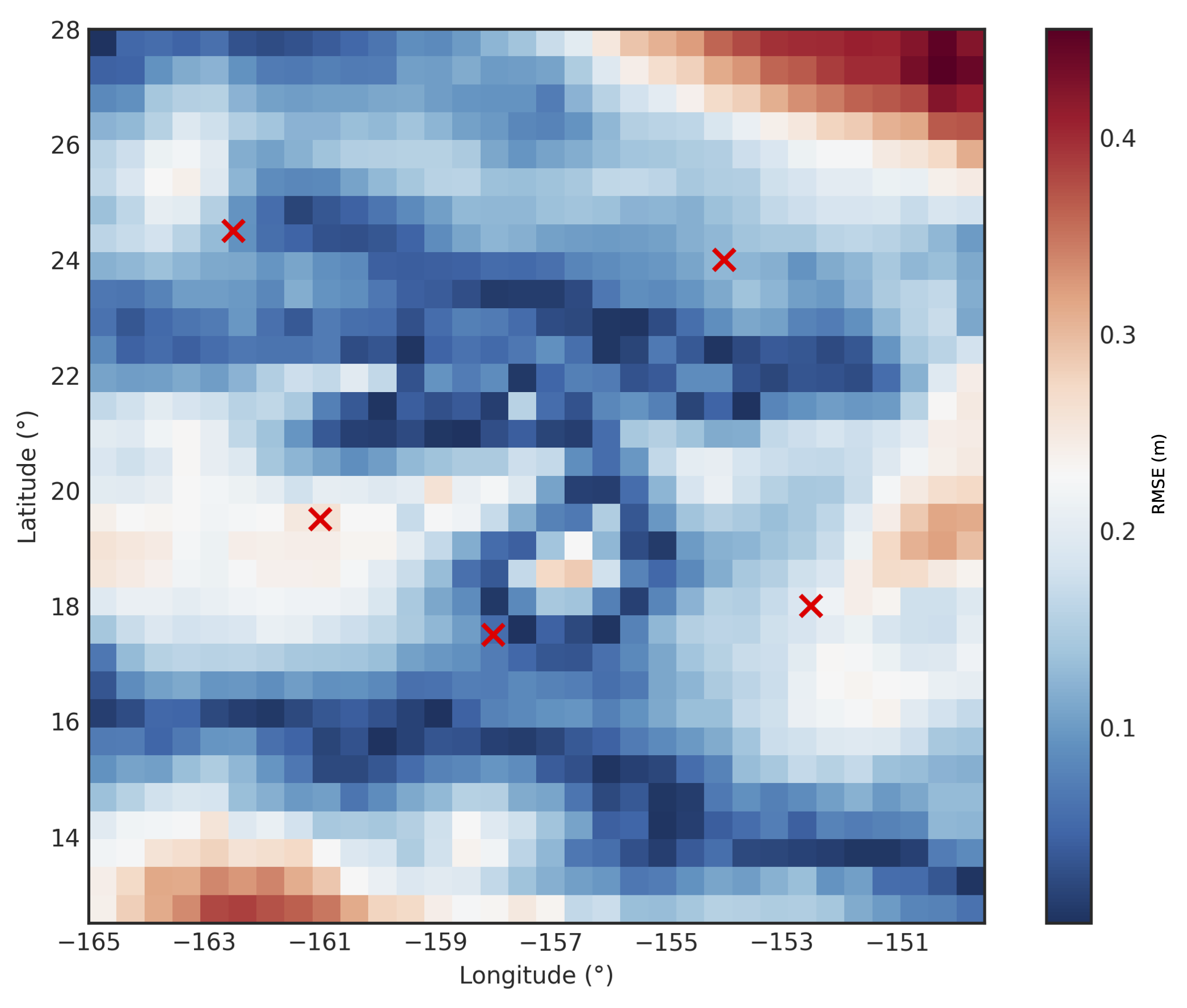}
      \innersubcaption{Error (RMSE=0.1647)}{fig:hawaii_e3_err}
    \end{minipage}
    \scriptsize \vspace{2pt}\par\itshape Snapshot at time step 3
  \end{subfigure}

  \begin{subfigure}[t]{\textwidth}
    \refstepcounter{subfigure}\label{fig:hawaii_ex4}
    \setcounter{innersubfigure}{0}
    \centering
    \begin{minipage}[t]{0.32\textwidth}
      \includegraphics[width=\linewidth]{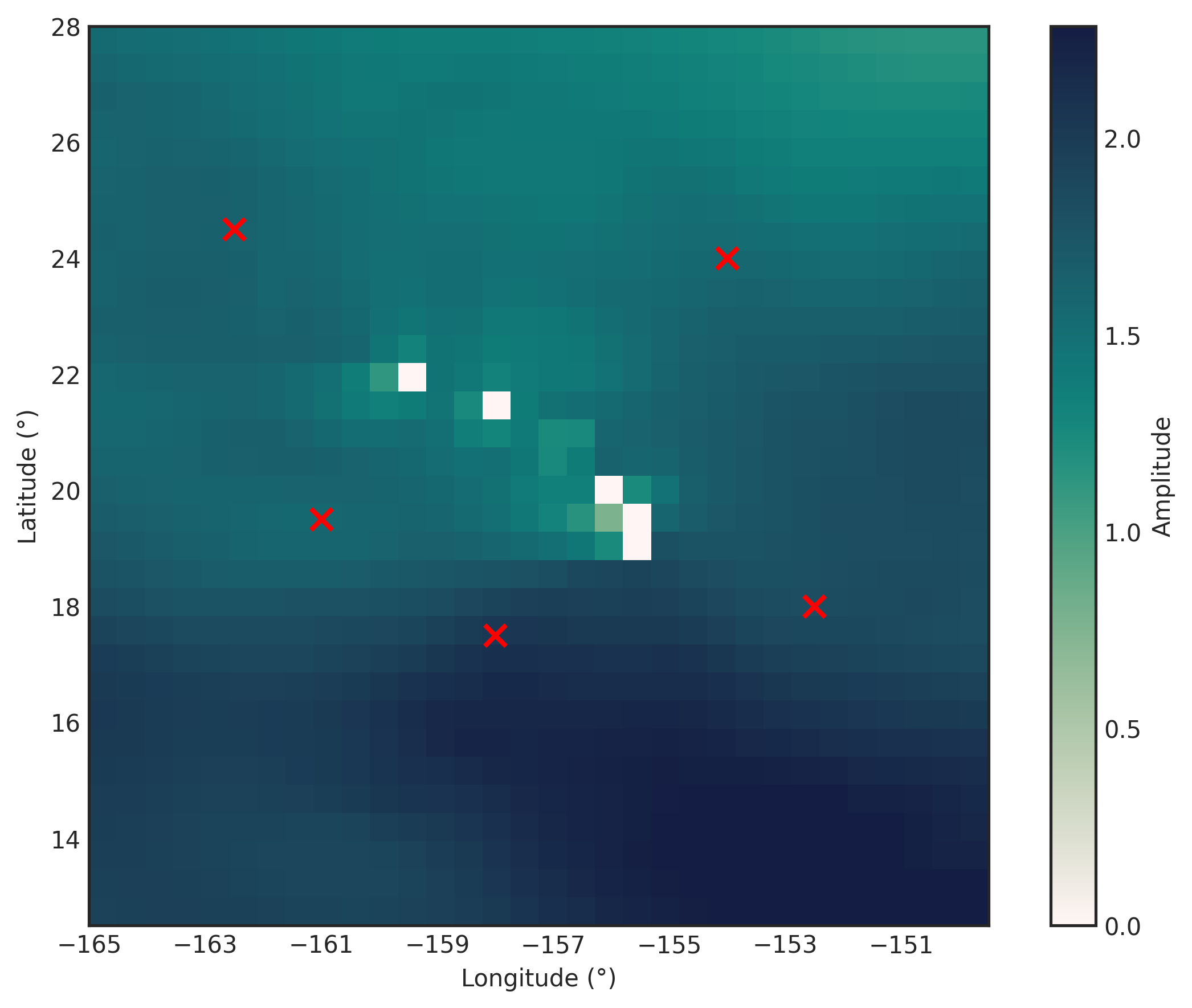}
      \innersubcaption{True}{fig:hawaii_e4_true}
    \end{minipage}\hfill
    \begin{minipage}[t]{0.32\textwidth}
      \includegraphics[width=\linewidth]{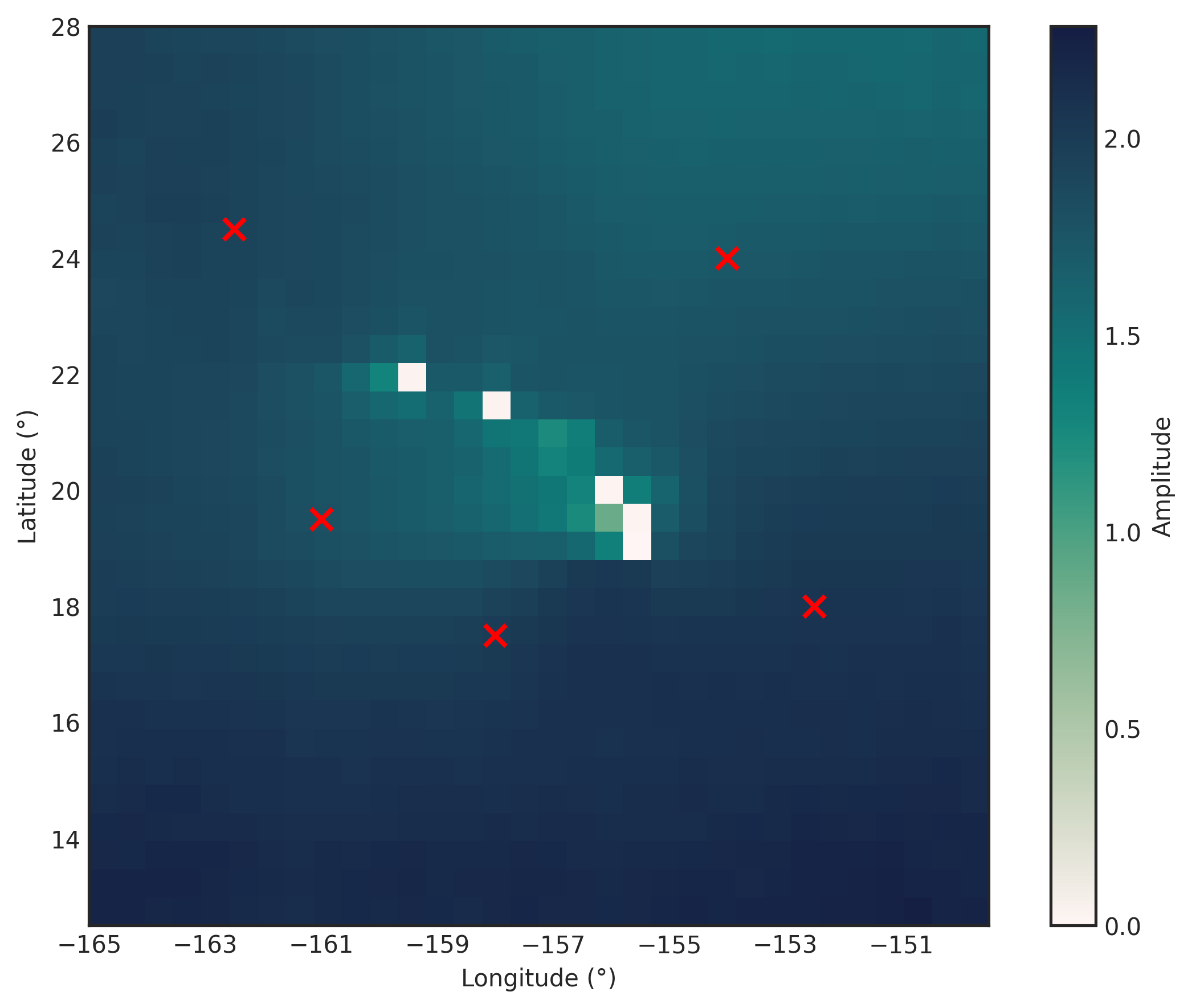}
      \innersubcaption{Predicted}{fig:hawaii_e4_pred}
    \end{minipage}\hfill
    \begin{minipage}[t]{0.32\textwidth}
      \includegraphics[width=\linewidth]{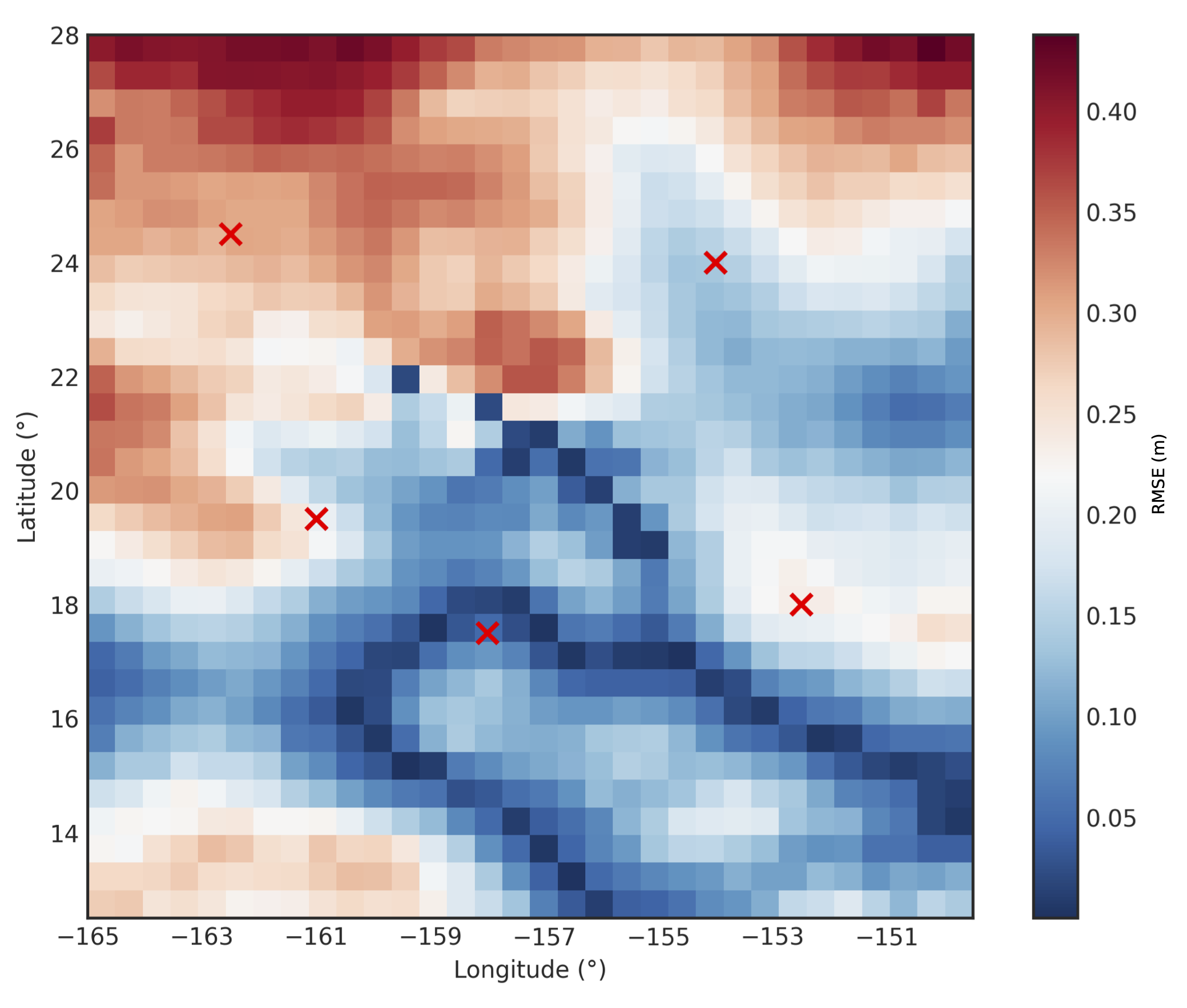}
      \innersubcaption{Error (RMSE=0.2183)}{fig:hawaii_e4_err}
    \end{minipage}
    \scriptsize \vspace{2pt}\par\itshape Snapshot at time step 4
  \end{subfigure}
  \caption{Comparison of true and predicted wave fields for Hawaii}
  \label{fig:hawaii_all}
\end{figure}

\begin{figure}
  \ContinuedFloat
  \begin{subfigure}[t]{\textwidth}
  \refstepcounter{subfigure}\label{fig:hawaii_ex5}
  \setcounter{innersubfigure}{0}
    \centering
    \begin{minipage}[t]{0.32\textwidth}
      \includegraphics[width=\linewidth]{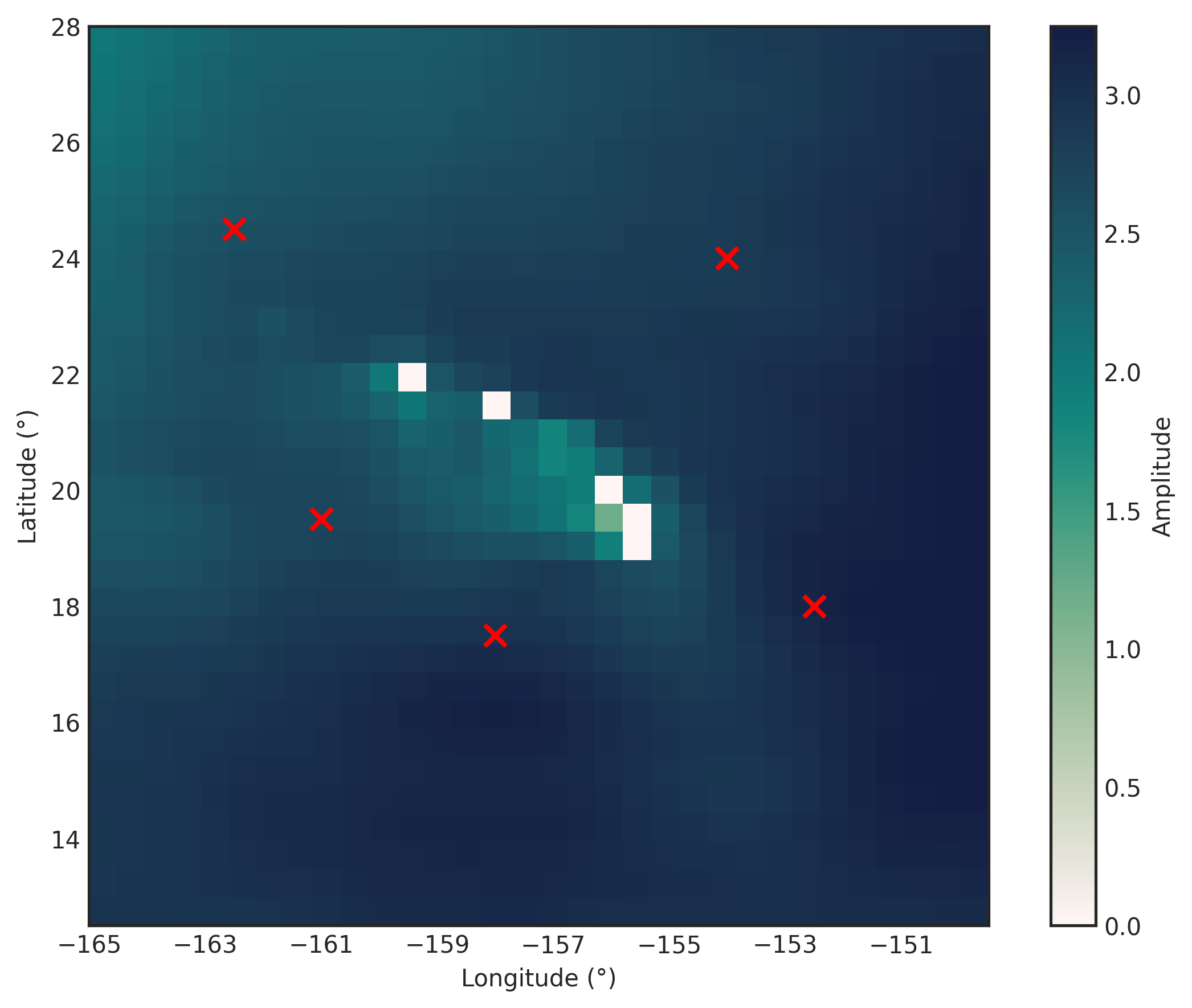}
      \innersubcaption{True}{fig:hawaii_e5_true}
    \end{minipage}\hfill
    \begin{minipage}[t]{0.32\textwidth}
      \includegraphics[width=\linewidth]{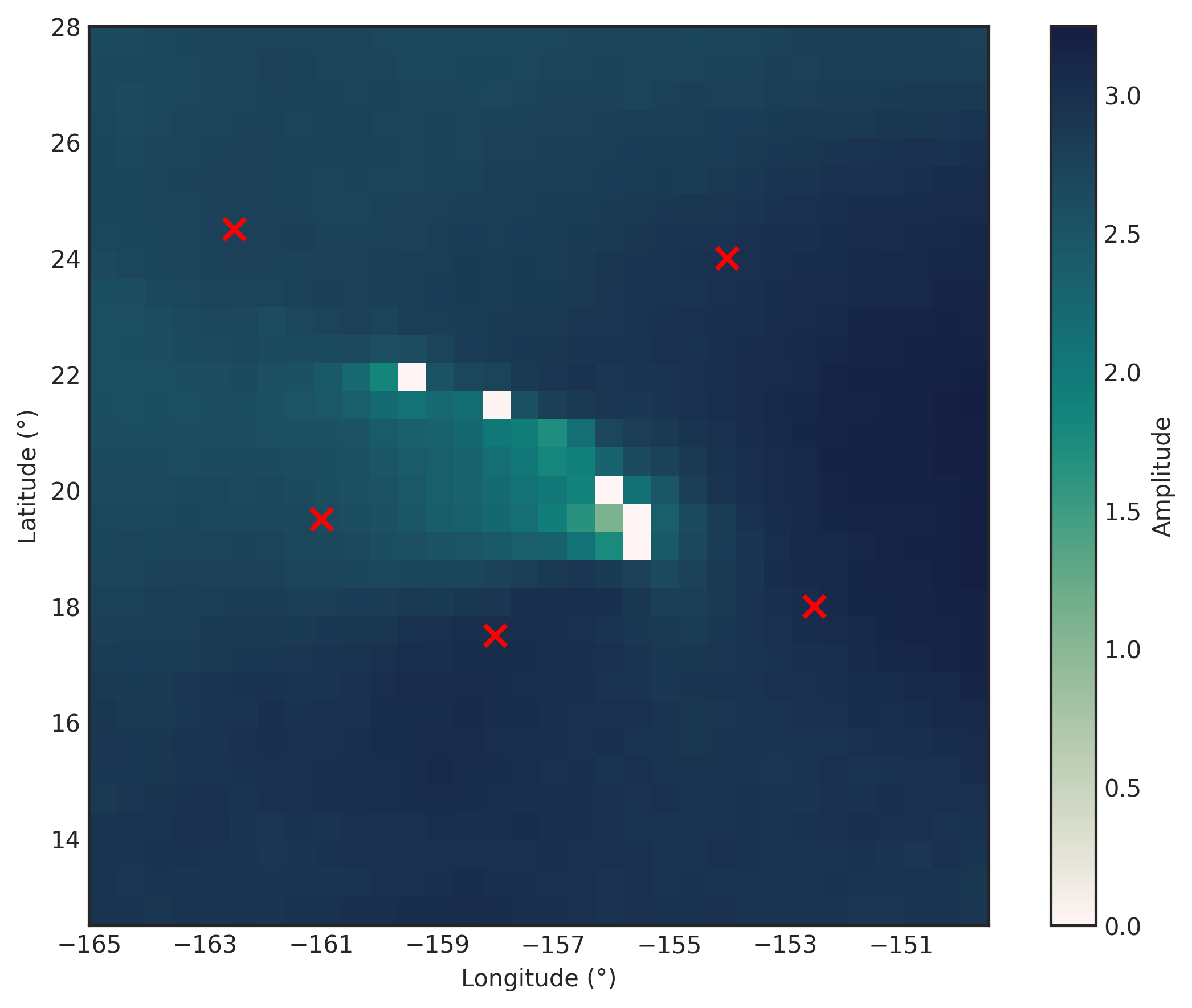}
      \innersubcaption{Predicted}{fig:hawaii_e5_pred}
    \end{minipage}\hfill
    \begin{minipage}[t]{0.32\textwidth}
      \includegraphics[width=\linewidth]{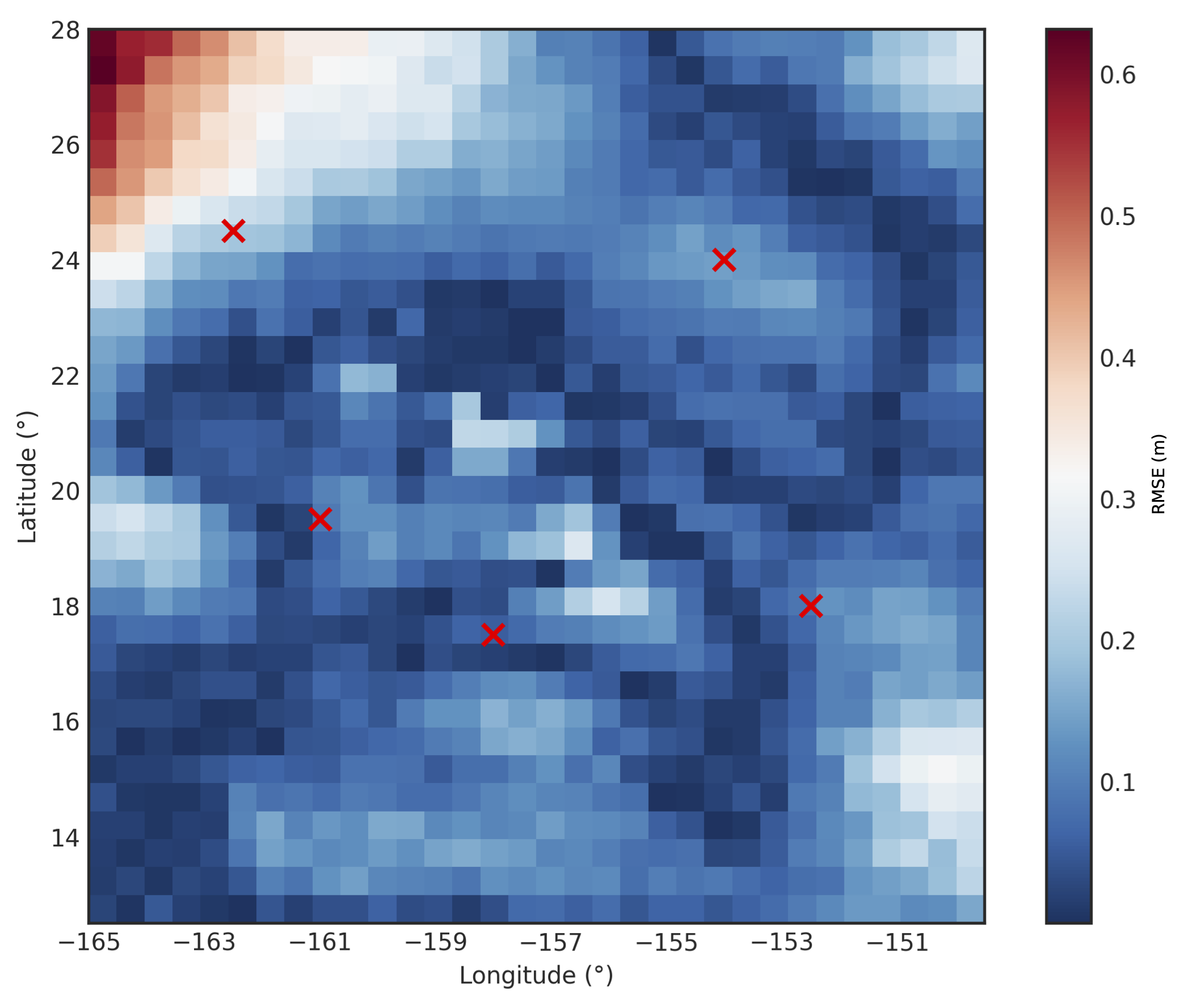}
      \innersubcaption{Error (RMSE=0.1463)}{fig:hawaii_e5_err}
    \end{minipage}
    \scriptsize \vspace{2pt}\par\itshape Snapshot at time step 5
  \end{subfigure}
  \caption{Comparison of true and predicted wave fields for Hawaii (continued)}
\end{figure}

\subsection{RMSE Distribution and Spatial Patterns}
\label{sec:rmse-distribution}
\begin{figure}[htbp]
  \begin{subfigure}[b]{0.59\textwidth}
    \centering
    \includegraphics[width=\textwidth]{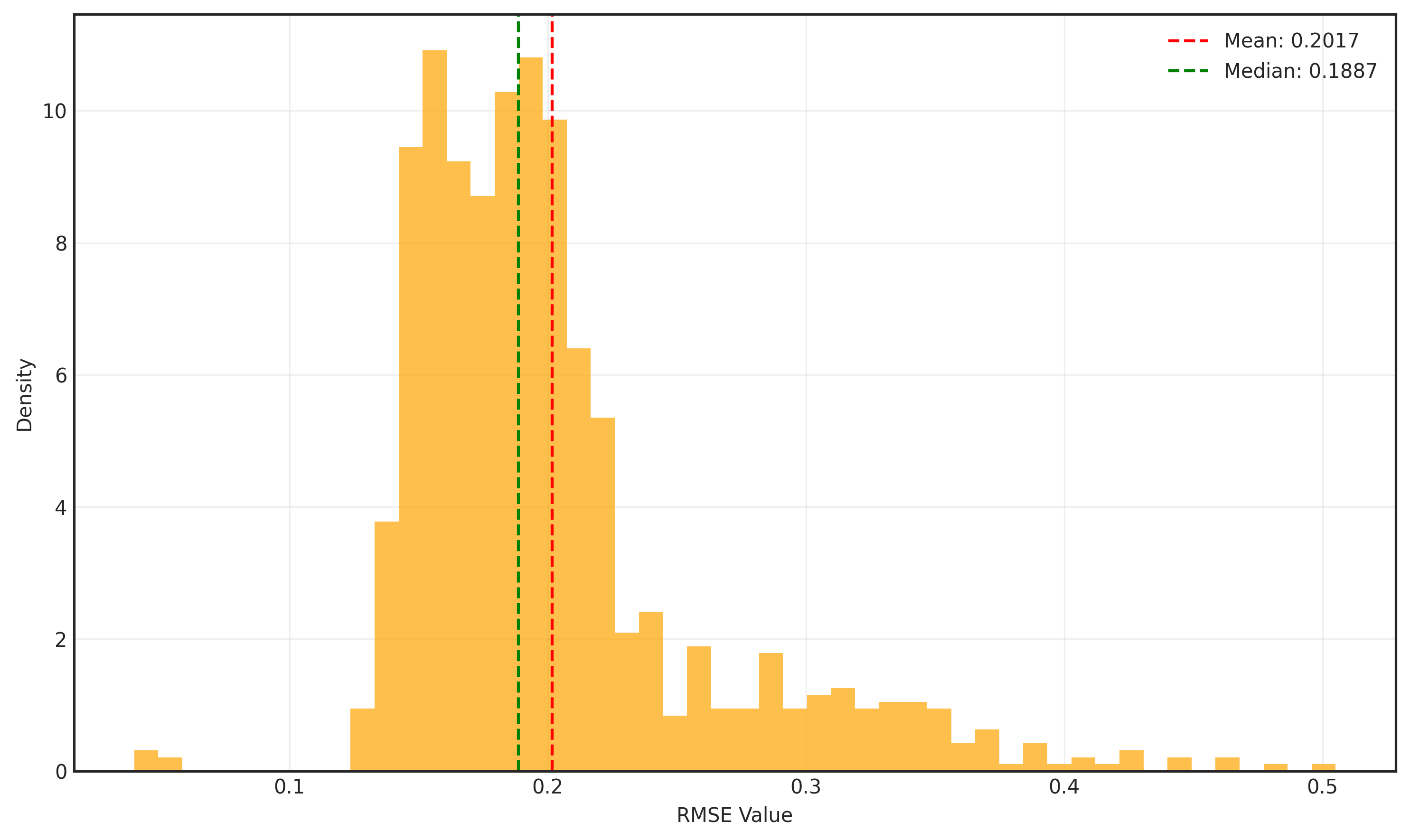}
    \caption{RMSE Distribution (Temporal)}
    \label{fig:rmse_distribution_hawaii_histogram}
  \end{subfigure}%
  \begin{subfigure}[b]{0.41\textwidth}
    \centering
    \includegraphics[width=\textwidth]{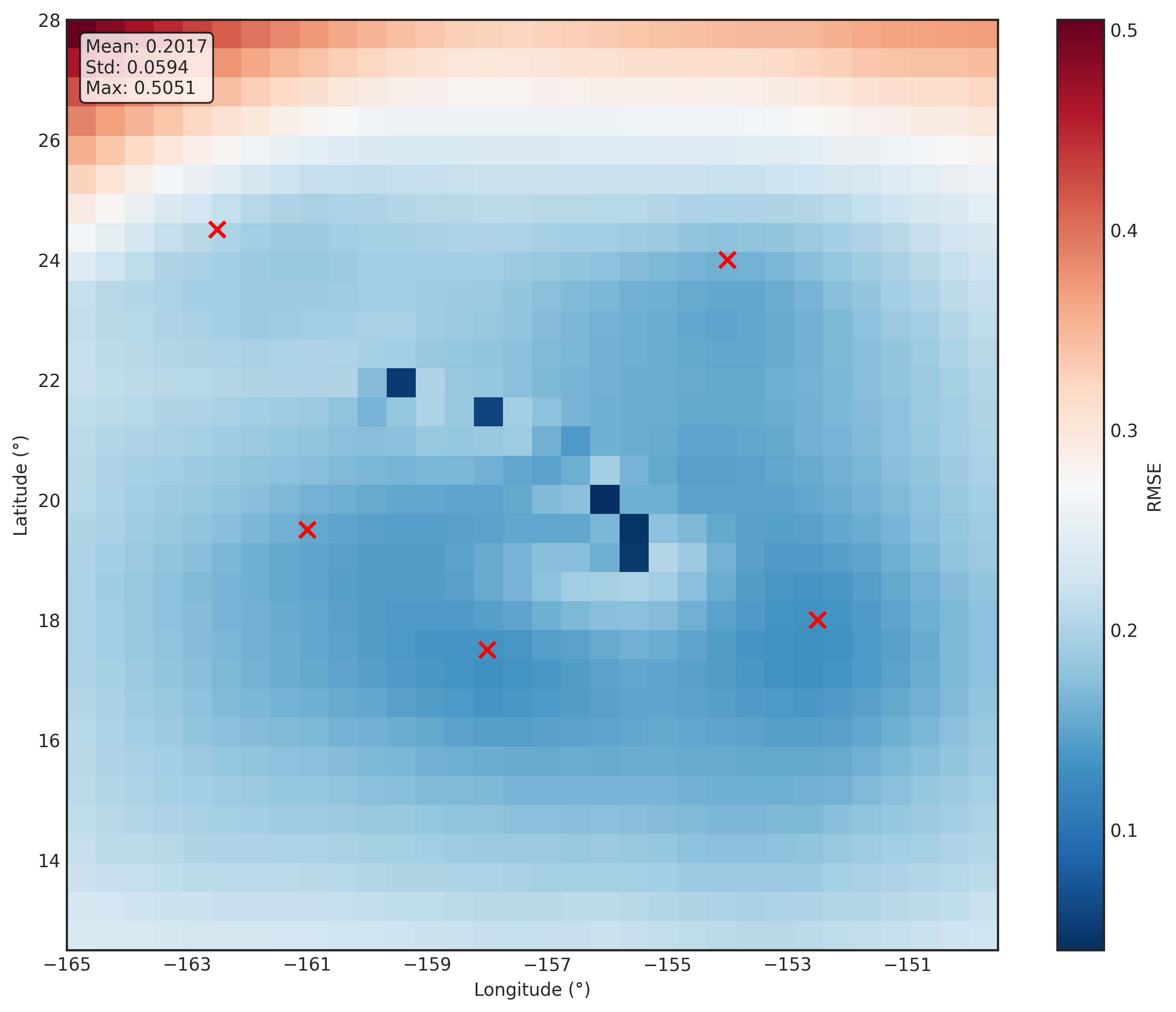}
    \caption{RMSE Distribution (Spatial)}
    \label{fig:rmse_distribution_hawaii_spatial}
  \end{subfigure}%

  \caption{RMSE distribution across the Hawaii dataset. (a)Histogram of RMSE values over all time steps, showing the frequency of different error magnitudes. (b) Spatial distribution of RMSE values at each grid point, with warmer colors indicating higher errors.}
  \label{fig:rmse_distribution_hawaii}
\end{figure}

In addition to individual examples, we analyze the overall RMSE distribution across the entire Hawaii dataset to understand the spatial and temporal error patterns. Figure~\ref{fig:rmse_distribution_hawaii} presents (a) the spatial distribution of RMSE values at each grid point, with warmer colors indicating higher errors, and (b) a histogram of RMSE values over all time steps, showing the frequency of different error magnitudes. This analysis helps identify areas and periods where the model performs well or struggles, providing insights for further improvements.

\rev{As shown in Figure~\ref{fig:rmse_distribution_hawaii_histogram}, the distribution of RMSE values is predominantly centered around lower error magnitudes. The mean RMSE across all samples is 0.2017, reflecting the overall average reconstruction error and its sensitivity to occasional high-error cases, while the median RMSE of 0.1887 represents the typical reconstruction accuracy achieved in the majority of time steps and is more robust to outliers. The slight right-skewness of the distribution, where the mean is slightly higher than the median, therefore indicates the presence of a tail towards higher RMSE values, corresponding to a limited number of challenging reconstruction scenarios where model performance was less optimal.}

Figure~\ref{fig:rmse_distribution_hawaii_spatial} presents the spatial distribution of the average RMSE for the Hawaii region. The overall mean RMSE for the spatial domain is consistent with the histogram at 0.2017, with a standard deviation of 0.0594 and a maximum observed RMSE of 0.5051 in localized regions. Spatially, the model generally achieves lower RMSE values (represented by darker blue regions) in proximity to the buoy locations, which serve as direct observation points. This suggests that the model effectively leverages the available sparse data. Conversely, higher RMSE values (lighter blues and occasional red patches) are more prevalent in regions further away from the buoys. These areas, characterized by greater data sparsity, present a larger challenge for accurate reconstruction through interpolation or extrapolation.

\subsection{Ablation and Sensitivity Analysis}
\label{sec:ablation-sensitivity}
\begin{figure}[htbp]
  \centering

  \begin{subfigure}[b]{0.9\textwidth}
    \centering
    \includegraphics[width=\linewidth]{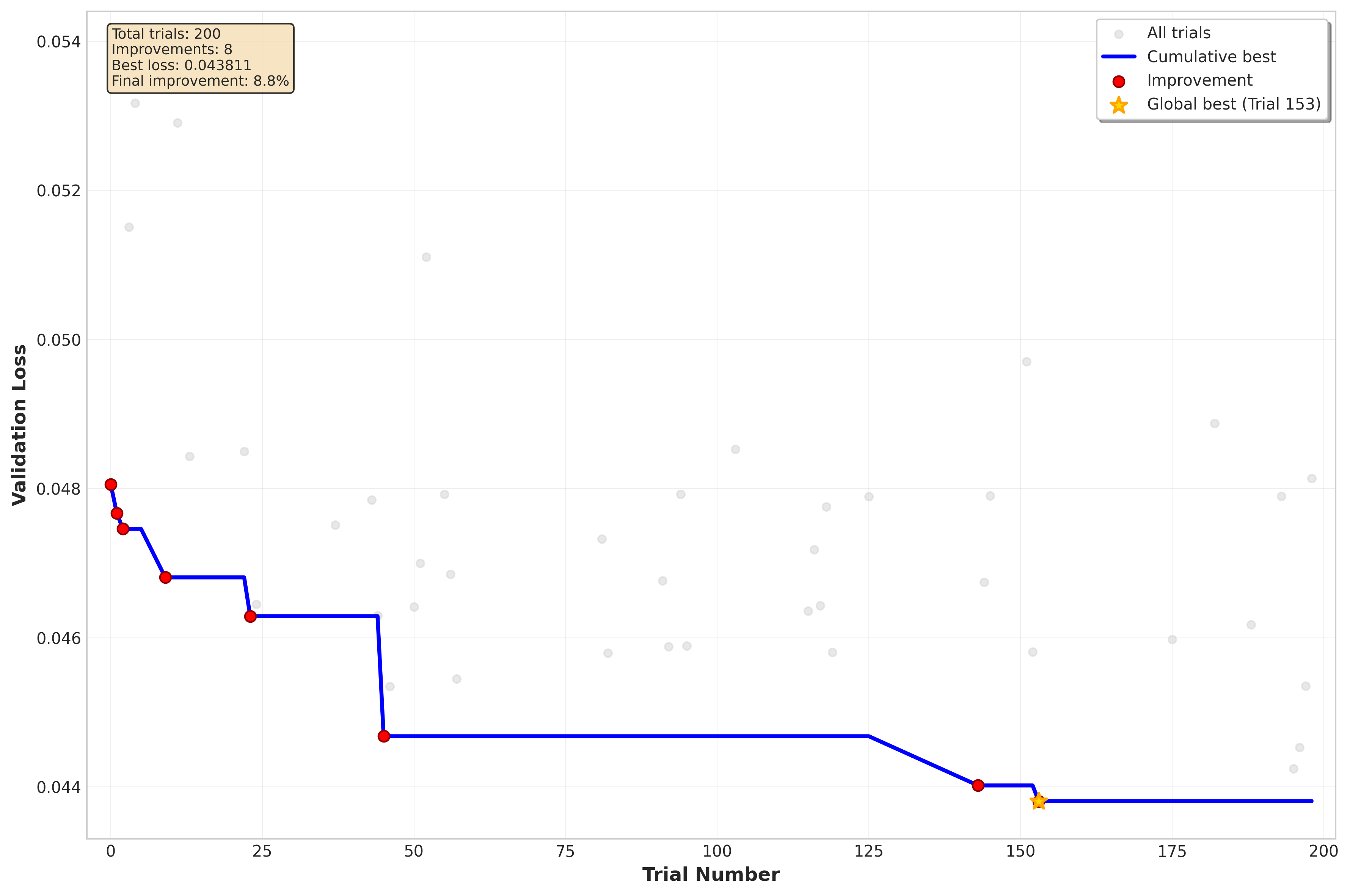}
    \caption{Hyperparameter Optimization History}
    \label{fig:hyperopt_history}
  \end{subfigure}

  \par

  \begin{subfigure}[b]{0.9\textwidth}
    \centering
    \includegraphics[width=\linewidth]{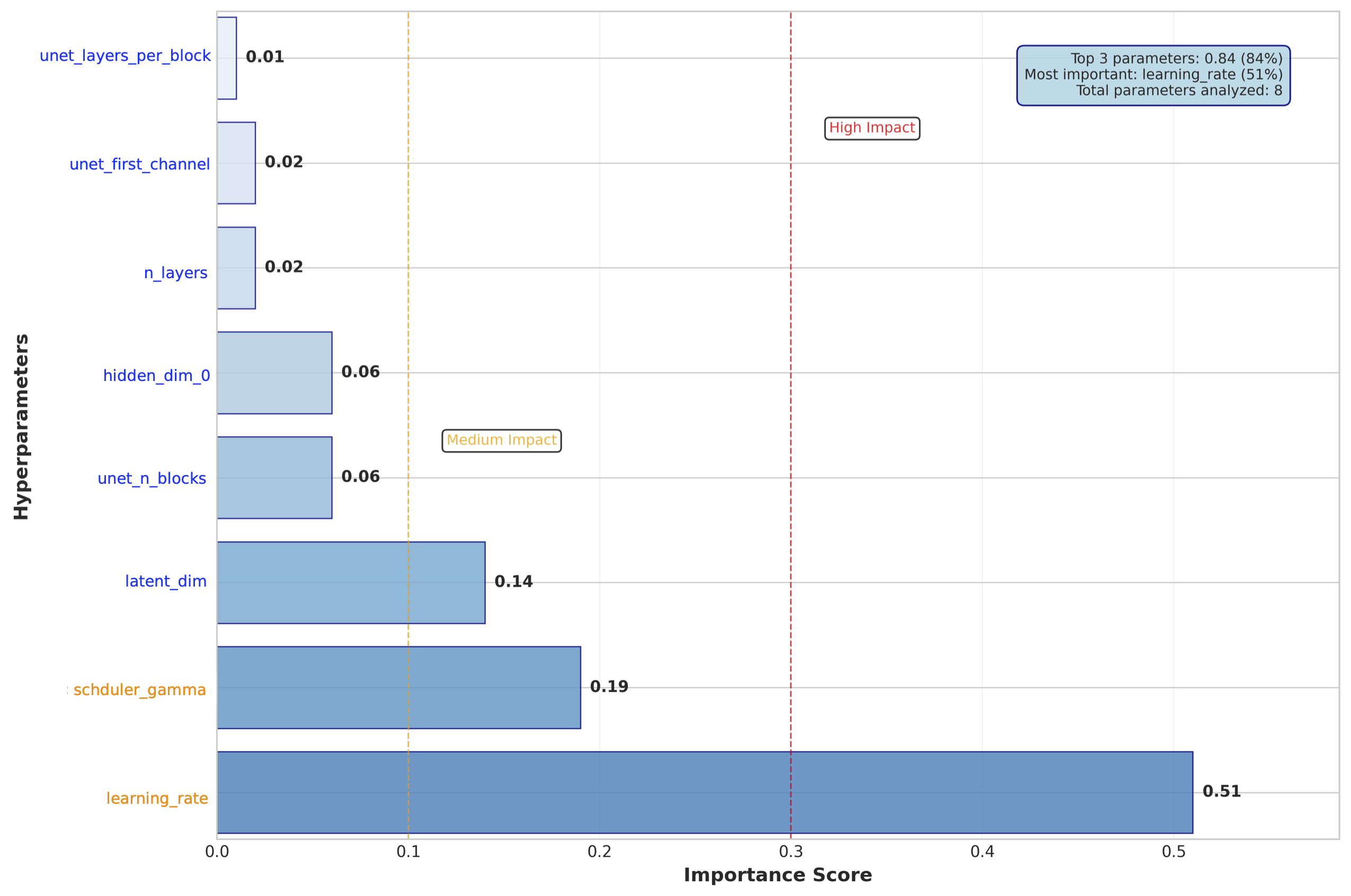}
    \caption{Hyperparameter Importance}
    \label{fig:hyperopt_importance}
  \end{subfigure}

  \caption{Hyperparameter optimization results using Optuna, illustrating the optimization history and the relative importance of model architecture and training hyperparameters. Blue bars denote model architecture hyperparameters, whereas orange bars indicate training hyperparameters.}
  \label{fig:optuna-results}
\end{figure}

To better understand the influence of individual hyperparameters on the model's performance, we visualize the optimization process and the parameter importance derived from 500 Optuna trials, as shown in Figure~\ref{fig:optuna-results}.

Figure~\ref{fig:optuna-results}(a) presents the optimization trajectory across trials. A clear downward trend in the objective value indicates effective convergence, with the best-performing configuration achieving a minimum validation loss of 0.0433.

Figure~\ref{fig:optuna-results}(b) reports the relative importance of hyperparameters as estimated by Optuna’s surrogate model. The learning rate contributed most significantly (\textasciitilde51\%), followed by the scheduler decay rate and latent dimension. Structural parameters such as U-Net depth and attention contributed comparatively less. These results informed the final model configuration.

\subsection{Impact of Buoy Configuration on Reconstruction Accuracy}
\begin{figure}[htbp]
  \begin{subfigure}[b]{0.49\textwidth}
    \centering
    \includegraphics[width=\textwidth]{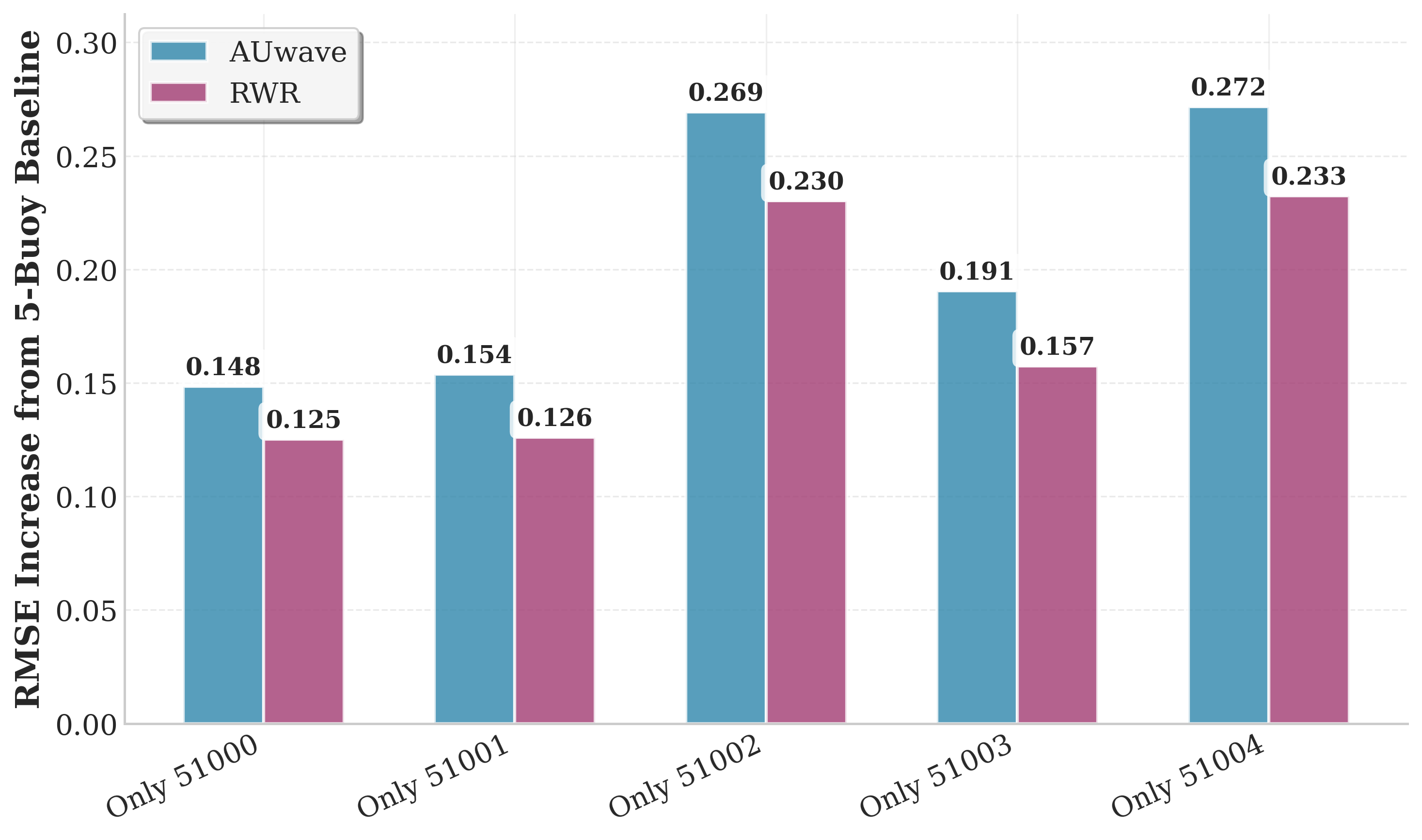}
    \caption{Single Buoy Input}
    \label{fig:single_buoy_input}
  \end{subfigure}
  \begin{subfigure}[b]{0.49\textwidth}
    \centering
    \includegraphics[width=\textwidth]{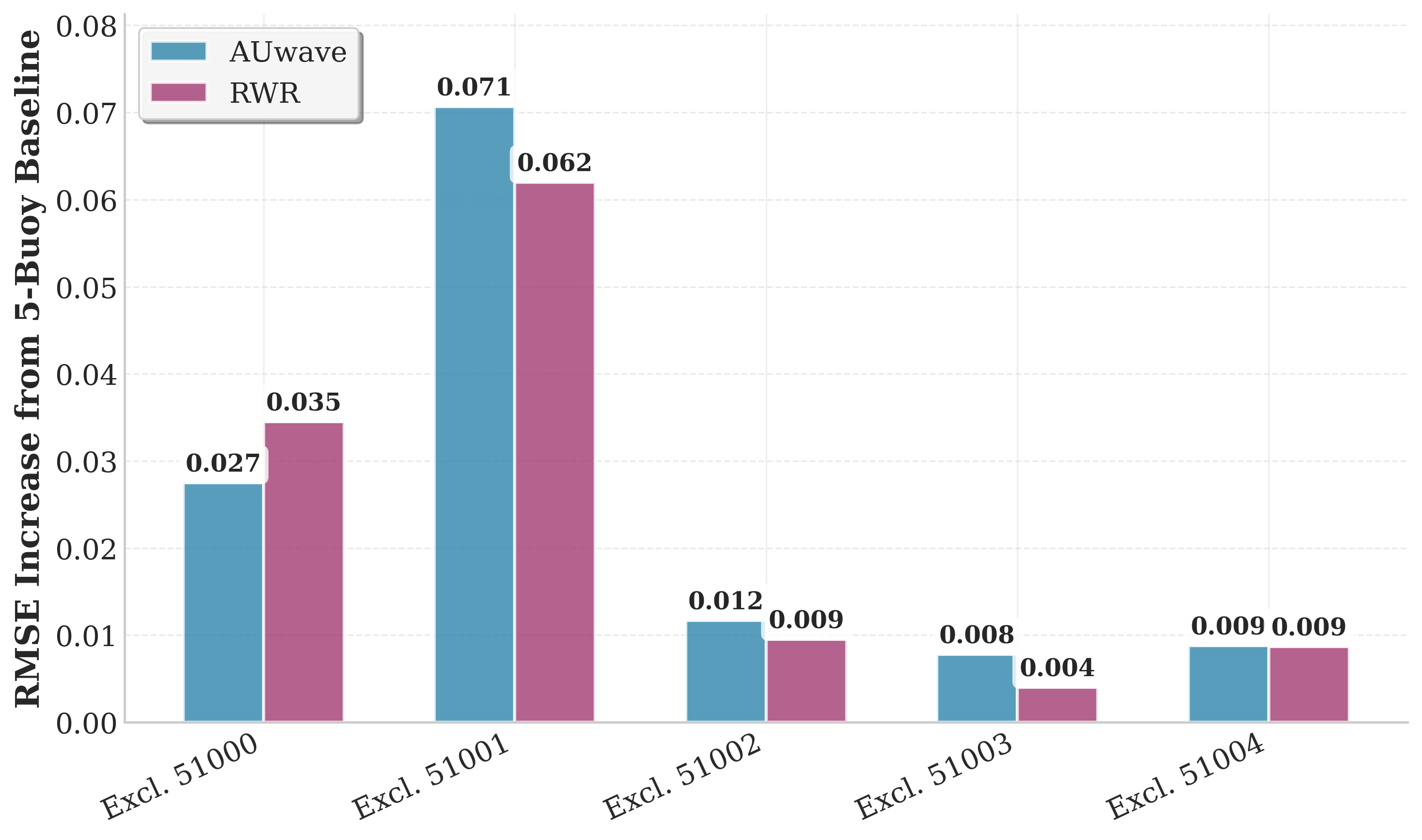}
    \caption{Four Buoys Input}
    \label{fig:four_buoys_input}
  \end{subfigure}
  \caption{Model Sensitivity to Buoy Configuration.}
  \label{fig:buoy_configuration_sensitivity}

\end{figure}

This section evaluates the performance of the AUWave and RWR models for ocean wave field reconstruction under varying data sparsity, with detailed results presented in Table \ref{tab:rmse_buoy_combinations_two_models} and Figure \ref{fig:buoy_configuration_sensitivity}. The results demonstrate that the AUWave model consistently outperforms the RWR model under data-rich conditions (5 and 4 buoys). In the 5-buoy baseline, for instance, AUWave achieved an RMSE of 0.2129 compared to RWR's 0.2472. Both models' performance was sensitive to buoy configuration, with the removal of Buoy 51001 causing the most significant negative impact. Under the extreme sparsity of single-buoy inputs, AUWave maintained its lead in three of the five cases. However, a notable exception occurred when using only Buoy 51002 or 51004, where RWR showed a marginal advantage. These two configurations also corresponded to the highest overall errors for both models.
\begin{table}[htbp]
  \centering
  \begin{tabular}{l r r}
    \toprule
    \textbf{Buoy Configuration}                     & {\textbf{AUWave}} & {\textbf{RWR}}  \\
    \midrule
    \multicolumn{3}{c}{\textit{--- Reference Configuration ---}}                          \\
    All 5 Buoys (51000, 51001, 51002, 51003, 51004) & \textbf{0.2129}            & 0.2472          \\
    \midrule
    \multicolumn{3}{c}{\textit{--- Removing One Buoy (4-Buoy Combinations) ---}}          \\
    Excluding Buoy 51000                            & \textbf{0.2404}   & 0.2817          \\
    Excluding Buoy 51001                            & \textbf{0.2836}   & 0.3092          \\
    Excluding Buoy 51002                            & \textbf{0.2246}   & 0.2567          \\
    Excluding Buoy 51003                            & \textbf{0.2207}   & 0.2512          \\
    Excluding Buoy 51004                            & \textbf{0.2217}   & 0.2559          \\
    \midrule
    \multicolumn{3}{c}{\textit{--- Using Single Buoy (Removing Four Buoys) ---}}          \\
    Buoy 51000 Only                                 & \textbf{0.3614}   & 0.3724          \\
    Buoy 51001 Only                                 & \textbf{0.3668}   & 0.3732          \\
    Buoy 51002 Only                                 & 0.4822            & \textbf{0.4774} \\
    Buoy 51003 Only                                 & \textbf{0.4035}   & 0.4046          \\
    Buoy 51004 Only                                 & 0.4846            & \textbf{0.4797} \\
    \bottomrule
  \end{tabular}
  \caption{RMSE of Ocean Wave Field Reconstruction Using Different Sparse Buoy Combinations for AUWave and RWR Models}
  \caption*{Note: The best (lowest) RMSE value for each configuration is shown in bold.} % 表格注释
  \label{tab:rmse_buoy_combinations_two_models}
\end{table}

\subsection{Cross-Regional Transfer Validation: Atlantic and Pacific}
\label{sec:transfer}
To assess the out-of-domain generalization ability of the Hawaii-trained AUWave model, we conducted transfer validation in the North Atlantic and the North Pacific without any re-tuning. Both qualitative case studies and quantitative RMSE distributions are reported, following the Hawaii evaluation protocol.

\begin{figure}[p]
  \centering

  \begin{subfigure}[t]{\textwidth}
    \refstepcounter{subfigure}\label{fig:atl_snapshot1}
    \setcounter{innersubfigure}{0}
    \centering

    \begin{minipage}[t]{0.32\textwidth}
      \centering
      \includegraphics[width=\linewidth]{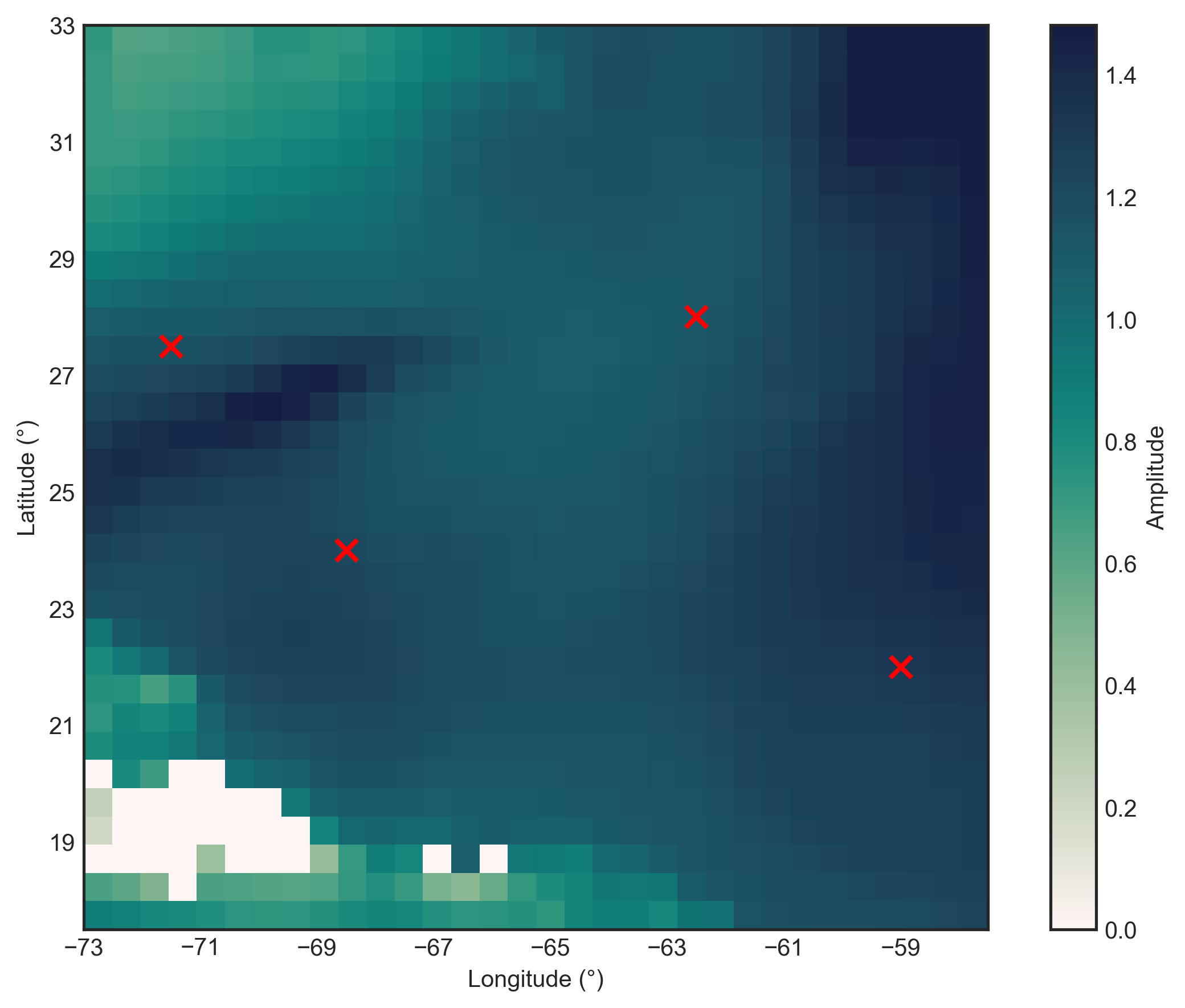}
      \innersubcaption{True}{fig:atl_e1_true}
    \end{minipage}\hfill
    \begin{minipage}[t]{0.32\textwidth}
      \centering
      \includegraphics[width=\linewidth]{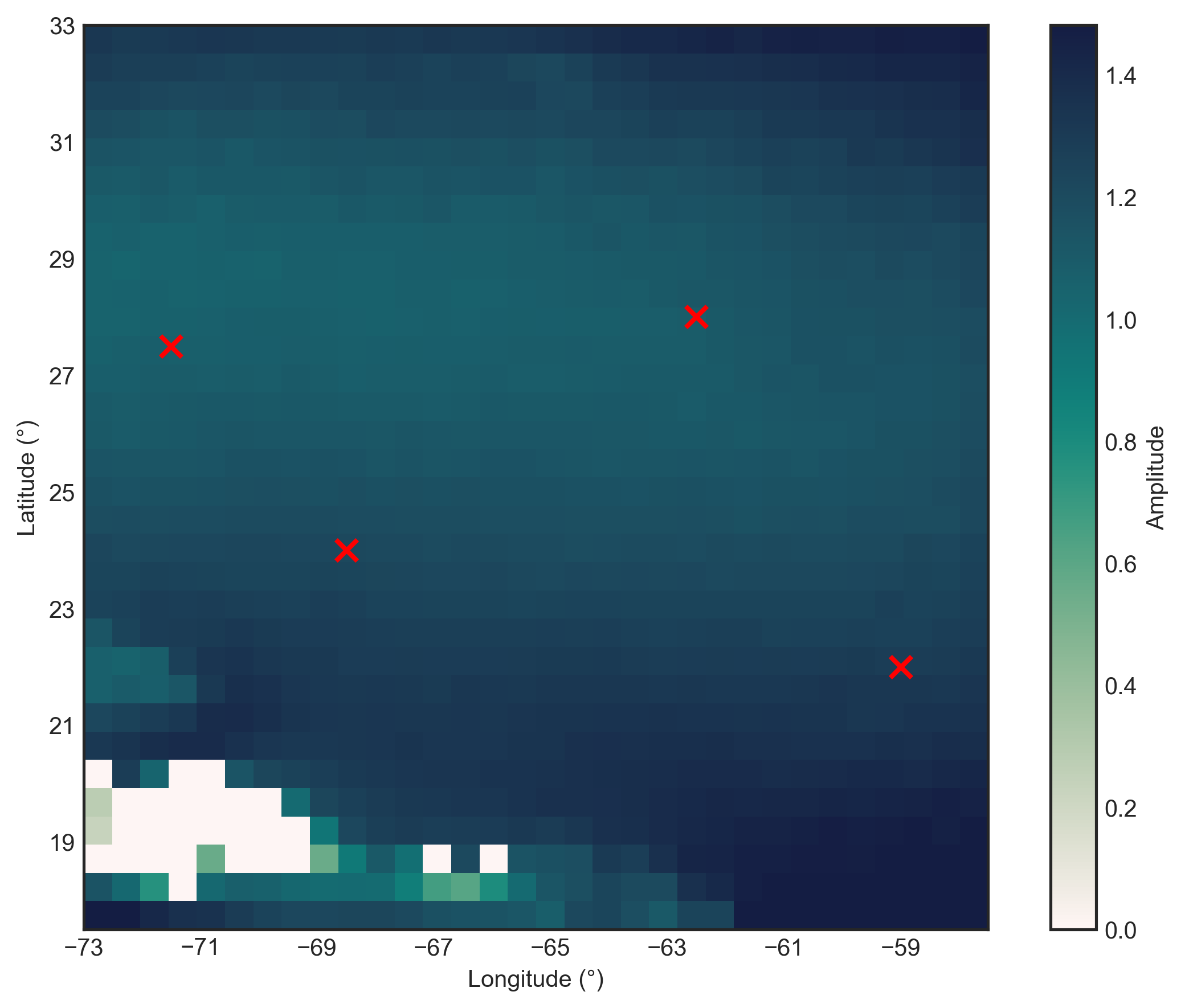}
      \innersubcaption{Predicted}{fig:atl_e1_pred}
    \end{minipage}\hfill
    \begin{minipage}[t]{0.32\textwidth}
      \centering
      \includegraphics[width=\linewidth]{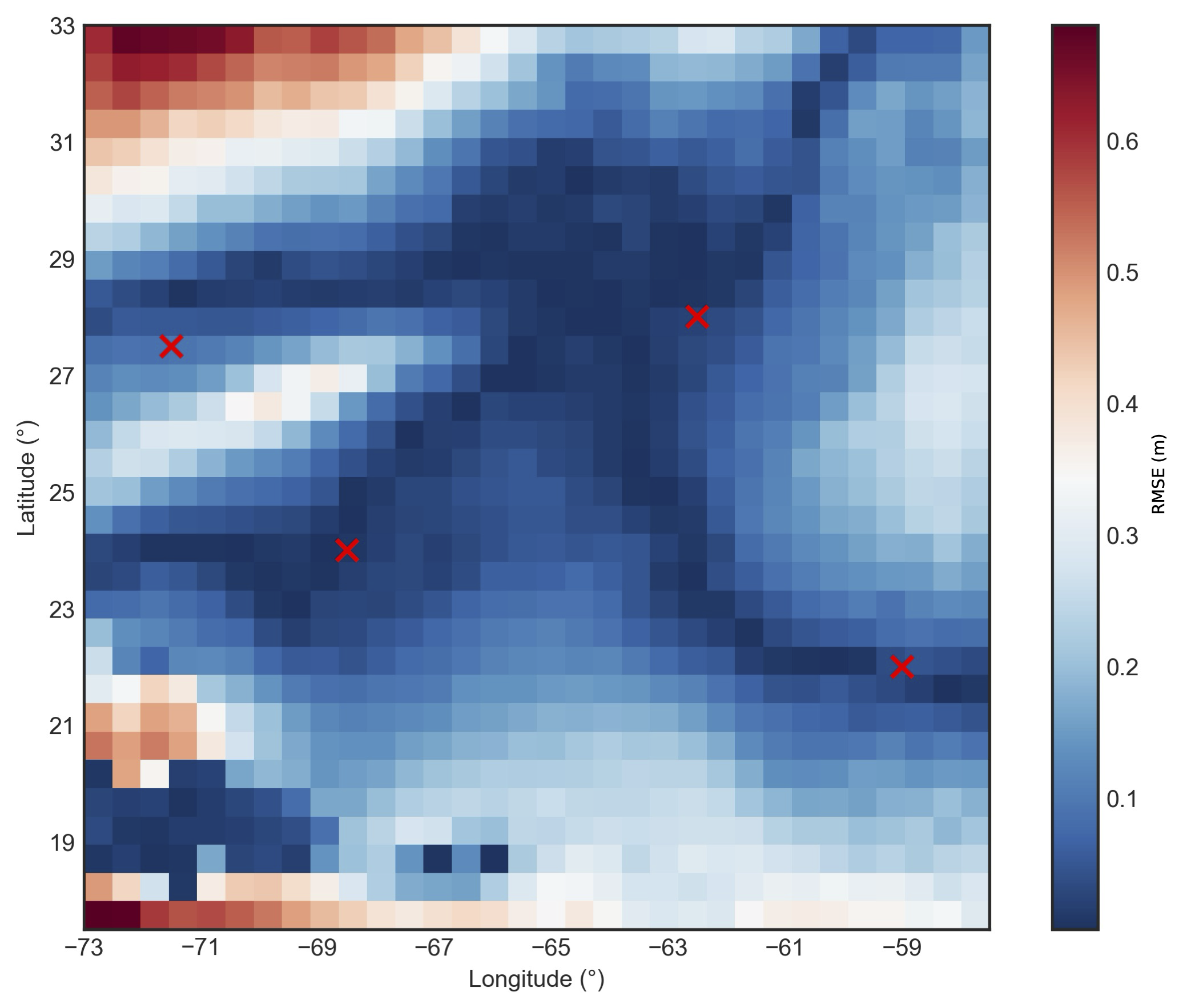}
      \innersubcaption{Error (RMSE=0.2075)}{fig:atl_e1_err}
    \end{minipage}

    \scriptsize \vspace{2pt}\par\itshape Snapshot at time step 1
  \end{subfigure}

  \begin{subfigure}[t]{\textwidth}
    \refstepcounter{subfigure}\label{fig:atl_snapshot2}
    \setcounter{innersubfigure}{0}
    \centering

    \begin{minipage}[t]{0.32\textwidth}
      \centering
      \includegraphics[width=\linewidth]{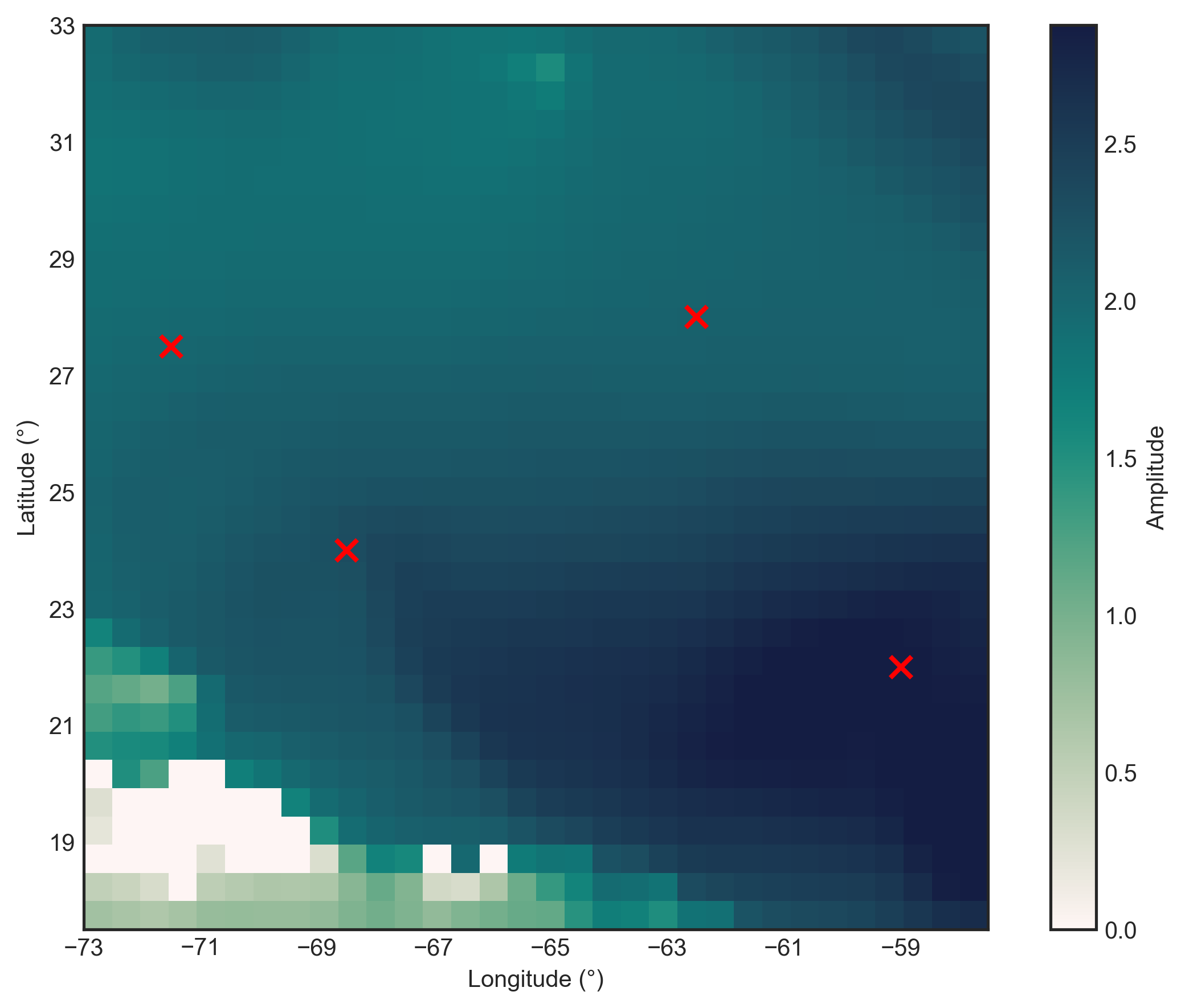}
      \innersubcaption{True}{fig:atl_e2_true}
    \end{minipage}\hfill
    \begin{minipage}[t]{0.32\textwidth}
      \centering
      \includegraphics[width=\linewidth]{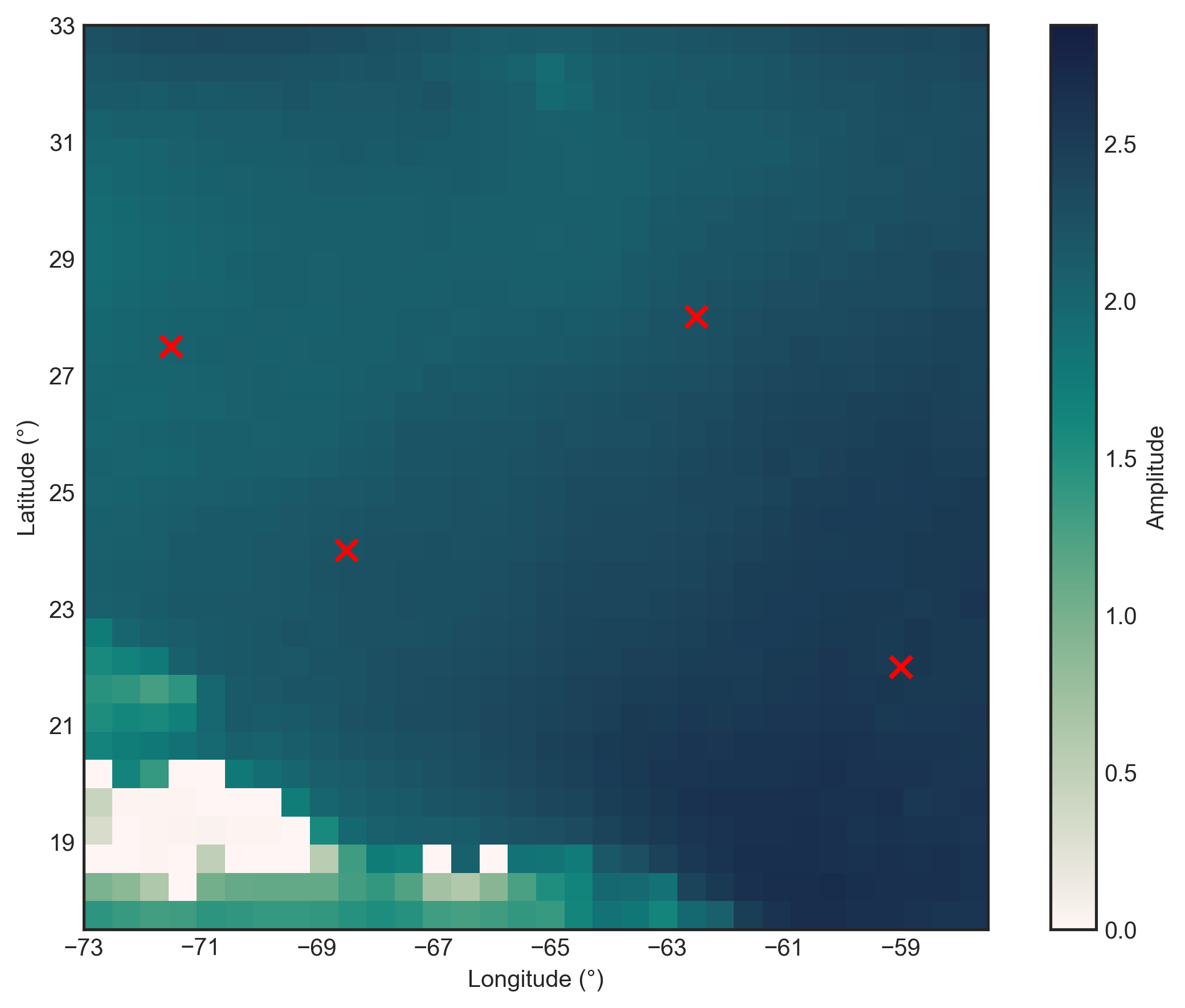}
      \innersubcaption{Predicted}{fig:atl_e2_pred}
    \end{minipage}\hfill
    \begin{minipage}[t]{0.32\textwidth}
      \centering
      \includegraphics[width=\linewidth]{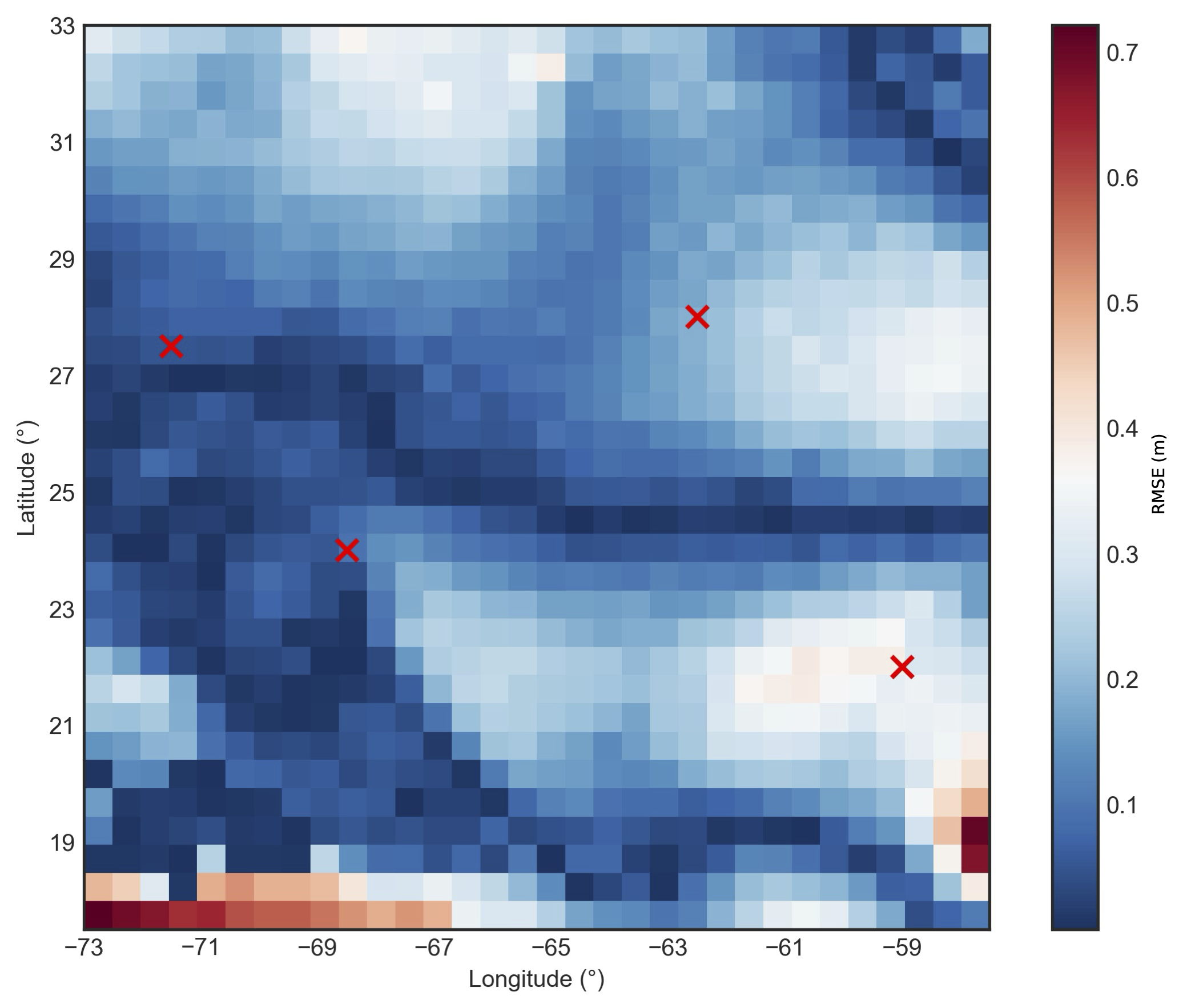}
      \innersubcaption{Error (RMSE=0.1937)}{fig:atl_e2_err}
    \end{minipage}

    \scriptsize \vspace{2pt}\par\itshape Snapshot at time step 2
  \end{subfigure}

  \begin{subfigure}[t]{\textwidth}
    \refstepcounter{subfigure}\label{fig:atl_snapshot3}
    \setcounter{innersubfigure}{0}
    \centering

    \begin{minipage}[t]{0.32\textwidth}
      \centering
      \includegraphics[width=\linewidth]{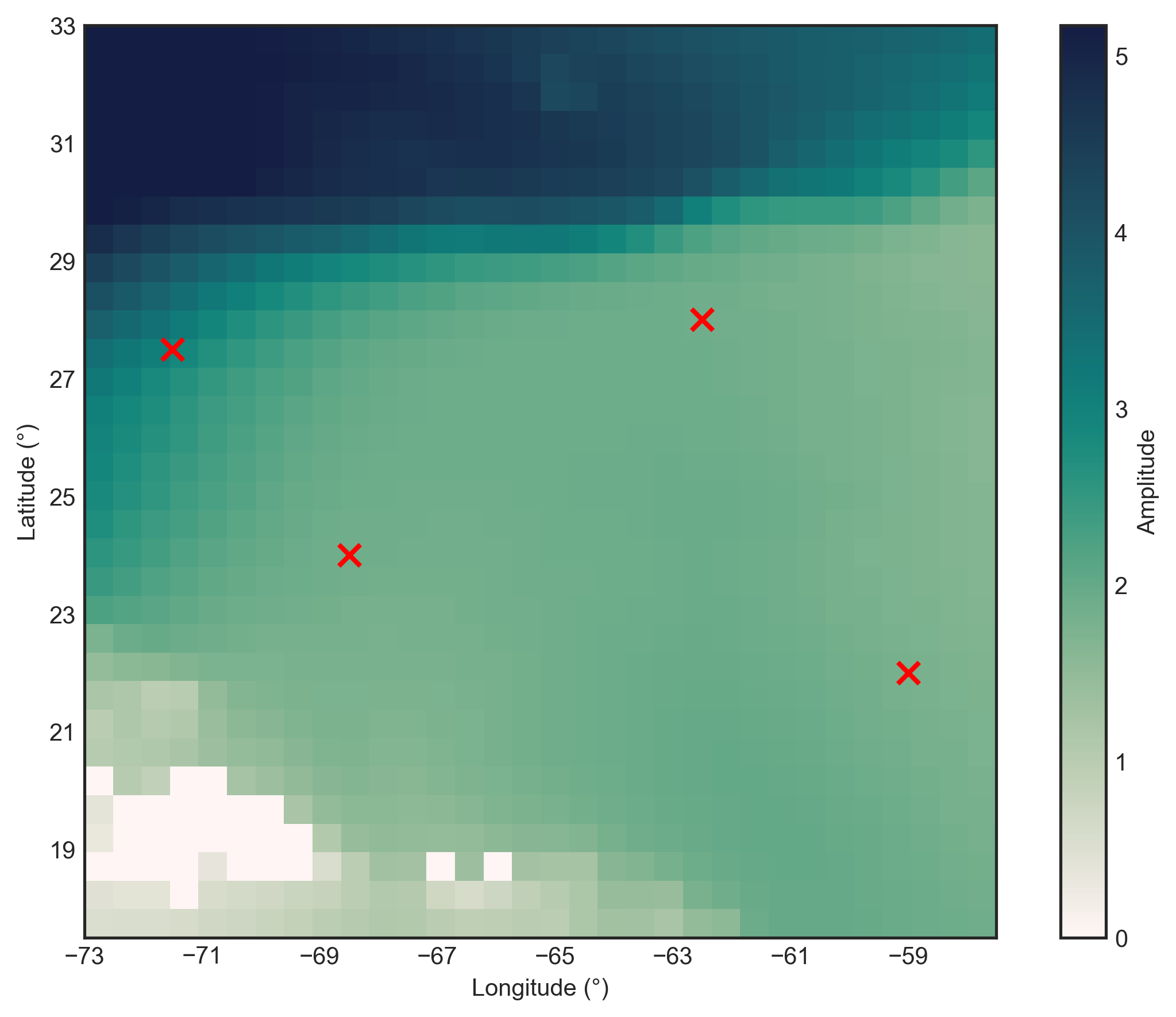}
      \innersubcaption{True}{fig:atl_e3_true}
    \end{minipage}\hfill
    \begin{minipage}[t]{0.32\textwidth}
      \centering
      \includegraphics[width=\linewidth]{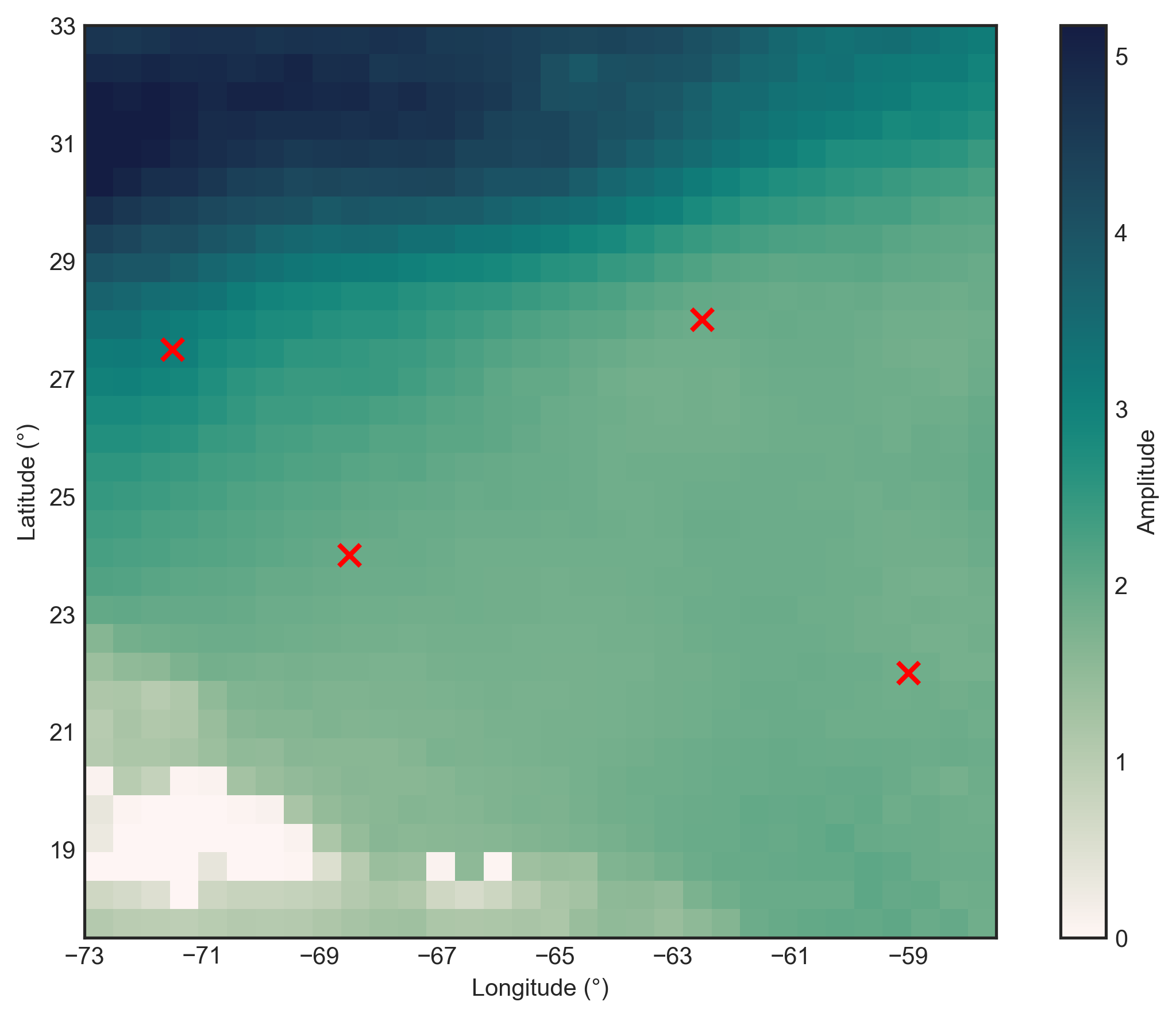}
      \innersubcaption{Predicted}{fig:atl_e3_pred}
    \end{minipage}\hfill
    \begin{minipage}[t]{0.32\textwidth}
      \centering
      \includegraphics[width=\linewidth]{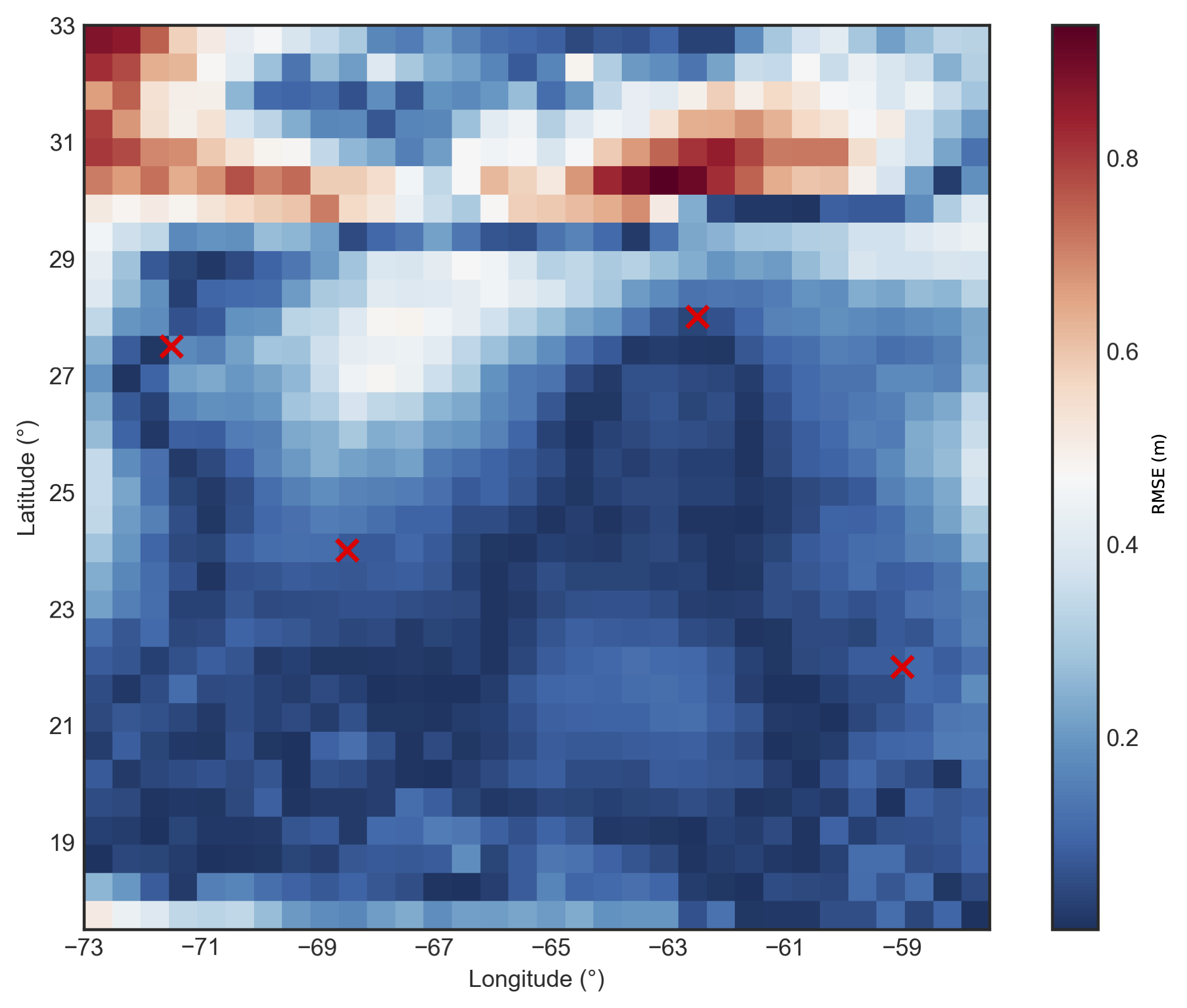}
      \innersubcaption{Error (RMSE=0.2646)}{fig:atl_e3_err} 
    \end{minipage}

    \scriptsize \vspace{2pt}\par\itshape Snapshot at time step 3
  \end{subfigure}

  \begin{subfigure}[t]{\textwidth}
    \refstepcounter{subfigure}\label{fig:atl_snapshot4}
    \setcounter{innersubfigure}{0}
    \centering
    \begin{minipage}[t]{0.32\textwidth}
      \centering
      \includegraphics[width=\linewidth]{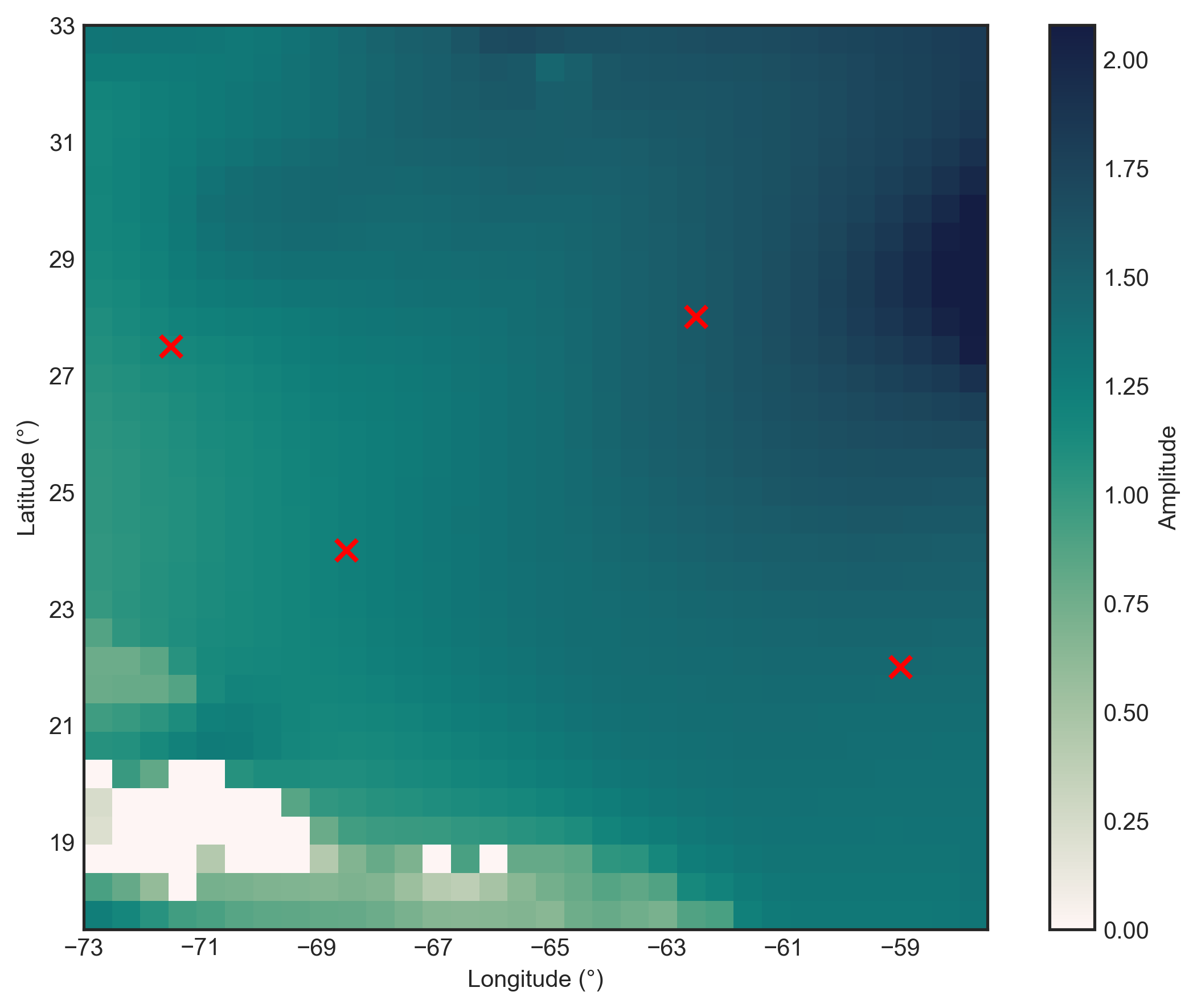}
      \innersubcaption{True}{fig:atl_e4_true}
    \end{minipage}\hfill
    \begin{minipage}[t]{0.32\textwidth}
      \centering
      \includegraphics[width=\linewidth]{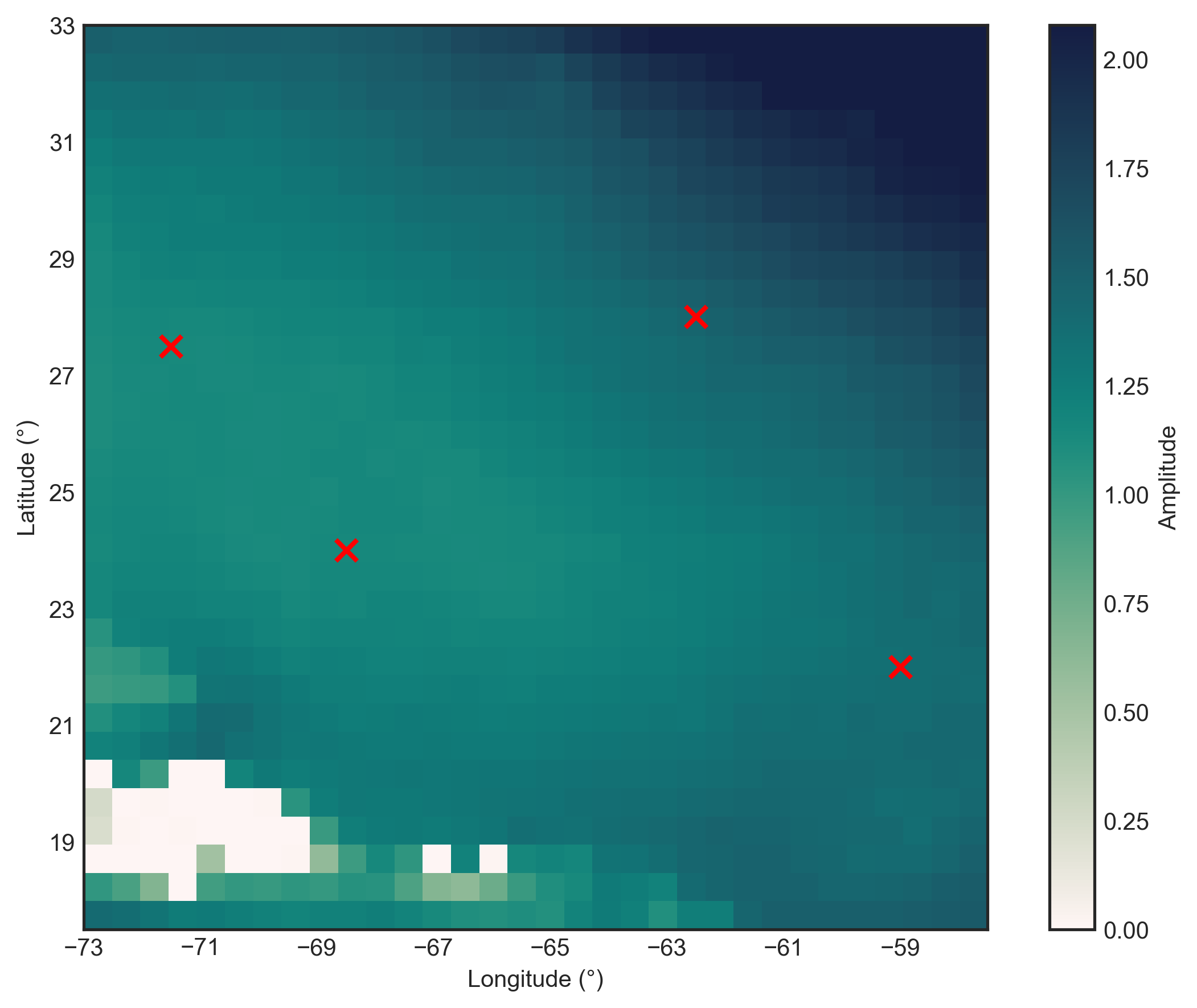}
      \innersubcaption{Predicted}{fig:atl_e4_pred}
    \end{minipage}\hfill
    \begin{minipage}[t]{0.32\textwidth}
      \centering
      \includegraphics[width=\linewidth]{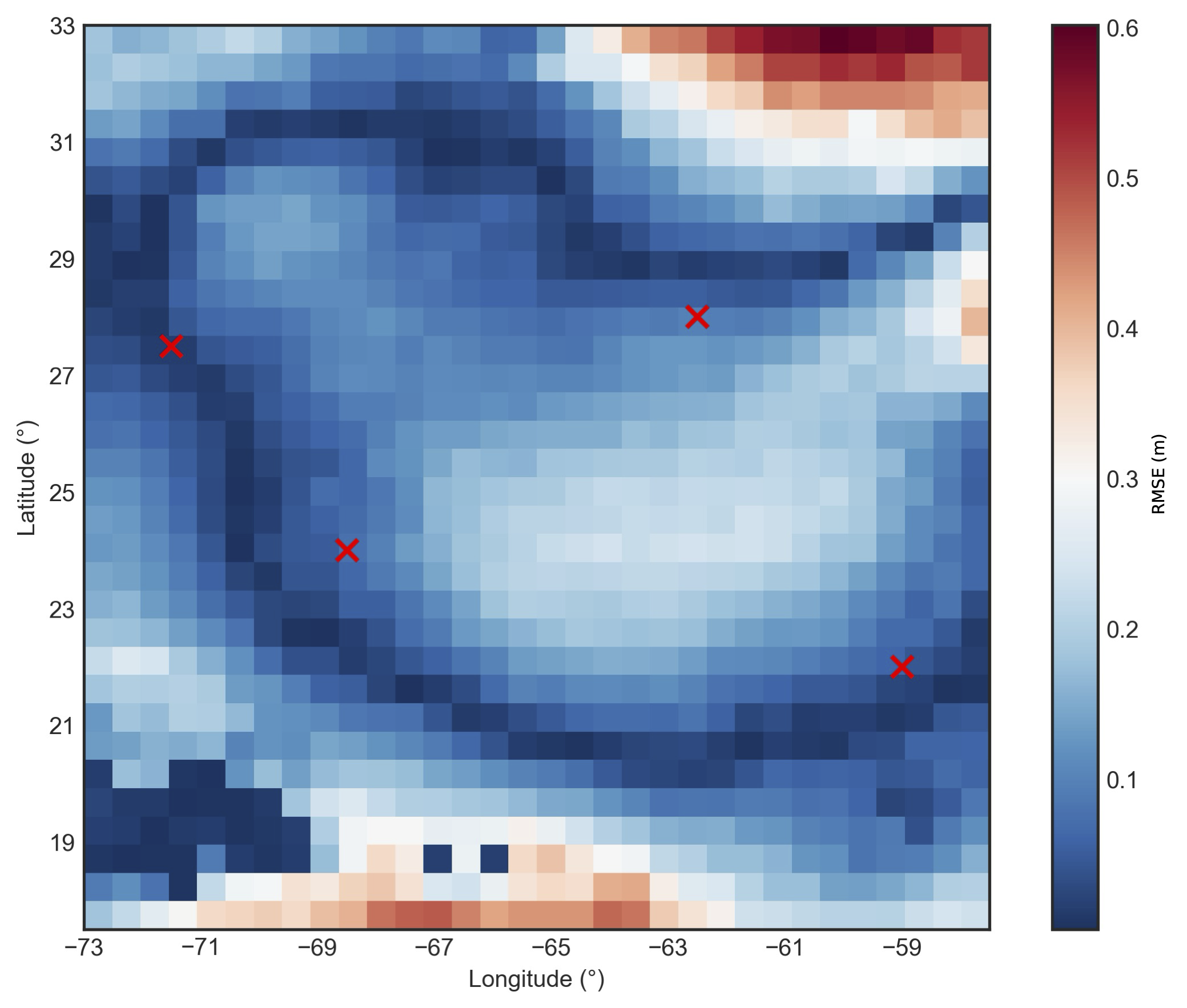}
      \innersubcaption{Error (RMSE=0.1778)}{fig:atl_e4_err}
    \end{minipage}
    \scriptsize \vspace{2pt}\par\itshape Snapshot at $t=3$
  \end{subfigure}
  \caption{Comparison of true, predicted, and error maps for the Atlantic}
  \label{fig:atlantic_all}
\end{figure}

\begin{figure}[htbp]
  \begin{subfigure}[b]{0.59\textwidth}
    \centering
    \includegraphics[width=\textwidth]{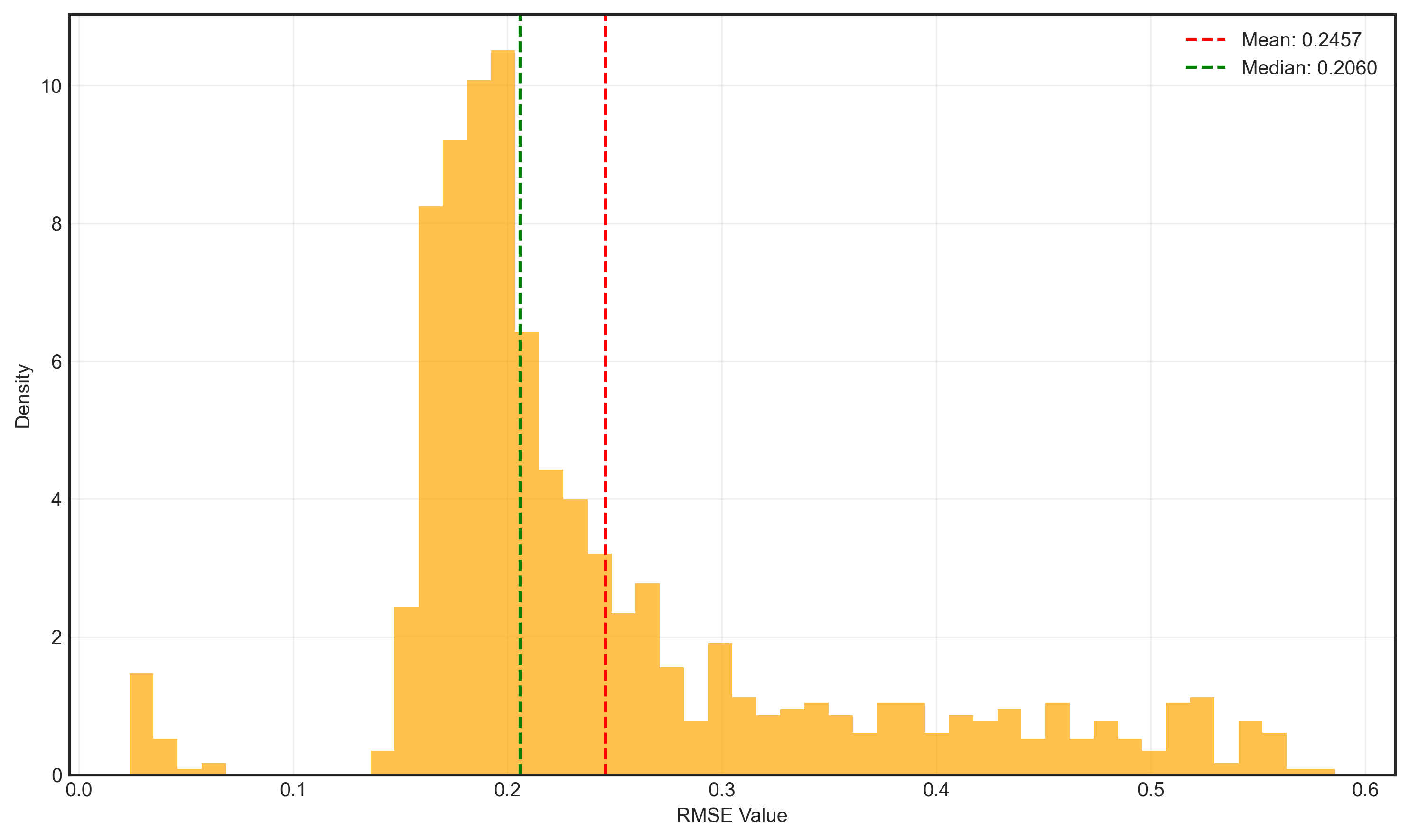}
    \caption{RMSE Distribution (Temporal)}
    \label{fig:rmse_distribution_atlantic_histogram}
  \end{subfigure}%
  \begin{subfigure}[b]{0.41\textwidth}
    \centering
    \includegraphics[width=\textwidth]{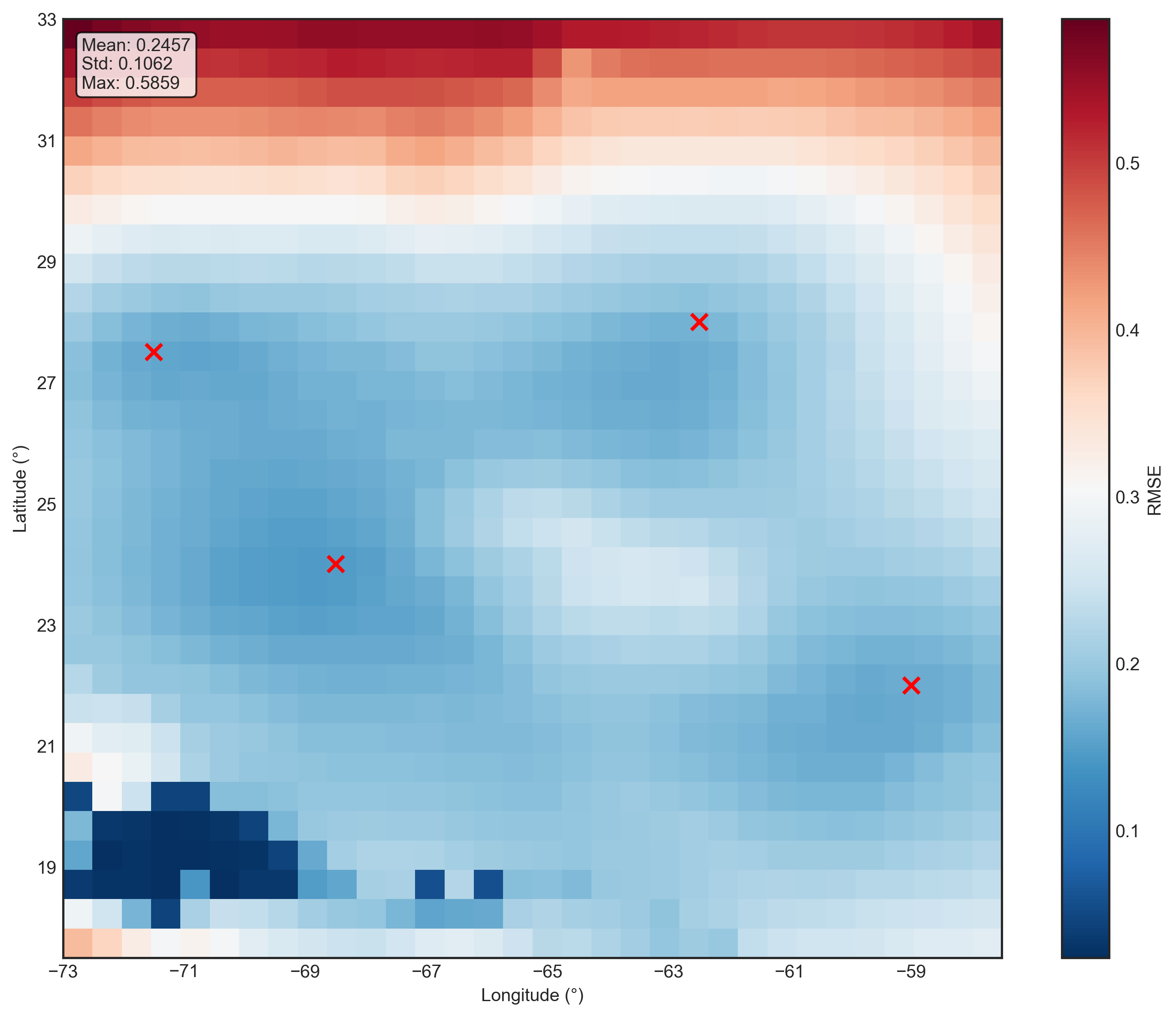}
    \caption{RMSE Distribution (Spatial)}
    \label{fig:rmse_distribution_atlantic_spatial}
  \end{subfigure}%

  \caption{RMSE distribution across the Atlantic dataset. (a) Histogram of RMSE values over all time steps. (b) Spatial distribution of RMSE values at each grid point.}
  \label{fig:rmse_distribution_atlantic}
\end{figure}

For the Atlantic, representative snapshots (Figures~\ref{fig:atl_snapshot1}--\ref{fig:atl_snapshot4}) show that errors are typically lower in buoy-proximate regions and increase with distance. The worst-case RMSE among these examples is 0.265 (Figure~\ref{fig:atl_snapshot3}), while several scenes achieve RMSE values of approximately 0.18--0.21. The dataset-wide RMSE histogram and spatial mean error map (Figure~\ref{fig:rmse_distribution_atlantic}) exhibit a slightly right-skewed distribution and geographically coherent error hotspots. Quantitatively, AUWave achieved a basin-wide RMSE of 0.286, outperforming RWR, which obtained 0.354.

\begin{figure}[p]
  \centering
  \begin{subfigure}[t]{\textwidth}
    \refstepcounter{subfigure}\label{fig:pac_snapshot1}
    \setcounter{innersubfigure}{0}
    \centering
    \begin{minipage}[t]{0.32\textwidth}
      \centering
      \includegraphics[width=\linewidth]{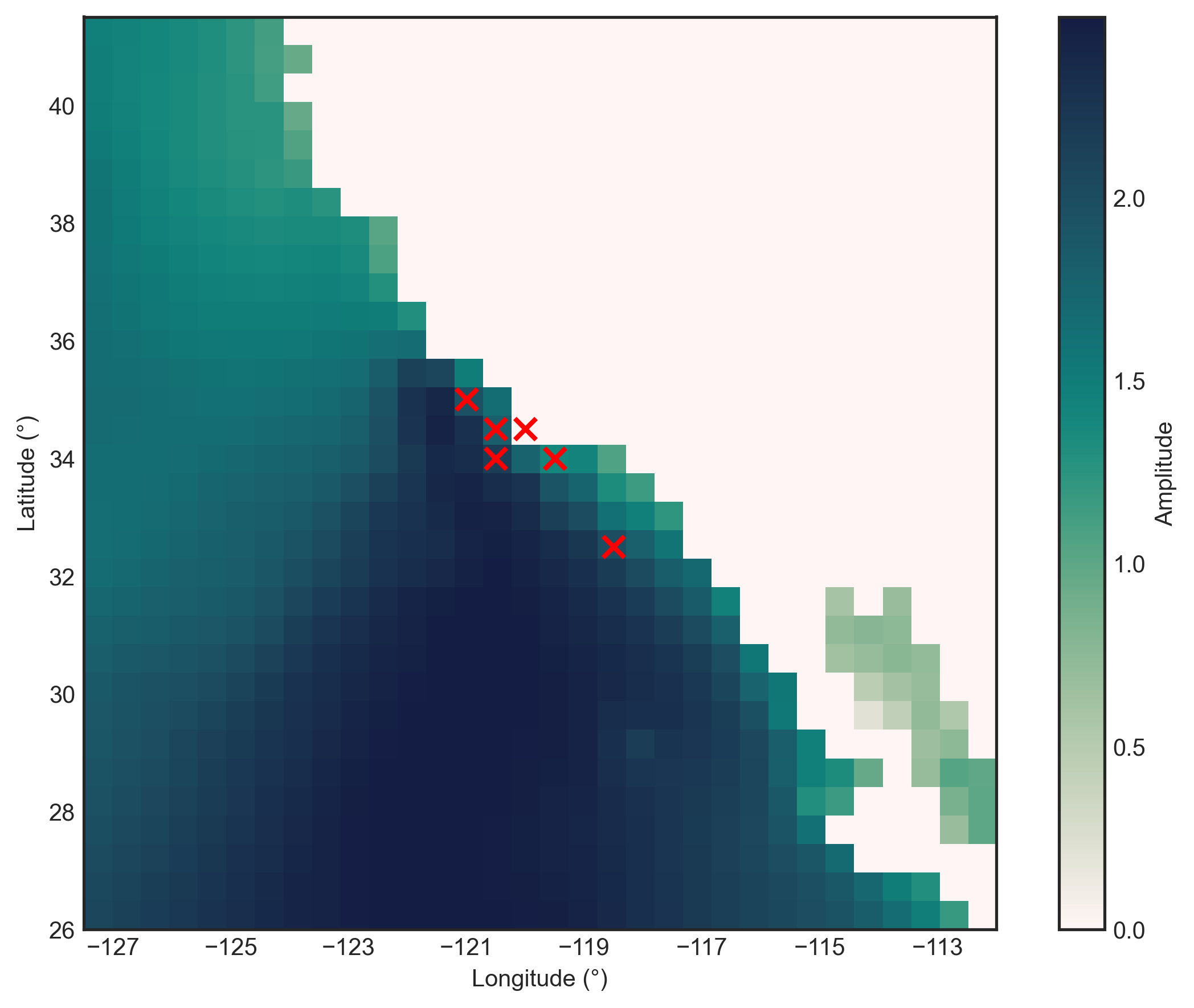}
      \innersubcaption{True}{fig:pac_e1_true}
    \end{minipage}\hfill
    \begin{minipage}[t]{0.32\textwidth}
      \centering
      \includegraphics[width=\linewidth]{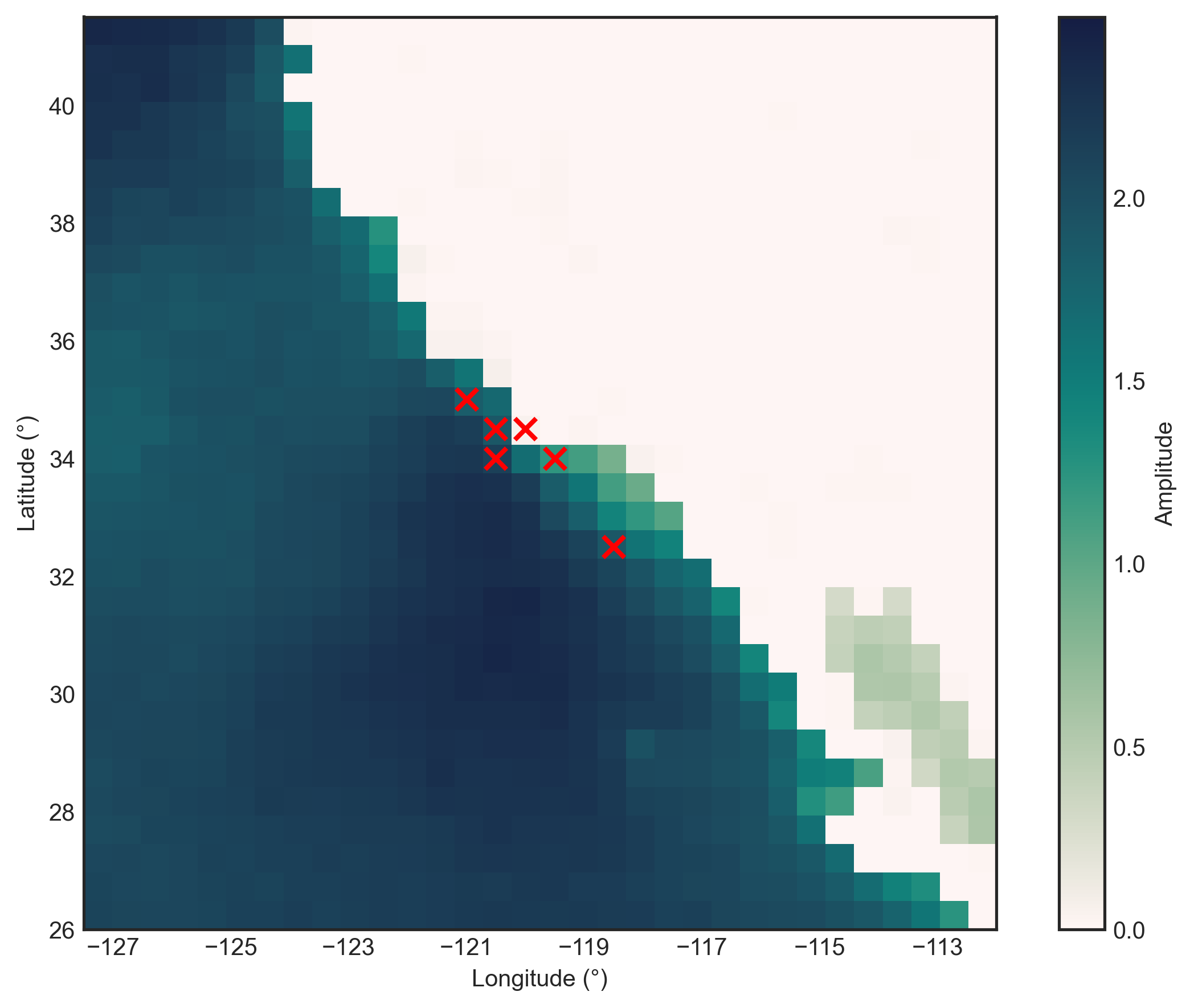}
      \innersubcaption{Predicted}{fig:pac_e1_pred}
    \end{minipage}\hfill
    \begin{minipage}[t]{0.32\textwidth}
      \centering
      \includegraphics[width=\linewidth]{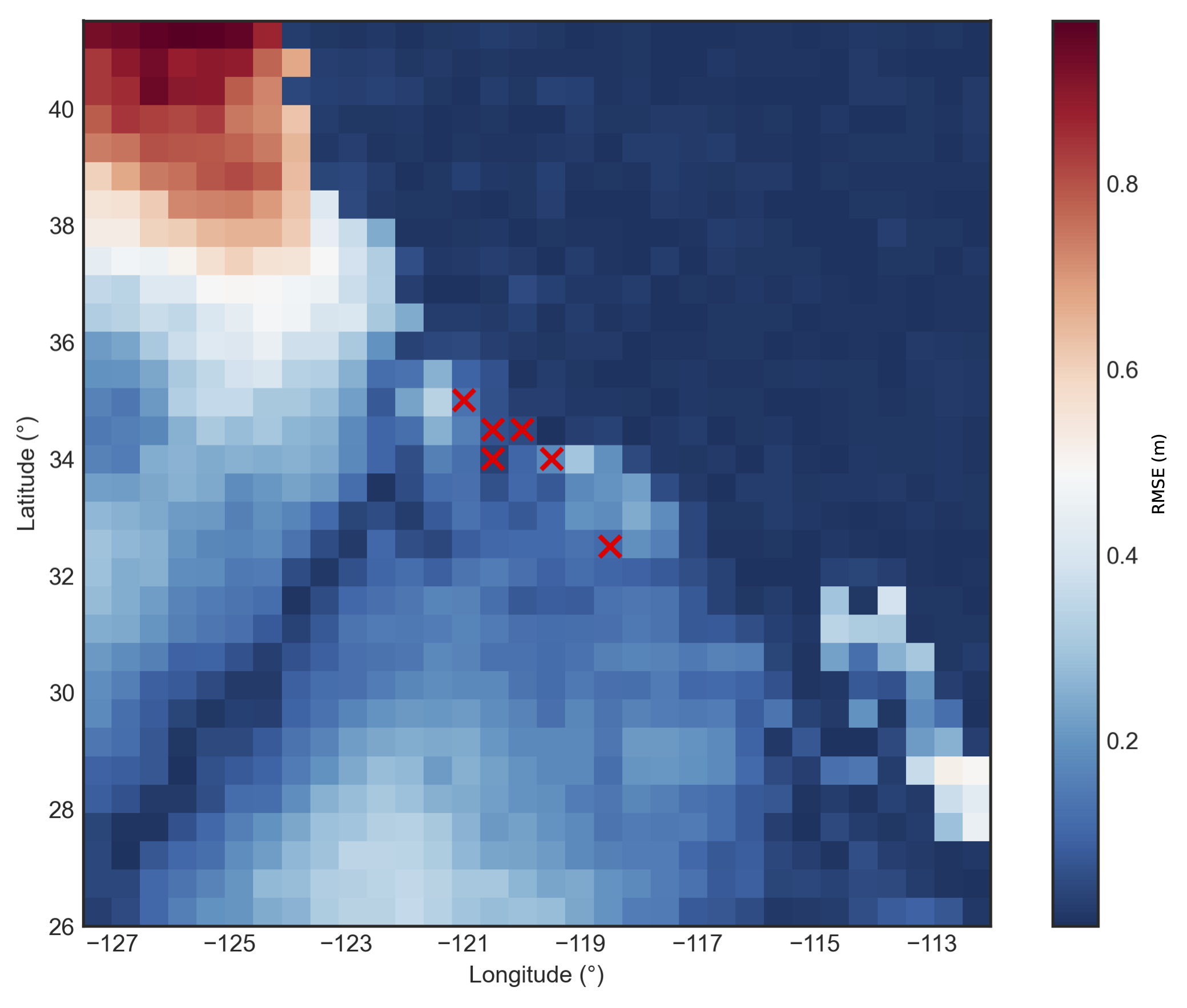}
      \innersubcaption{Error (RMSE=0.2474)}{fig:pac_e1_err}
    \end{minipage}
    \scriptsize \vspace{2pt}\par\itshape Snapshot at time step 1
  \end{subfigure}

  \begin{subfigure}[t]{\textwidth}
    \refstepcounter{subfigure}\label{fig:pac_snapshot2}
    \setcounter{innersubfigure}{0}
    \centering
    \begin{minipage}[t]{0.32\textwidth}
      \includegraphics[width=\linewidth]{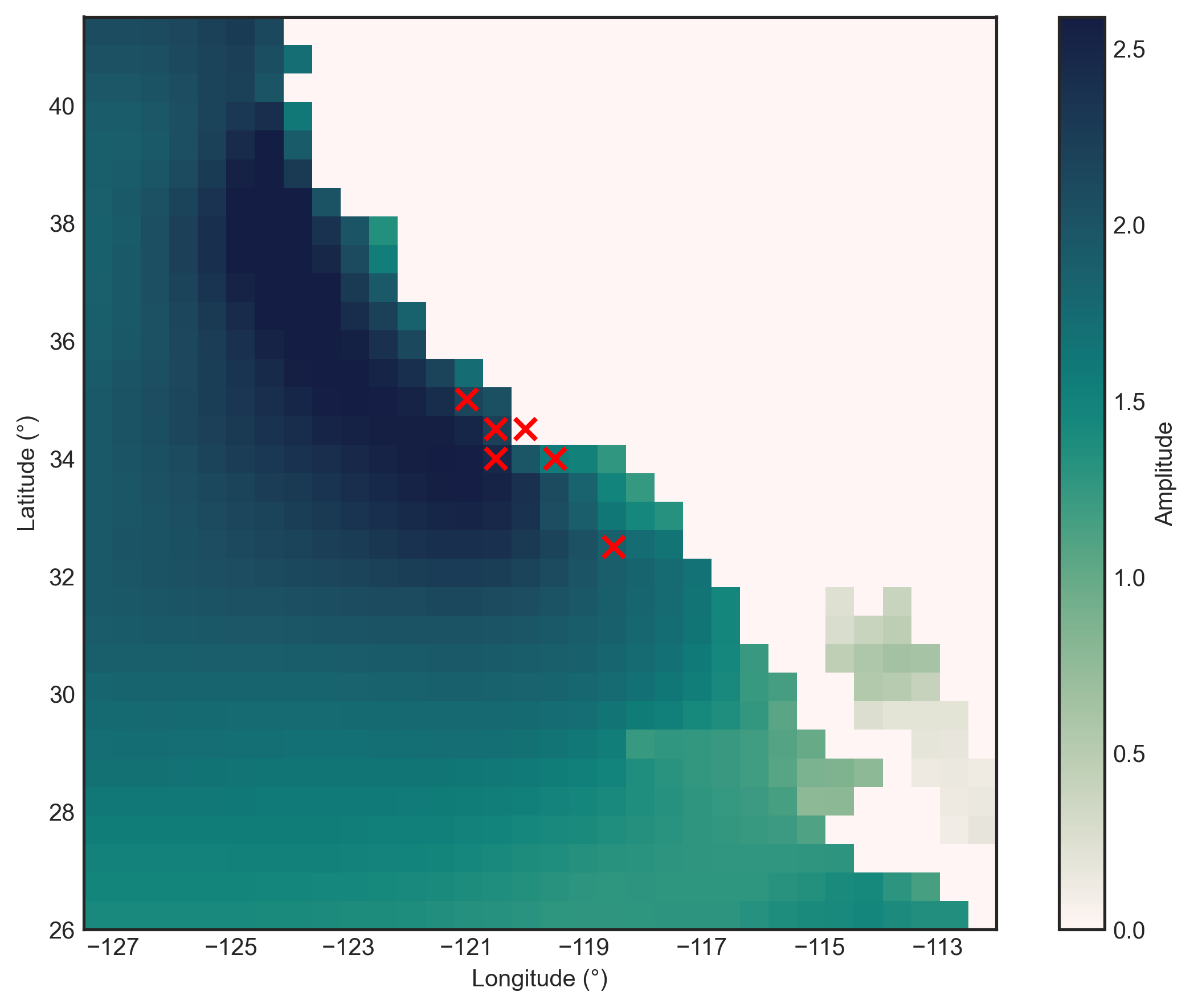}
      \innersubcaption{True}{fig:pac_e2_true}
    \end{minipage}\hfill
    \begin{minipage}[t]{0.32\textwidth}
      \includegraphics[width=\linewidth]{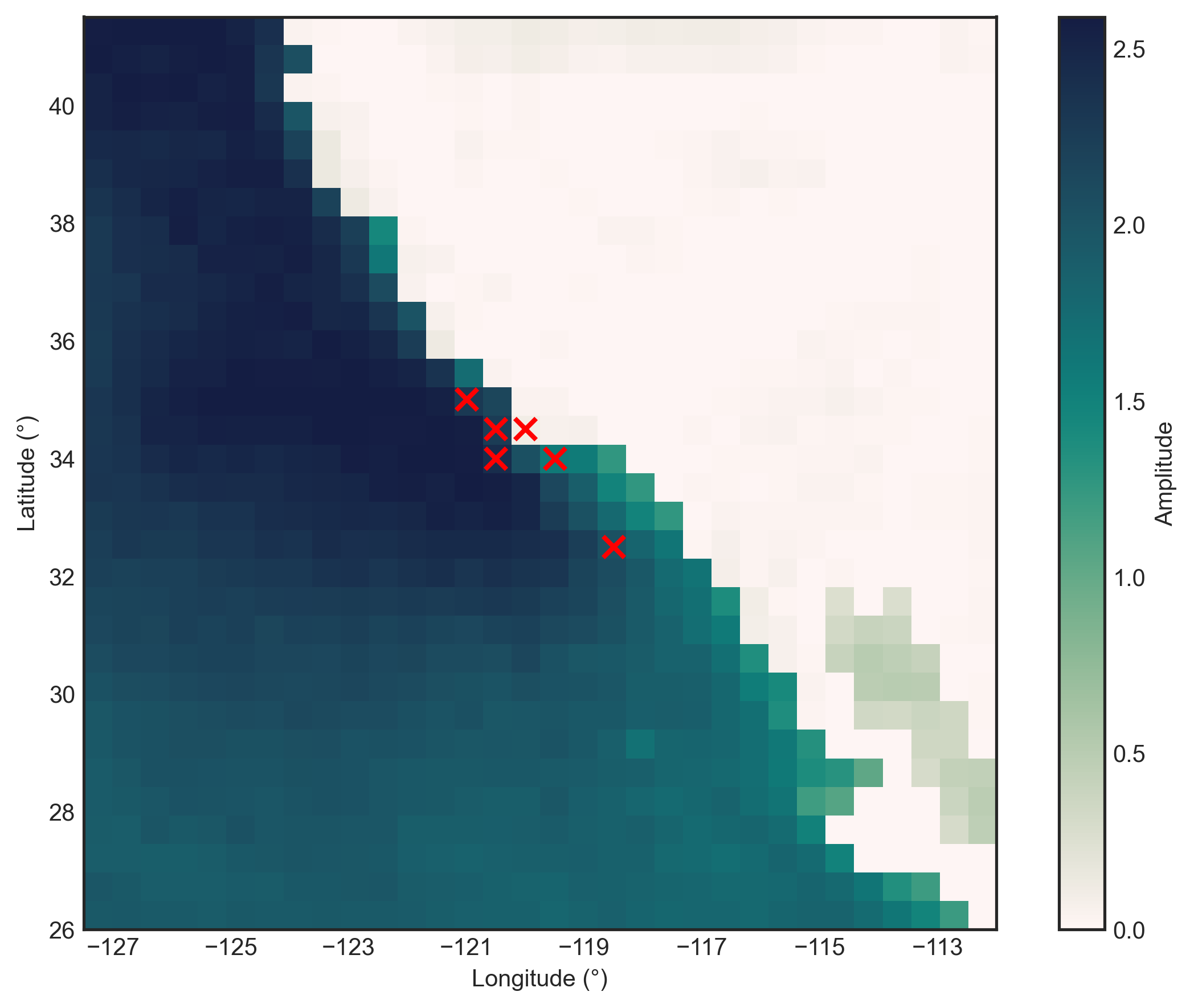}
      \innersubcaption{Predicted}{fig:pac_e2_pred}
    \end{minipage}\hfill
    \begin{minipage}[t]{0.32\textwidth}
      \includegraphics[width=\linewidth]{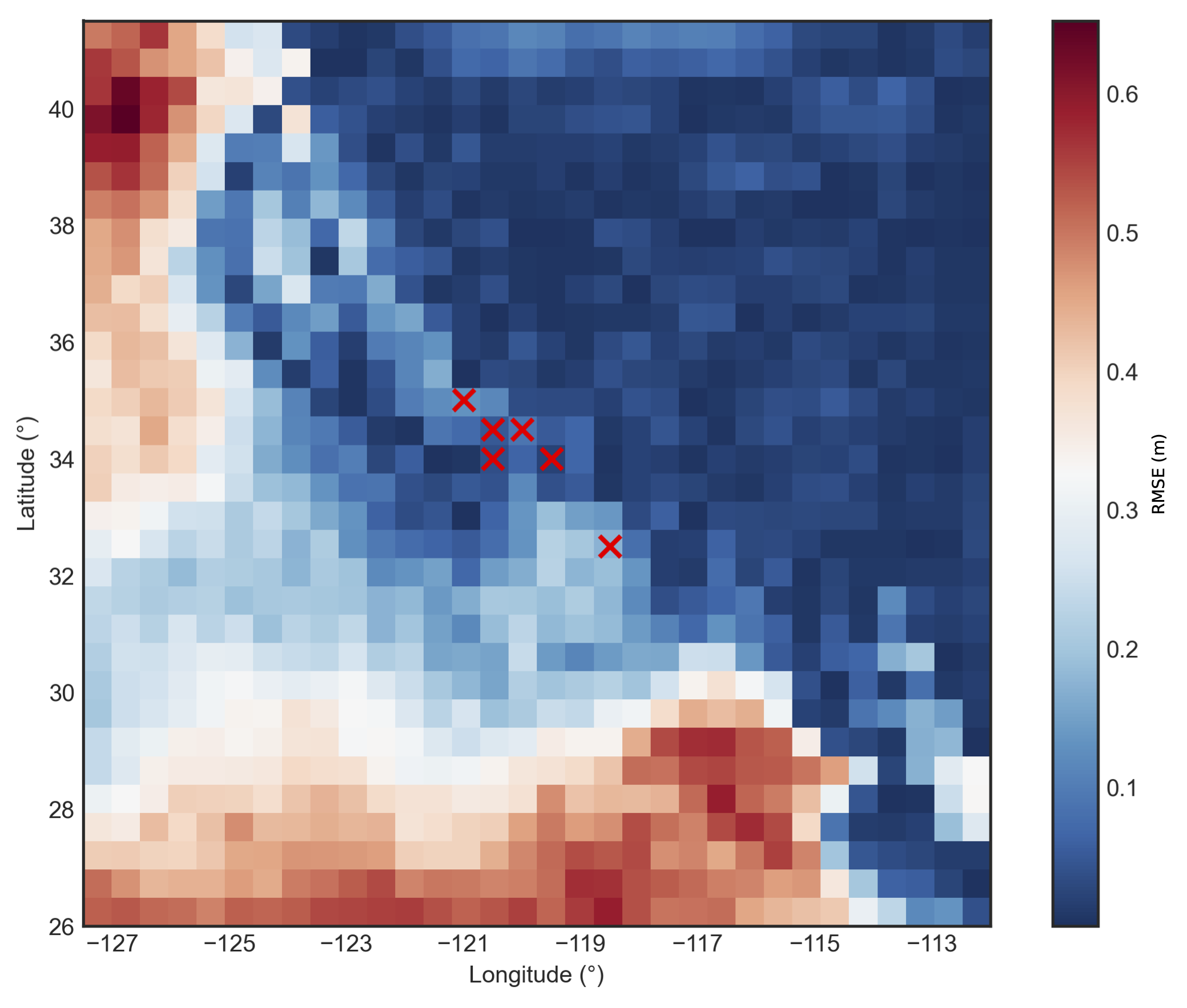}
      \innersubcaption{Error (RMSE=0.2545)}{fig:pac_e2_err}
    \end{minipage}
    \scriptsize \vspace{2pt}\par\itshape Snapshot at time step 2
  \end{subfigure}

  \begin{subfigure}[t]{\textwidth}
    \refstepcounter{subfigure}\label{fig:pac_snapshot3}
    \setcounter{innersubfigure}{0}
    \centering
    \begin{minipage}[t]{0.32\textwidth}
      \includegraphics[width=\linewidth]{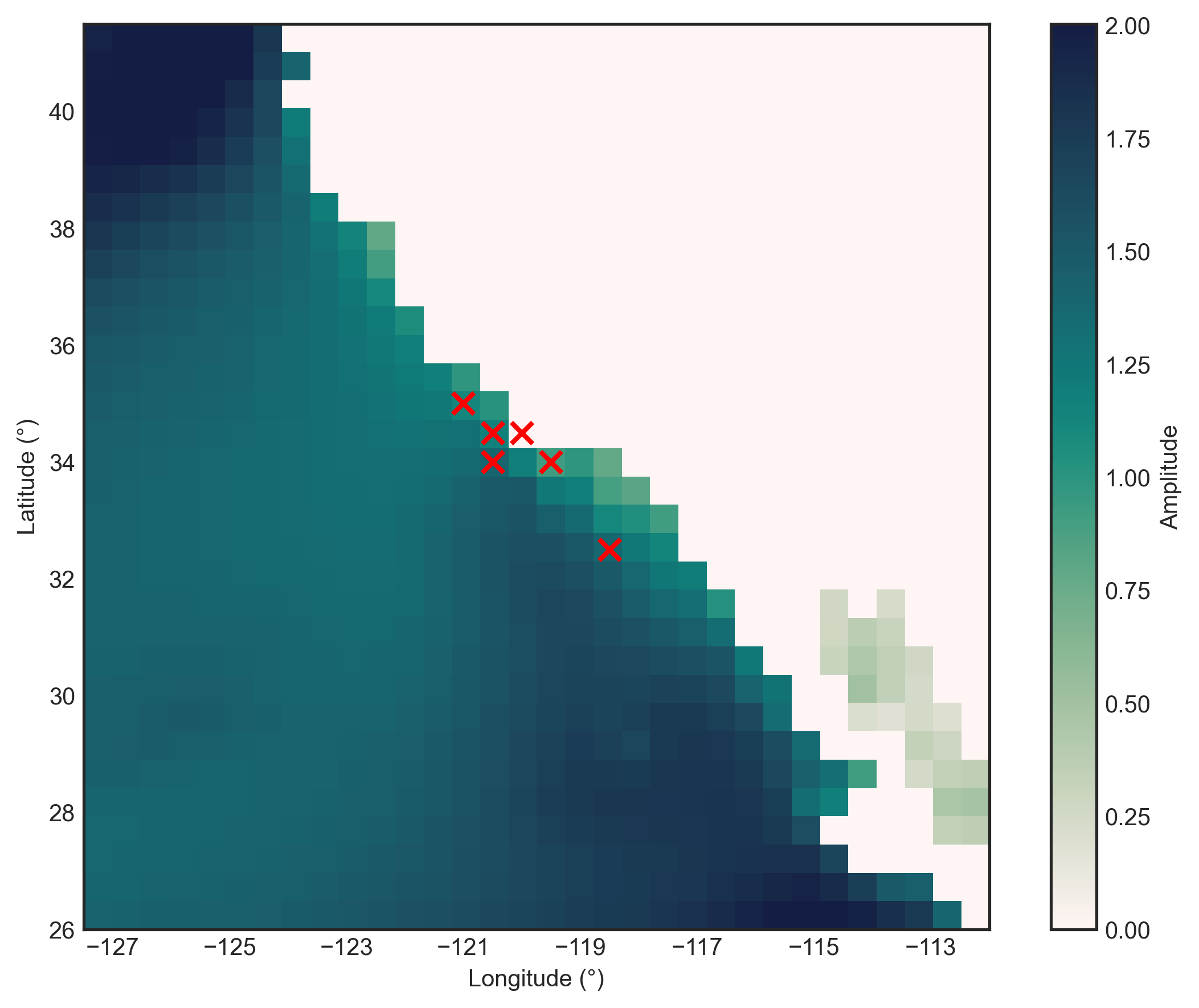}
      \innersubcaption{True}{fig:pac_e3_true}
    \end{minipage}\hfill
    \begin{minipage}[t]{0.32\textwidth}
      \includegraphics[width=\linewidth]{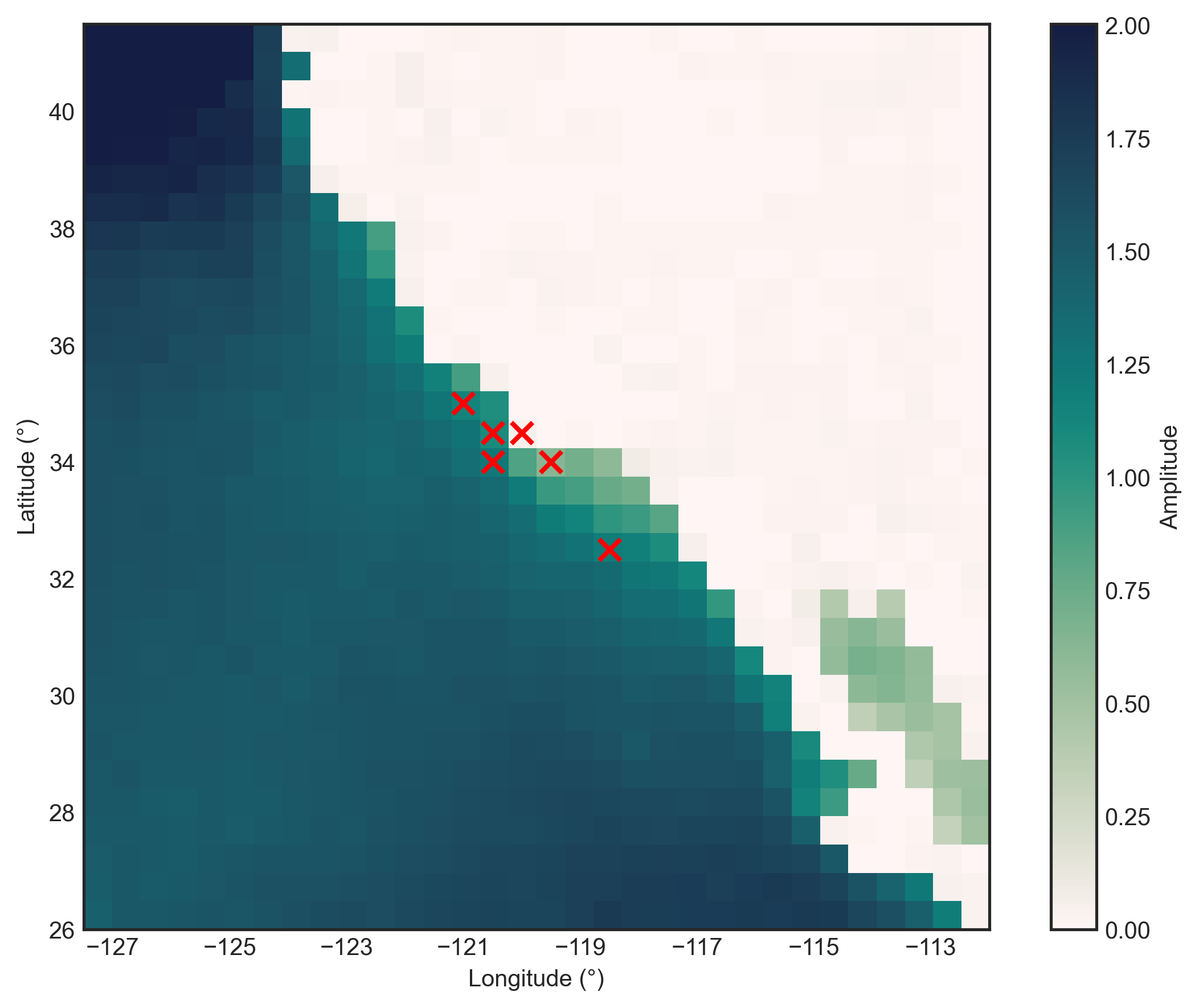}
      \innersubcaption{Predicted}{fig:pac_e3_pred}
    \end{minipage}\hfill
    \begin{minipage}[t]{0.32\textwidth}
      \includegraphics[width=\linewidth]{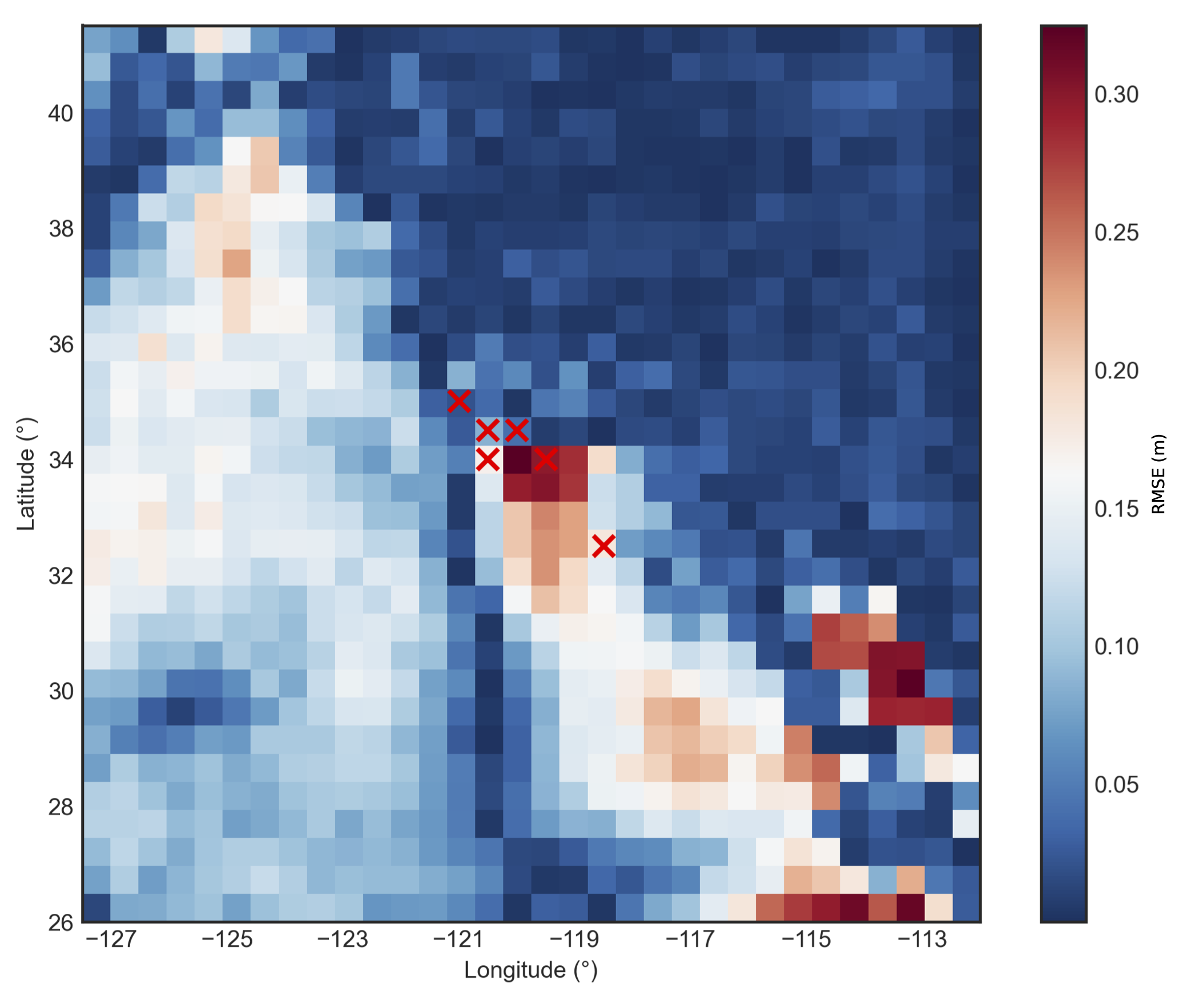}
      \innersubcaption{Error (RMSE=0.1020)}{fig:pac_e3_err}
    \end{minipage}
    \scriptsize \vspace{2pt}\par\itshape Snapshot at time step 3
  \end{subfigure}

  \begin{subfigure}[t]{\textwidth}
    \refstepcounter{subfigure}\label{fig:pac_snapshot4}
    \setcounter{innersubfigure}{0}
    \centering
    \begin{minipage}[t]{0.32\textwidth}
      \includegraphics[width=\linewidth]{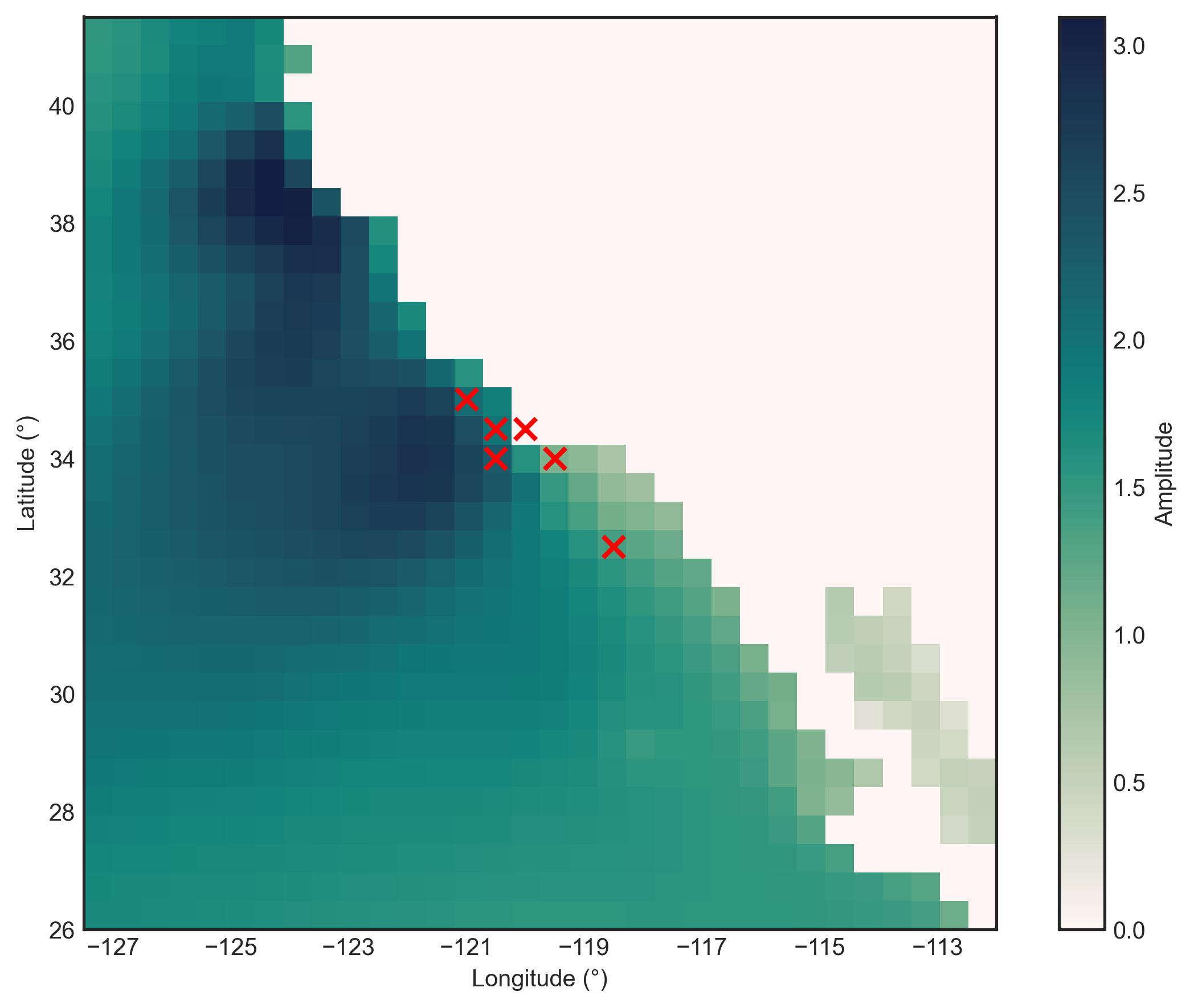}
      \innersubcaption{True}{fig:pac_e4_true}
    \end{minipage}\hfill
    \begin{minipage}[t]{0.32\textwidth}
      \includegraphics[width=\linewidth]{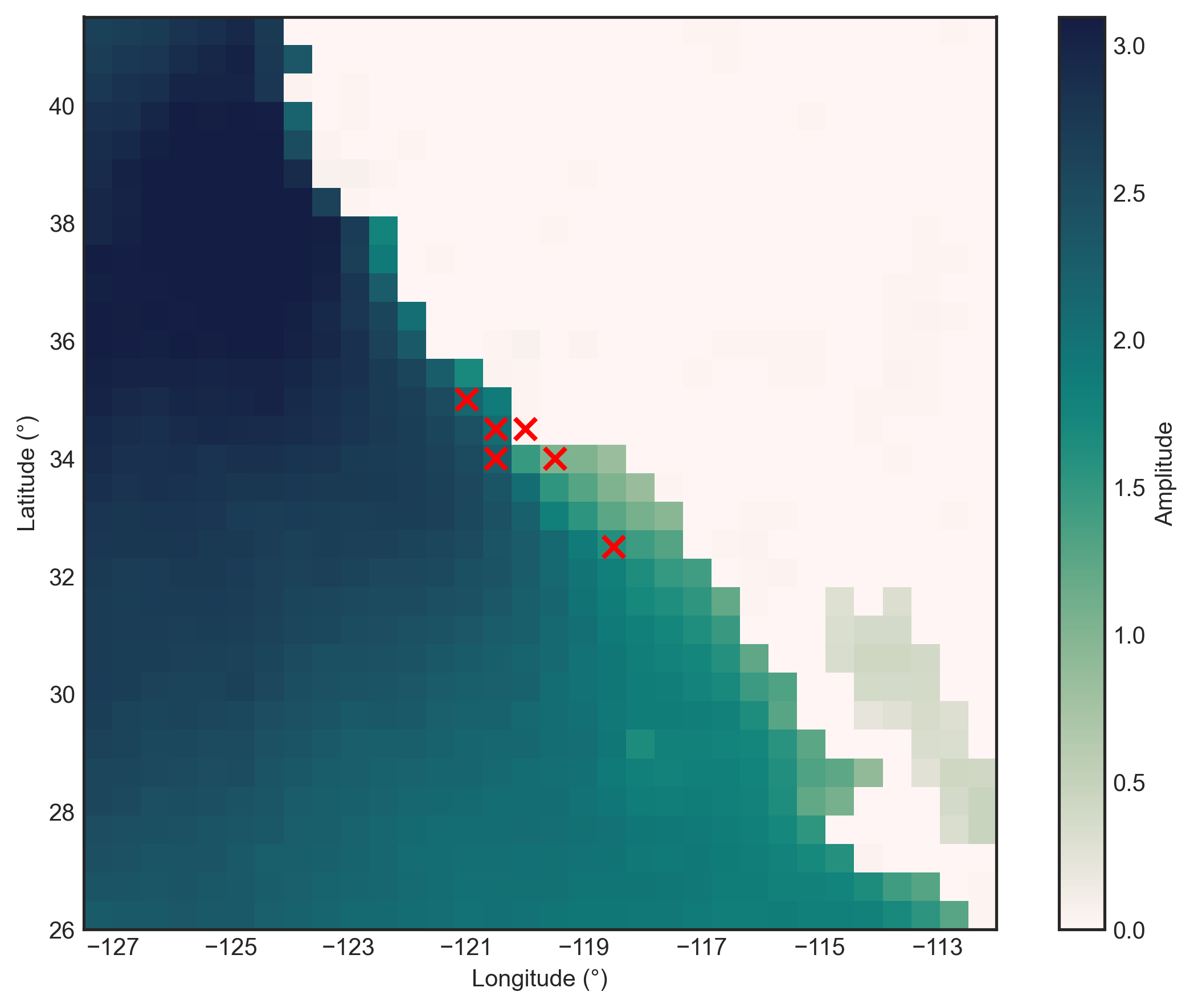}
      \innersubcaption{Predicted}{fig:pac_e4_pred}
    \end{minipage}\hfill
    \begin{minipage}[t]{0.32\textwidth}
      \includegraphics[width=\linewidth]{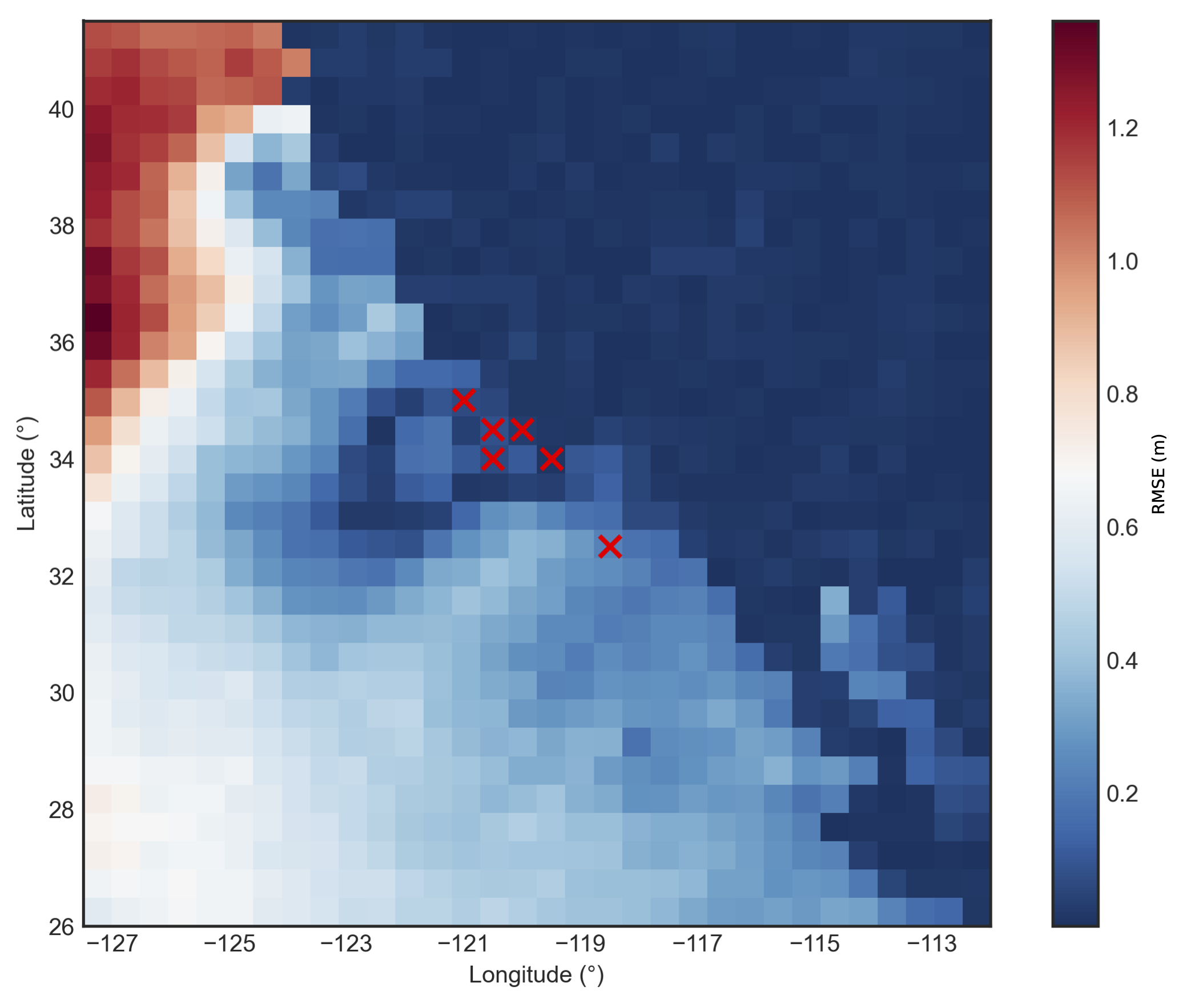}
      \innersubcaption{Error (RMSE=0.4110)}{fig:pac_e4_err}
    \end{minipage}
    \scriptsize \vspace{2pt}\par\itshape Snapshot at time step 4
  \end{subfigure}

  \caption{Comparison of true and predicted wave fields for Pacific}
  \label{fig:pacific_all}
\end{figure}

\begin{figure}[htbp]
  \begin{subfigure}[b]{0.59\textwidth}
    \centering
    \includegraphics[width=\textwidth]{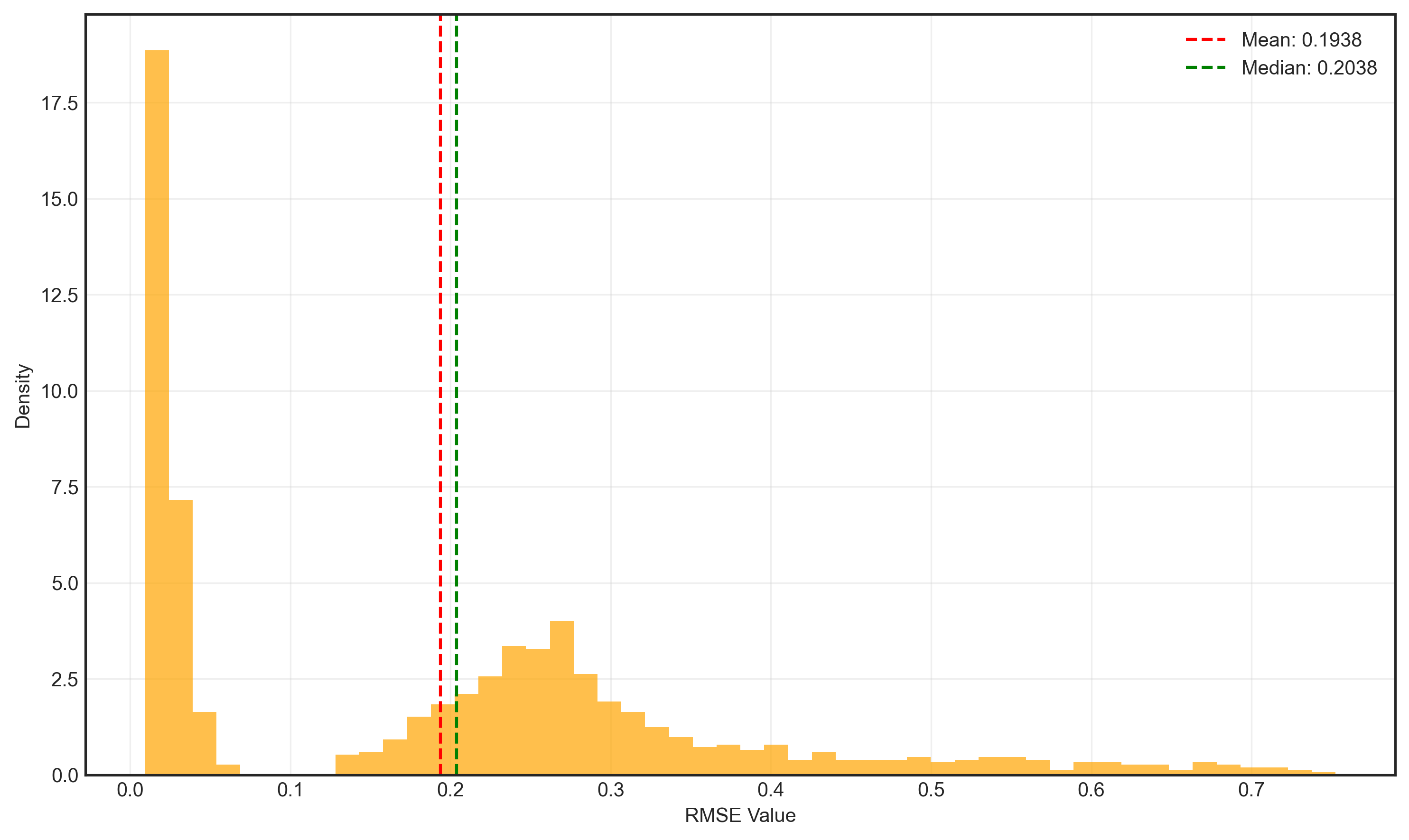}
    \caption{RMSE Distribution (Temporal)}
    \label{fig:rmse_distribution_pacific_histogram}
  \end{subfigure}%
  \begin{subfigure}[b]{0.41\textwidth}
    \centering
    \includegraphics[width=\textwidth]{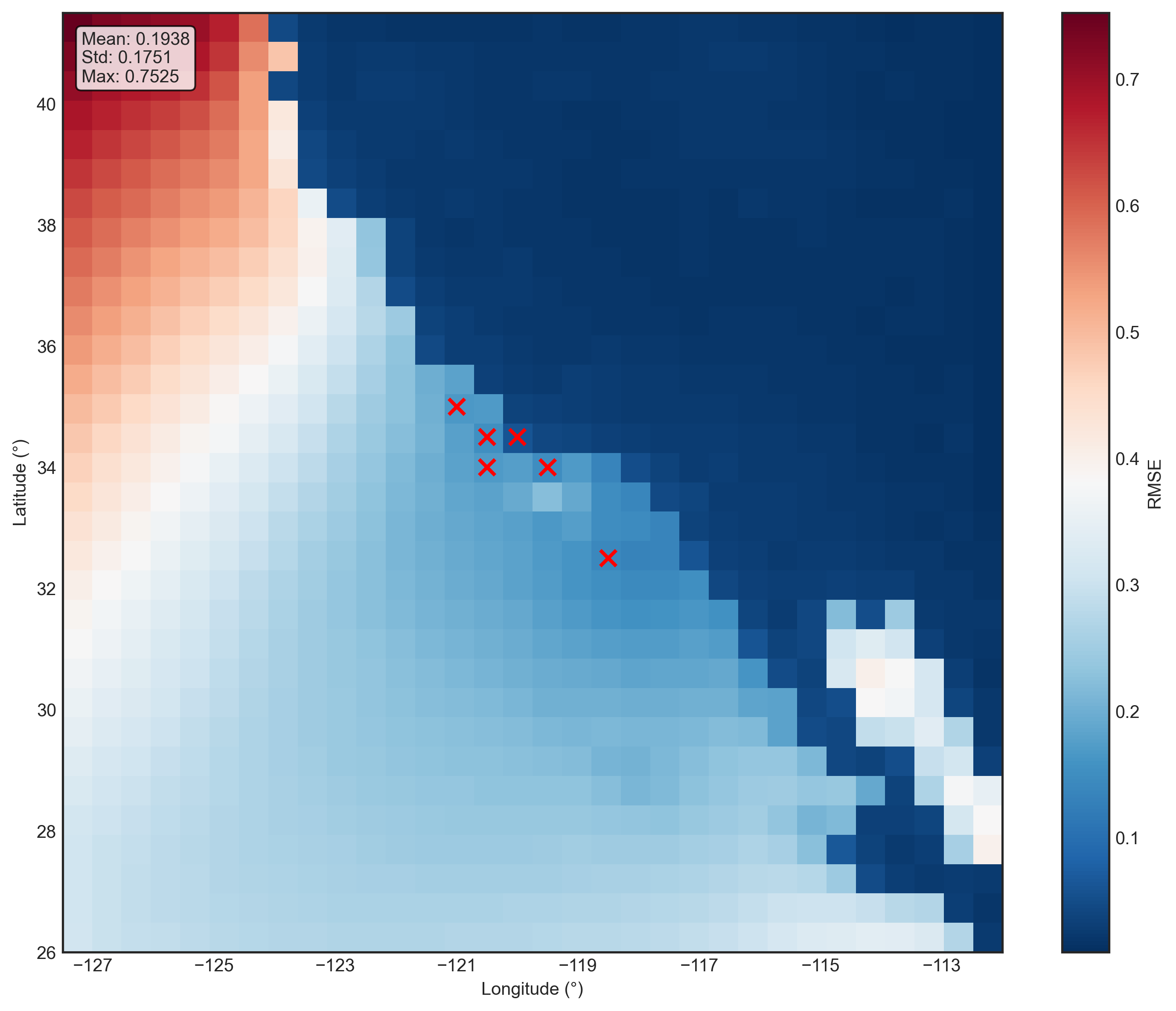}
    \caption{RMSE Distribution (Spatial)}
    \label{fig:rmse_distribution_pacific_spatial}
  \end{subfigure}%

  \caption{RMSE distribution across the Pacific dataset. (a) Histogram of RMSE values over all time steps. (b) Spatial distribution of RMSE values at each grid point.}
  \label{fig:rmse_distribution_pacific}
\end{figure}

For the Pacific, a similar evaluation procedure was applied. Four representative examples are presented in Figures~\ref{fig:pac_snapshot1}--\ref{fig:pac_snapshot4}, with dataset-level error distributions summarized in Figure~\ref{fig:rmse_distribution_pacific}. In this basin, AUWave reached an RMSE of 0.307 compared to RWR's 0.333, again confirming the relative advantage of AUWave.

Overall, the results indicate that the Hawaii-trained AUWave model retains strong transferability across basins. While both models experience some performance degradation in data-sparse regions, AUWave consistently achieves lower errors than RWR, highlighting its robustness across diverse oceanic domains.

\section{Discussion}
\label{sec:discussion}
This study investigated the problem of reconstructing ocean wave fields from sparse buoy observations and introduced AUWave, an attention-enhanced U-Net framework with station-wise encoding. Across extensive experiments, AUWave demonstrated robust accuracy, outperforming the RWR baseline under most buoy configurations and maintaining transferability across distinct ocean basins. Beyond performance metrics, the analyses revealed several important insights: (1) reconstruction errors are strongly shaped by buoy geometry, with anchor placement emerging as a critical determinant of accuracy; (2) hyperparameters, particularly learning rate schedules, play a larger role than architectural variations, suggesting that optimization strategies outweigh complex structural redesign; and (3) AUWave captures both large-scale coherence and local gradients more effectively than RWR, enabling superior reconstructions in data-rich settings while retaining resilience under sparsity. Together, these findings position AUWave as a practical and generalizable approach for filling observational gaps, supporting operational forecasting, and guiding the design of future buoy networks.

\subsection{Model performance and error characteristics}
This study demonstrates that AUWave can robustly reconstruct $32 \times 32$ regional SWH fields from sparse buoy observations. With systematic hyperparameter optimization (Figure~\ref{fig:optuna-results}), the model reached a minimum validation loss of 0.0433. Over the Hawaii test domain, the RMSE distribution is slightly right-skewed, with a mean of 0.2017 and a median of 0.1887 (Figure~\ref{fig:rmse_distribution_hawaii}). Case studies (Figures~\ref{fig:hawaii_ex1}--\ref{fig:hawaii_ex5}) further illustrate that AUWave effectively reproduces dominant spatial structures. In low-error cases (RMSE $\approx$0.15–0.16), predictions closely align with the ground truth; even in the highest-error case (RMSE $\approx$0.36), large-scale wave patterns persist, though local amplitudes deviate systematically. Spatially, errors are lowest near buoy locations and increase away from anchors (Figure~\ref{fig:rmse_distribution_hawaii_spatial}), reflecting the identifiability limits imposed by sparse sensing. The long tail of the RMSE histogram corresponds to a small subset of difficult scenes, typically associated with unfavorable sampling geometry, short-term anomalies, or extreme conditions.

\subsection{Role of hyperparameters and architecture}
Our hyperparameter importance analysis (Figure~\ref{fig:optuna-results}) confirms that training dynamics—particularly the learning rate schedule—are the dominant factors shaping reconstruction accuracy. This finding is consistent with widely reported results in neural network optimization. By contrast, architecture-related parameters such as the number of U-Net blocks, layers per block, and base channel width exerted negligible influence on performance. This suggests that the selected U-Net backbone is already sufficiently expressive for the task and does not require substantial structural modification.

Two arise from this result. First, it reduces the necessity of costly architecture searches, since most performance gains can be realized through refined optimization strategies rather than network redesign. Second, it highlights that AUWave is robust across a wide range of architectural configurations, which makes the model practical for deployment under varying computational budgets.

\subsection{Comparison with the RWR baseline}
Architecturally, AUWave differs fundamentally from the RWR baseline in its approach to fusing sparse observations. AUWave decouples temporal and station-wise encoding from hierarchical spatial refinement: a station-wise MLP first maps heterogeneous buoy observations into a spatially aligned latent grid, which is then refined by a residual U-Net with attention layers. This design enables the model to recover both broad swell patterns and localized gradients. In contrast, RWR mixes sparse inputs at an early stage in a feedforward CNN stack, enforcing a stronger smoothness prior and promoting homogeneous receptive field growth. While such bias can be beneficial under extreme sparsity by reducing variance, it limits detail recovery when richer observation sets are available.

Empirically, AUWave consistently achieves lower reconstruction errors in multi-buoy configurations (e.g., RMSE of 0.2129 vs.\ 0.2472 for RWR in the 5-buoy case), while RWR shows marginal advantages in a few single-buoy scenarios. The superiority of AUWave can be attributed to several factors: its decoupled fusion strategy improves optimization stability and gradient flow; its multi-scale pathways and bottleneck self-attention preserve high-frequency details while capturing long-range coherence; and its convolutional weight sharing offers greater sample efficiency compared to the larger FC-CNN front-end of RWR. Furthermore, the station encoder in AUWave learns station-specific weights, improving robustness under suboptimal buoy layouts and partial station dropout. Finally, AUWave incorporates a physics-aligned inductive bias, since swell propagation requires both non-local coupling and local gradient preservation—properties that are naturally captured by an attention-enhanced U-Net, but less so by RWR’s smoother isotropic CNN prior. Nevertheless, RWR retains relative advantages under extreme sparsity, suggesting that its smoothing bias could provide complementary inductive priors for hybrid future designs.

\rev{Under extreme single-station sparsity, the reconstruction task becomes strongly underdetermined, and model performance is therefore primarily constrained by the information content of the observation rather than architectural complexity, which explains the limited performance separation between AUWave and RWR in this regime.}

\subsection{Practical implications}
The results emphasize the importance of observational geometry (Table~\ref{tab:rmse_buoy_combinations_two_models}, Figure~\ref{fig:buoy_configuration_sensitivity}). AUWave fuses multi-point observations more effectively than RWR, especially with 4–5 buoys. However, both models degrade significantly when central anchors are removed; buoy 51001, in particular, proves critical for maintaining reconstruction fidelity. These findings yield practical design guidance: prioritize the maintenance of key anchor buoys, and consider relocating or adding buoys in high-RMSE zones to maximize error reduction per unit cost. Beyond guiding network design, AUWave can complement reanalysis and operational forecasting systems by (1) gap-filling sparse buoy networks, (2) providing high-resolution priors for data assimilation, and (3) enabling emergency reconstructions during observational outages.

\subsection{Cross-regional transferability}
Cross-basin experiments were conducted by re-training AUWave in the North Atlantic and North Pacific using the same architecture and hyperparameters selected from the Hawaii study, without further tuning. This setting evaluates the stability and generalizability of the model design under distinct oceanic regimes.

In the Atlantic, qualitative case studies (Figures~\ref{fig:atl_snapshot1}--\ref{fig:atl_snapshot4}) and dataset-level diagnostics (Figure~\ref{fig:rmse_distribution_atlantic}) reveal error structures consistent with Hawaii: a right-skewed RMSE distribution and spatial error gradients aligned with buoy density. Example-wise RMSEs of $\approx$0.18-0.26 suggest that the learned architecture and training recipe remain effective under different wind-wave regimes. Basin-level results further show that AUWave achieves an RMSE of 0.286, outperforming RWR's 0.354.

In the Pacific, AUWave similarly outperforms RWR, achieving 0.307 versus 0.333 RMSE. While the overall distribution follows the same trends as Hawaii and Atlantic, errors exhibit larger variance during extreme events, which likely reflects regime differences in storm track variability, fetch, and a sparser buoy network.
These results imply that AUWave's architecture and optimization strategy, once selected, can be reused across basins without the need for costly re-tuning. Although physical regime differences still introduce variability, the model demonstrates consistent performance advantages over RWR, highlighting its robustness and practical portability across diverse geophysical domains.

\subsection{Limitations}
Several limitations must be acknowledged. First, ERA5 reanalysis is treated as ground truth, despite known systematic biases and a 0.5$^\circ$ resolution that underrepresents mesoscale and coastal variability. Second, although evaluation extends beyond Hawaii to the Atlantic and Pacific, spatial and temporal coverage remains limited and does not exhaust seasonal or extreme conditions. Third, the evaluation metrics emphasize single-time RMSE and do not explicitly account for dynamical or spectral consistency. Finally, the AUWave framework imposes no explicit physical constraints, so non-physical oscillations may emerge under extreme scenarios. In addition, as a purely data-driven model, AUWave provides limited interpretability compared to physics-based approaches, which may hinder operational adoption.

\subsection{Future directions}
Future work will extend AUWave along several axes:
\begin{enumerate}
    \item \textbf{Incorporate} physics priors (e.g., spectral consistency, PDE-based soft constraints) to mitigate MSE bias and enforce dynamical plausibility.
    \item \textbf{Adopt} spatiotemporal architectures that explicitly model wave evolution, enabling forecasting as well as reconstruction.
    \item \textbf{Quantify} uncertainty using Bayesian inference, ensembles, or Monte Carlo dropout, to better inform risk-sensitive applications.
    \item \textbf{Fuse} multimodal inputs (e.g., winds, spectra, altimetry, SAR) to enhance generalization and improve reconstructions under sparse conditions.
    \item \textbf{Explore} domain adaptation strategies to improve robustness across regions and sea-state regimes, including transfer learning and meta-learning.
    \item \textbf{Couple} active learning with error hotspot maps to guide buoy deployment strategies, closing the loop between observation and modeling.
\end{enumerate}

\section{Conclusion}
\label{sec:conclusion}

In this study, we introduced AUWave, a hybrid deep learning framework designed to reconstruct high-resolution regional wave fields from sparse buoy observations. By integrating a sequence-encoding MLP with a U-Net featuring bottleneck self-attention, AUWave effectively fuses temporal information from individual stations with their broader spatial context. Applied to the Hawaiian region using NDBC and ERA5 data, the optimized model demonstrated robust and accurate performance in reconstructing SWH fields.

The model achieved high-fidelity reconstructions, with errors that were consistently low and stable across the study domain. Our analysis revealed several key findings: (1) Reconstruction accuracy is highest near observation points and predictably degrades with distance, highlighting the inherent limits of sparse sensing. (2) AUWave substantially outperformed a representative baseline method, particularly in multi-buoy configurations, demonstrating the value of its architectural design. (3) The study underscored the critical influence of buoy placement on overall accuracy, identifying key “anchor” buoys essential for domain-wide performance. Additionally, when retrained with fixed hyperparameters in the Atlantic and Pacific basins, AUWave consistently outperformed the RWR baseline, confirming the stability and portability of the proposed design across distinct oceanic regimes.

These findings carry significant practical implications. AUWave can serve as a valuable tool to complement existing ocean observation and forecasting systems by: (i) gap-filling data in sparse buoy networks, (ii) providing high-resolution initial fields for data assimilation models, and (iii) enabling rapid reconstruction during satellite or sensor outages. Furthermore, the model's learned error patterns can offer actionable guidance for optimizing the design and maintenance of future ocean observing networks.

This work is primarily limited by its reliance on ERA5 reanalysis as ground truth, its regional focus, and the use of a pointwise loss objective without explicit physical constraints. Future research will aim to incorporate physics-informed learning objectives, develop spatiotemporal models for wave evolution forecasting, quantify reconstruction uncertainty, and fuse multi-modal data such as satellite altimetry and wind fields. Exploring cross-region transfer learning and coupling the model with active learning for guided buoy deployment are also promising directions. Taken together, these results demonstrate that AUWave provides a robust, generalizable, and practically relevant framework for regional wave field reconstruction under sparse observations.
 
% \appendix
% \section{Example Appendix Section}
% \label{app1}

% Appendix text.

\bibliographystyle{elsarticle-harv}
\bibliography{manuscript.bib}

\section{Acknowledgments}
\label{sec:acknowledgments}
The author(s) declare financial support was received for the research, authorship, and/or publication of this article. This research was supported by the Shandong Provincial Lab and Talent Program ("Double Hundred Plan for Oversees Experts" Talent Category, Grant No. WSR2024073 and No. WSR2023026). The authors would like to thank the anonymous reviewers for their valuable feedback and suggestions, which greatly improved the quality of this paper. We also acknowledge the support of the research community and institutions that provided access to the datasets used in this study.

\section{Author Contributions}
\label{sec:author-contributions}
Yilin Zhai was responsible for the overall conceptualization, data analysis, experimental design, and drafting of the manuscript; Hongyuan Shi participated in research discussions, provided valuable guidance and revisions; Chao Zhan reviewed the overall direction of the paper and offered constructive comments;  Qing Wang carefully reviewed the manuscript and provided feedback and suggestions;  Zaijin You assisted with data discussions and contributed to several sections of the manuscript; and Ping Dong participated in reviewing and revising the final draft.  All authors have read and approved the final version of the manuscript.

\section{Conflict of Interest}
\label{sec:conflict-of-interest}
The authors declare that the research was conducted in the absence of any commercial or financial relationships that could be construed as a potential conflict of interest.
\end{document}